\documentclass[]{aastex631}
\let\tablenum\relax
\usepackage{amsmath, multirow} 
\usepackage{mathrsfs}
\usepackage[scr=rsfs,cal=boondox]{mathalfa}

\usepackage{CJKutf8}
\usepackage{threeparttable}

\begin{document}

\title{A Multi-Species Atmospheric Escape Model with Excited Hydrogen and Helium: Application to HD209458b}

\author[0009-0003-5882-9663]{Anna Ruth Taylor} 
\affiliation{Lunar and Planetary Laboratory, University of Arizona, Tucson, AZ 85721, USA}

\author[0000-0003-3071-8358]{Tommi T. Koskinen}
\affiliation{Lunar and Planetary Laboratory, University of Arizona, Tucson, AZ 85721, USA}

\author[0000-0003-1093-4379]{Luca Argenti}
\affiliation{The University of Central Florida, Department of Physics, Orlando, FL 32816, USA}

\author[0000-0001-9580-3487
]{Nicholas Lewis}
\affiliation{The University of Central Florida, Department of Physics, Orlando, FL 32816, USA}

\author[0000-0001-9446-6853]{Chenliang Huang \begin{CJK*}{UTF8}{gbsn}(黄辰亮)\end{CJK*}}
\affiliation{Shanghai Astronomical Observatory, Chinese Academy of Sciences, Shanghai 200030, China}

\author[0000-0001-5097-4784]{Anthony Arfaux}
\affiliation{Groupe de Spectrométrie Moléculaire et Atmosphérique, Université de Reims Champagne Ardenne, Reims, France}

\author[0000-0002-5360-3660]{Panayotis Lavvas}
\affiliation{Groupe de Spectrométrie Moléculaire et Atmosphérique, Université de Reims Champagne Ardenne, Reims, France}

\begin{abstract}

Atmospheric escape shapes exoplanet evolution and star-planet interactions, with He I 10830 Å absorption serving as a key tracer of mass loss in hot gas giants. However, transit depths vary significantly across observed systems for reasons that remain poorly understood. HD209458b, the archetypal hot-Jupiter, exhibits relatively weak He I 10830 \AA\ and H$\alpha$ absorption, which has been interpreted as evidence for a high H/He ratio (98/2), possibly due to diffusive separation. To investigate this possibility and other processes that control these transit depths, we reassess excitation and de-excitation rates for metastable helium and explore the impact of diffusion processes, stellar activity, and tidal forces on the upper atmosphere and transit depths using a model framework spanning the whole atmosphere. Our model reproduces the observed He I transit depth and H$\alpha$ upper limit, showing strong diffusive separation. We match the observations assuming a photoelectron efficiency of 20-40\%, depending on the composition of the atmosphere, corresponding to mass-loss rates of $1.9-3\times10^{10}$ g/s. We find that the He I 10830 \AA\ transit depth is sensitive to both stellar activity and diffusion processes, while H$\alpha$ is largely unaffected due to its strong dependence on Lyman-$\alpha$ excitation. These differences may help explain the system-to-system scatter seen in population-level studies of the He I line. While He I data alone may not tightly constrain mass-loss rates or temperatures, they do confirm atmospheric escape and help narrow the viable parameter space when interpreted with physically motivated models. Simultaneous observations of He I, H$\alpha$, and stellar activity indicators provide powerful constraints on upper atmosphere dynamics and composition, even in the absence of full transmission spectra. 

\end{abstract}

\keywords{}

\section{Introduction} \label{sec:intro}

\noindent
Exoplanets orbiting close to their host stars receive high-energy stellar radiation that causes their atmospheres to escape. Understanding atmospheric escape allows us to gain insights into the chemical processes occurring within exoplanetary atmospheres, and escape significantly influences the distribution and characteristics of exoplanets in terms of size, composition, and atmospheric retention. The population of known exoplanets detected by the Kepler Space Telescope and the Transiting Exoplanet Survey Satellite (TESS) includes features such as the gap in planet occurrence near 1.8 R$\Earth$, known as the radius valley \citep{Fulton2017AJ....154..109F}, and the absence of Neptunes and Mini-Neptunes at close-in orbits, known as the Hot Neptune desert \citep{Mazeh2016A&A...589A..75M}. Models of atmospheric escape and radius evolution show that these features can be produced by photoevaporation driven by stellar X-ray and extreme ultraviolet (EUV/XUV) radiation and envelope loss due to rapid Roche lobe overflow \citep[e.g.,][]{Owen2017ApJ...847...29O, 2022ApJ...929...52K}. To understand the evolution of the exoplanet population, we must understand the escape of the upper atmospheres of exoplanets because atmospheric loss shapes their characteristics over time. 

Among the exoplanets discovered to date, Hot Jupiters and Warm Neptunes are especially useful for investigating atmospheric escape because of their relatively large size and close proximity to the host stars. On these exoplanets, models of escaping exoplanet upper atmospheres can be validated by comparisons with observations. In the past, observations of exoplanet upper atmospheres were achieved from space with the Hubble Space Telescope (HST) and were limited to far-ultraviolet (FUV) wavelengths where signal-to-noise (S/N) is relatively low and the interpretation of the observations has proven challenging \citep[e.g.,][]{v-m2003Natur.422..143V,Ben-Jaffel2010ApJ...709.1284B, Vidal2004ApJ...604L..69V, Linksy2010ApJ...717.1291L, Ballester2015ApJ...804..116B}. Recently, however, new observations have been accumulating, including ground and space-based detections of excited H and He, as well as various heavy elements and metals across UV to near-IR wavelengths \citep[e.g.,][]{jenesen2012ApJ...751...86J, Spake2018Natur.557...68S, Casasayas2020A&A...640C...6C}. Simultaneous observations of H and He are especially valuable because they probe the dominant neutral species of giant planet atmospheres and constrain the thermal structure and non-local-thermodynamic-equilibrium (non-LTE) level populations in the upper atmosphere. These observations also probe altitudes from the middle atmosphere to the exosphere, providing powerful constraints on the energy balance of the upper atmosphere and mass loss rates.

Absorption by He I at 10830 Å and the H I Balmer lines (6563, 4861, 4340 Å) have been detected on many Hot Jupiter-type exoplanets \citep[e.g.,][]{Spake2018Natur.557...68S, Krishnamurthy_2024, masson2024probing}.  The He I 10830~Å line corresponds to the transition between the n = 2 and n = 3 electronic levels of the helium atom. The lower level of this transition is the metastable 2$^3$S state, which has a half-life of $\sim$7800 seconds \citep{DrakePhysRevA.3.908}, a lifetime that allows He I (2$^3$S) to persist long enough for substantial density to build up and absorption to occur. The H I Balmer lines result from transitions of electrons in the hydrogen atom from the n = 2 energy level to higher energy levels. None of these lines are strongly absorbed by the interstellar medium and they can both be observed from the ground \citep{Indriolo2009ApJ...703.2131I}. This means that they are more easily observable than FUV absorption lines such as the H Lyman~$\alpha$ line that is strongly absorbed by the ISM and can only be observed from space \citep[e.g.,][]{v-m2003Natur.422..143V, Linksy2010ApJ...717.1291L, Koskinen2013Icar..226.1678K}. 

Studies that have investigated trends in the He I 10830 \AA\ transit depths across many systems have concluded that they correlate with stellar XUV flux \citep{Krishnamurthy_2024, Sanz-Forcada2025A&A...693A.285S} or mass-loss rates \citep{Ballabio2025MNRAS.537.1305B} but there is still a significant scatter of the observations deviating from these proposed trends \citep{Linssen2024arXiv241003228L, Allan2025arXiv250402578A}. This suggests that additional factors—such as atmospheric composition, heating structure, and planetary dynamics—affect He I 10830 \AA\ absorption in ways that are not captured by simple scaling relations. In some cases, the observed absorption probes relatively low altitudes in the thermosphere \citep[e.g.,][]{alonso2019A&A...629A.110A} while in others, excited helium is observed far from the planet in leading and trailing tails \citep[e.g.,][]{Spake2018Natur.557...68S, wANG2021ApJ...914...98W, WANG2021ApJ...914...99W, Zhang_Knutson_Wang_Dai_Barragan_2022}, and there is no straightforward explanation for the differences. Even for the archetypal hot Jupiter HD209458b, which orbits a relatively quiet Sun-like G-type star at 0.047 AU \citep{Southworth2010MNRAS.408.1689S}, models have had to impose a relatively low H/He ratio in the upper atmosphere (around 98/2) to match the relatively low He I 10830 \AA\ transit depth. This has led to the suggestion that the helium abundance in many exoplanetary atmospheres is substantially lower than solar \citep{Biassoni2024A&A...682A.115B}, or that the helium is experiencing diffusive separation \citep{Lampon2020A&A...636A..13L}. 

Given these challenges, we revisit HD209458b to systematically explore the key factors controlling He I 10830 \AA\ transit depths. HD209458b serves as an ideal benchmark for studying atmospheric escape due to its well-characterized system parameters, extensive observational data, and prior transit depth studies probing the upper atmosphere. Past studies have, however, predicted a wide range of mass-loss rates $1–10 \times 10^{10}$ g/s \citep{oklop2018ApJ...855L..11O, Lampon2020A&A...636A..13L, masson2024probing} based on He I 10830~\AA~absorption, and previous models of this absorption have often relied on ad-hoc assumptions about helium abundance to match the observations. Additionally, HD209458b’s debated FUV transit signals highlight the need to better understand how He I 10830 \AA\ absorption relates to other probes of atmospheric escape. By reanalyzing this system with an updated treatment of some of the key physics, we aim to clarify why the He I 10830 \AA\ absorption by the planet's upper atmosphere is lower than might be expected based on other observations and what this implies for broader trends in observations of exoplanet atmospheres.

The models that are often used to fit He I 10830 \AA\ transit observations make several assumptions about thermal structure, escape mechanisms, and atmospheric composition that affect the predicted transit depths. For example, He I (2$^3$S) observations are often interpreted using 1D Parker wind models, such as the publicly available \texttt{p-winds} model \citep{santos2022A&A...659A..62D} to describe the structure of the escaping atmosphere. These models often assume an isothermal outflow and do not consider much of the detailed physics of atmospheric escape. They result in degenerate solutions for temperature, mass-loss rate, density, and velocity, meaning that multiple combinations of these parameters can fit a given transit depth. Additionally, they do not properly account for the coupling between the lower/middle atmosphere and the thermosphere, which can play an important role in setting the boundary conditions for atmospheric escape \citep{Chadney2017A&A...608A..75C,Huang__2023}. Many current models also do not explicitly include diffusion, leading to ad-hoc assumptions about mixing, particularly in the case of helium-to-hydrogen separation. Finally, many models do not have the ability to simultaneously model hydrogen and helium absorption, despite the fact that analyzing these two features together potentially enables cross-validation of derived thermal structures and non-LTE level populations when observations of both lines are available. 

While diffusive separation of helium has been proposed as a possible explanation for the relatively weak He I 10830~\AA\ absorption observed on HD209458b, only a handful of models have explicitly included this process \citep[e.g.,][]{Xing2023ApJ...953..166X, Schulik2024arXiv241205258S}. These models solve separate momentum equations while we show in this work that a simpler approach can reproduce similar results. Additionally, previous models did not account for the effects of eddy diffusion, which is expected to impact helium mixing and transit depths. Finally, He I 10830 \AA\ transit depths depend on the choice of excitation and de-excitation rates, including photoionization cross-sections. Past studies have relied on lower-resolution photoionization cross-sections for the 2$^3$S state of He that neglect strong resonances in the EUV that overlap with prominent stellar emission lines \citep{waldrop2005JGRA..110.8304W, shefov2009Ge&Ae..49...93S, Lampon2020A&A...636A..13L, oklop2018ApJ...855L..11O, santos2022A&A...659A..62D}. We provide updated cross sections based on \citet{Argenti_2006, Argenti_2008, Argenti_2023} here. We also update the Penning ionization rate to a temperature-dependent value using available cross sections \citep{1971CPL....10..623C, 1979JPhB...12.1805M}. Finally, motivated by studies of He I (2$^3$S) in Earth’s atmosphere \citep{waldrop2005JGRA..110.8304W, shefov2009Ge&Ae..49...93S}, we explore the impact of photoelectron collisions on helium excitation, but find that their effect is negligible for HD209458b. 

To address the open questions about how differences in model atmospheres and reaction rates impact He I 10830 \AA\ and H$\alpha$ transit depths, we use a self-consistent, one-dimensional model of the thermosphere-ionosphere and hydrodynamic escape for close-in exoplanets to compute the composition and temperature in the upper atmosphere \citep{Koskinen2013Icar..226.1678K, 2022ApJ...929...52K}. We combine the results for the upper atmosphere with a radiative transfer-photochemistry model of the lower/middle atmosphere \citep{10.1093/mnras/stab456} to simulate the full atmospheric structure of HD209458b. We systematically test and validate our model in the following steps:
(1) We first reproduce the results of an isothermal Parker wind model for direct comparison with past studies based on this approach. (2) We then introduce our full thermosphere-ionosphere model, which accounts for non-isothermal temperature structures and diffusive separation. (3) We incorporate the lower/middle atmosphere model to fully couple the deeper layers with the escaping atmosphere, and present our best-fit model to the He I 10830 \AA\ and H$\alpha$ transit depth observations. (4) We analyze the impact of stellar activity variations, molecular and eddy diffusion, tidal effects, and the incorporation of metals on atmospheric structure, He I 10830 \AA\ absorption, and H$\alpha$ transit depths. This methodology enables us to separate the effects of different physical processes and determine which factors most significantly influence transit observations. Our results confirm that stellar activity variations, molecular diffusion, heating structure, and tidal effects all play a role in shaping He I 10830 \AA\ absorption. We show that multiple competing processes—beyond just XUV flux scaling—control He I 10830 \AA\ transit depths, likely explaining why some exoplanets appear as outliers in population studies.


\section{Methods}\label{sec:methods}

\noindent
Here, we summarize our methods to simulate the atmosphere and calculate transit depths. The stellar and planetary parameters we use for the HD209458b system are given in Table \ref{tab:params}. Section \ref{sec:lowmid} describes the lower/middle atmosphere model that we use to set the lower boundary conditions for the upper atmosphere. In Section \ref{sec:uppermodel}, we describe the multi-species model of hydrodynamic escape that we use to model the upper atmosphere. Section \ref{sec:stellar} describes the inputs we use to model HD209458's stellar flux. In Section \ref{sec:chem}, we describe the updates to the He I (2$^3$S) excitation/de-excitation rates. In Section \ref{sec:scat}, we describe the detailed balance model used to compute the excited state H populations. Finally, In Section \ref{sec:radtran}, we describe the methods that we use to simulate transit depths based on the output of the lower/middle atmosphere model and the upper atmosphere model.

\begin{deluxetable*}{ccc}[!h]
\centering
\tablecaption{HD209458b System Parameters \citep{Southworth2010MNRAS.408.1689S}}
\tablewidth{30pt}
\tablenum{1}
\tablehead{
\colhead{Parameter} & 
\colhead{Symbol} & 
\colhead{Value}  
}
\label{tab:params}
\startdata
Stellar Mass & $M_*$ & 1.148 $M_\odot$  \\
Stellar Radius & $R_*$ & 1.162 $R_\odot$ \\ 
Planetary Mass & $M_p$ & 0.714 $M_J$  \\
Planetary Radius & $R_p$ & 1.380 $R_J$ \\ 
Orbital Period & $P_{orb}$ & 3.524749 days \\
Semi-major axis & $a$ &  0.04747 AU \\
\enddata  
\end{deluxetable*}

\subsection{Lower and Middle Atmosphere} \label{sec:lowmid}

\noindent
Consistency with the properties of the lower and middle atmosphere can be important to correctly simulating the upper atmosphere and escape \citep[e.g.,][]{Chadney2017A&A...608A..75C, Huang__2023}. It is also required to connect the transit depths probing the upper atmosphere to the transit depths at adjacent wavelengths. To achieve this, we employ the lower and middle-atmosphere model \citep{10.1093/mnras/stab456} that computes the composition and temperature at pressures ranging from 1000 bar to 10$^{-6}$ bar where the latter pressure is the lower boundary of our upper atmosphere model. This model calculates photochemistry self-consistently with the temperature profile, encompassing approximately 150 molecules, atoms, and ions with more than 1600 reactions.  Based on the predicted atmospheric structure and composition, the model also calculates the theoretical transit spectrum of the planet, accounting for Rayleigh scattering, gaseous absorption, and Mie scattering by aerosols when applicable. Key contributors include \(\text{H}_2\), \(\text{H}_2\text{O}\), \(\text{CH}_4\), \(\text{NH}_3\), \(\text{HCN}\), \(\text{H}_2\text{S}\), \(\text{CO}_2\), Na, K, collision-induced absorptions by \(\text{H}_2-\text{H}_2\) and \(\text{He}-\text{H}_2\), and many metal atomic lines. Opacity sources follow \citet{Lavvas2021MNRAS.502.5643L}. For a more detailed model description, see \cite{Arfaux_2022}.

The lower and middle atmosphere model sets the lower boundary temperature and altitude for our upper atmosphere escape model at the pressure of 1 $\mu$bar. The transit `continuum' around the He I 10830 \AA\ and H$\alpha$ lines generally probes higher pressures at altitudes below the thermosphere and tends to be set by water absorption as well as extinction by clouds and hazes. Our model includes aerosol extinction due to photochemical hazes, based on a production rate of $1 \times 10^{-14}$ g cm$^{-2}$ s$^{-1}$ in the upper atmosphere that improves the fit to the observed transmission spectrum (see Section \ref{sec:fullatmo}). It is also necessary to extend the level population calculations below the lower boundary of the upper atmosphere model. To calculate the density of metastable helium in the lower and middle atmosphere, we assume a balance between production by recombination and loss to radiative decay and Penning ionization, guided by the results of our detailed balance model (see Section \ref{sec:chem}). We investigated other excitation mechanisms that might be important for populating excited helium in the lower/middle atmosphere, including Penning ionization with H$_2$, but find that these rates are not high enough to make a significant difference in the metastable helium population in the lower/middle atmosphere. For excited H, we assume LTE in the lower/middle atmosphere, multiplying the neutral hydrogen density by a Boltzmann distribution based on the temperature of the atmosphere. Both of these methods show smooth density transitions from the lower/middle atmosphere to the upper atmosphere. 

\subsection{Upper Atmosphere Escape Model} \label{sec:uppermodel}

\noindent
We use a 1D model of the thermosphere-ionosphere and hydrodynamic escape for close-in exoplanets \citep{Koskinen2013Icar..226.1678K, 2013Icar..226.1695K}. It solves the time-dependent coupled equations of continuity, momentum, and energy in the radial direction. The model includes photoionization by stellar XUV radiation, thermal ionization, recombination, charge exchange, advection and thermal escape of ions and neutrals, diffusion, viscous drag, heating by photoionization, adiabatic heating and cooling, heat conduction, and viscous dissipation. As we explain below, we only include H and He, their ions, and electrons in the reference simulations for this work, simplifying the model to highlight the impact of atmospheric structure and revised excitation and de-excitation rates on He I 10830 \AA\ transit depths. For radiative cooling, we include recombination and H I line cooling as described by \cite{Huang__2023}. Photochemistry and helium excitation/de-excitation are treated by the flexible open-source Kinetic Preprocessor KPP-3.0.0 chemical model that is fully coupled to the escape model \citep{KPP}. For completeness, further details of the model are given in the appendix of \cite{2022ApJ...929...52K}.

We consider two classes of upper atmosphere models in this work. Most of our simulations focus on hydrogen and helium only, including the species H, H(n=2), He, H$^+$, He$^+$, He$^{2+}$, He($2^3$S), and electrons. These models allow for direct comparisons with many previous studies of HD209458b and isolate the effects of updated model physics and parameter variations on the He I 10830~\AA\ transit depths. The reactions for these species are listed in Table \ref{table:rxns}, with the exception of H(n=2), which is handled separately (see below). In Section \ref{sec:metals}, we expand on this framework by including heavy elements and ions such as O, C$^+$, Si$^{2+}$, and Fe$^+$, for which observational evidence in HD209458b’s upper atmosphere has been reported \citep{Vidal2004ApJ...604L..69V, Linksy2010ApJ...717.1291L, Vidal-Madjar2013A&A...560A..54V, Cubillos2020AJ....159..111C}. These species can influence the energy balance and composition of the atmosphere, and we assess their impact on both He I 10830 \AA\ and H$\alpha$ transit depths. The relevant reaction rates for metal chemistry follow the implementation described by \citet{Huang__2023}.

\subsection{Stellar Flux}
\label{sec:stellar}

\noindent HD209458 is a G0-type inactive star with a coronal temperature comparable to that of the inactive Sun \citep{Czesla_2017}. The Sun’s quiescent X-ray luminosity ranges from $10^{27}$ to $10^{28}$ erg/s \citep{Judge2003ApJ...593..534J}, and the X-ray luminosity of HD2098458 sits at the lower end of that range at around $1.6 \times 10^{27}$ ergs/s \citep{Czesla_2017}. Therefore to match observations, we use a solar minimum spectrum from 2008, accessed from LISIRD\footnote{\url{https://lasp.colorado.edu/lisird/}} and scaled to the radius of HD209458 as our default input stellar spectrum.  To investigate how stellar activity level impacts the He I (2$^3$S) and H$\alpha$ Balmer line transit depths, we also employ a solar average and solar maximum spectrum, which are based on appropriate TIMED SEE spectra scaled to the radius of HD209458. Note that for each activity level, we use the LISIRD/TIMED SEE spectra at wavelengths shorter than 185 nm and a PHOENIX model (T$_\text{eff}$ = 6075 K, log($g$) = 4.38) \citep{Husser2013A&A...553A...6H} at longer wavelengths. To calculate the H I level populations (see Section \ref{sec:scat}), the H Ly$\alpha$ flux is needed for resonant scattering calculations. For each stellar activity level, we use the solar Ly$\alpha$ profiles given by \cite{LEMAIRE2005384} matched to the appropriate stellar activity.

\subsection{Non-LTE Rate Coefficients for He I (2$^3$S)} \label{sec:chem}

\begin{figure}[h]
    \centering
    \includegraphics[scale=0.4]{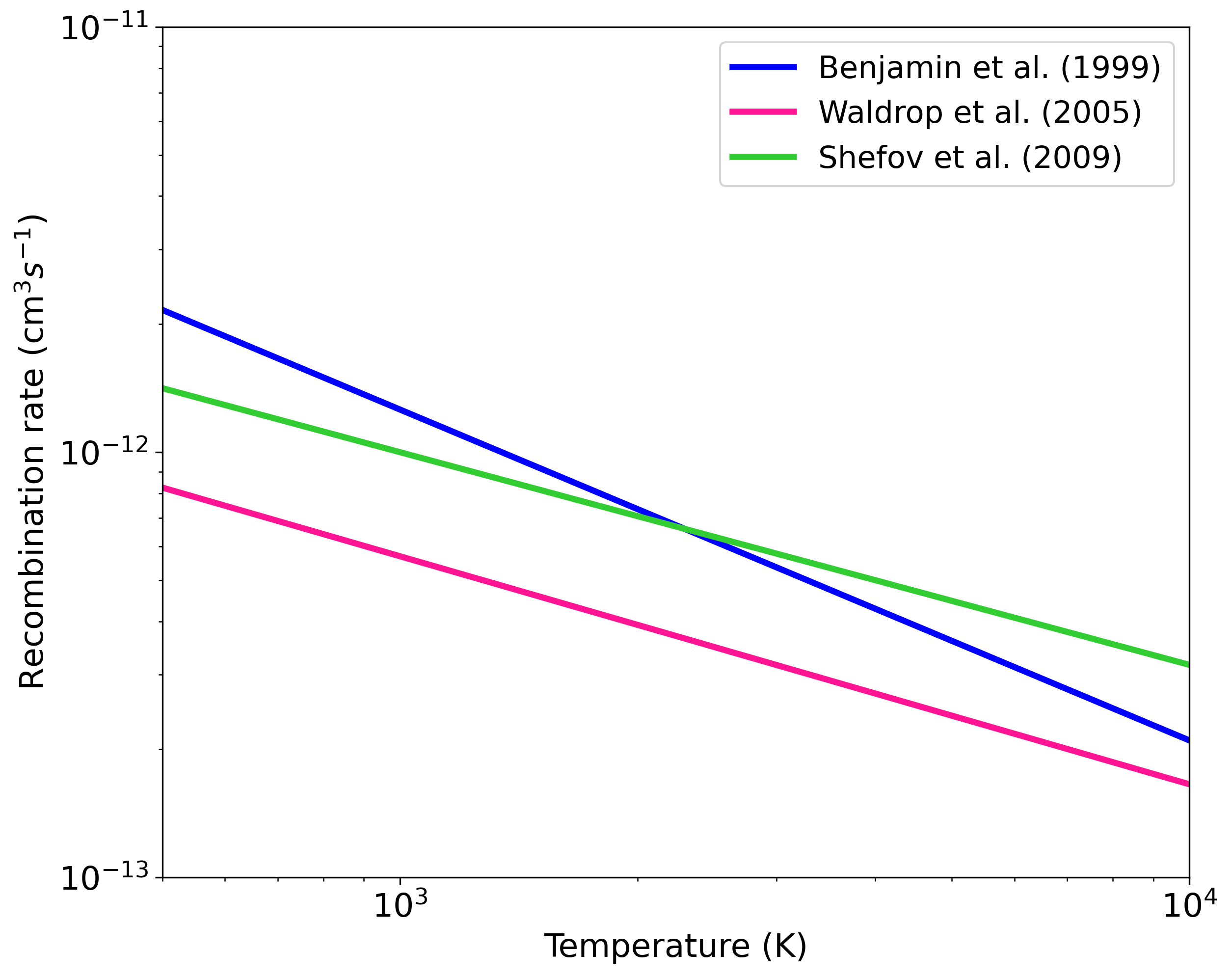}
    \caption{Recombination rates to the metastable He I (2$^3$S) triplet state versus electron temperature from different sources.}
    \label{fig:recomb}
\end{figure}

\noindent
Building on the work presented by \cite{Taylor2024Exo5}, we refine some of the rate coefficients that govern the population of metastable helium. The He I 10830~\AA\ transit depth is particularly sensitive to the excitation and de-excitation processes that control the abundance of the metastable 2$^3$S state. Among these processes, recombination of He$^+$ plays a dominant role in producing metastable helium in hot exoplanet atmospheres \citep{Saeger2000ApJ...537..916S}. As shown in Figure \ref{fig:recomb}, recombination rate coefficients used in some previous studies of metastable helium in different environments, including the Earth's upper atmosphere, vary by a factor of 2--3. These total recombination rates are based on the assumption that recombination from all higher-lying triplet states contributes to the population of the metastable 2$^3$S state, given the rapid radiative decay of these states. For example, in studies of metastable helium emission in the Earth's atmosphere, \citet{waldrop2005JGRA..110.8304W} used tables of partial recombination rates from \citet{1996ApJS..103..467V} while \citet{shefov2009Ge&Ae..49...93S} appears simply to have halved the total recombination rate to estimate the recombination rate to the triplet state. Exoplanet models, on the other hand, typically use the total case B triplet recombination rate from \citet{Benjamin1999ApJ...514..307B}. The discrepancies seen in Figure \ref{fig:recomb} largely result from incomplete summation over the recombination pathways in the Earth models. In particular, the tables from \cite{1996ApJS..103..467V} exclude recombination to many higher-lying triplet states in the triplet branch, leading to a significantly underestimated rate. In line with previous exoplanet work, we therefore use the recombination rates from \citet{Benjamin1999ApJ...514..307B}. In order to calculate the recombination rate to the ground state, we assume that recombination to all singlet states rapidly decays to the ground state and thus contributes to the total ground-state recombination rate. To avoid double counting, we compute the effective ground-state recombination rate by subtracting the triplet case B rate from the total recombination rate given by \citet{Benjamin1999ApJ...514..307B}. The resulting rate coefficients are listed in Table \ref{table:rxns}. 


\begin{figure}[!h]
  \centering
  \begin{minipage}[b]{0.43\textwidth}
     \includegraphics[width=\textwidth]{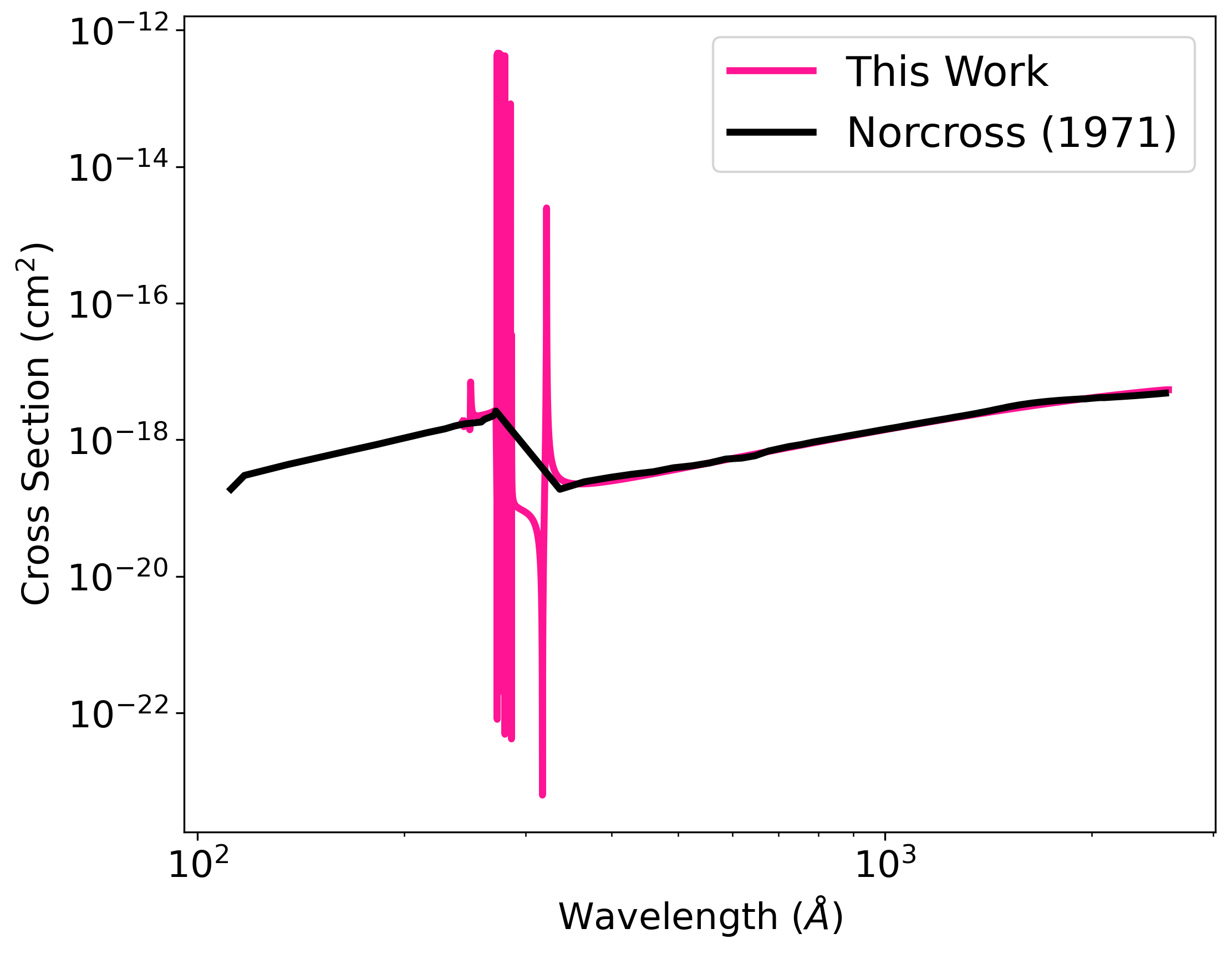}
  \end{minipage}
  \hfill
  \begin{minipage}[b]{0.45\textwidth}
    \includegraphics[width=\textwidth]{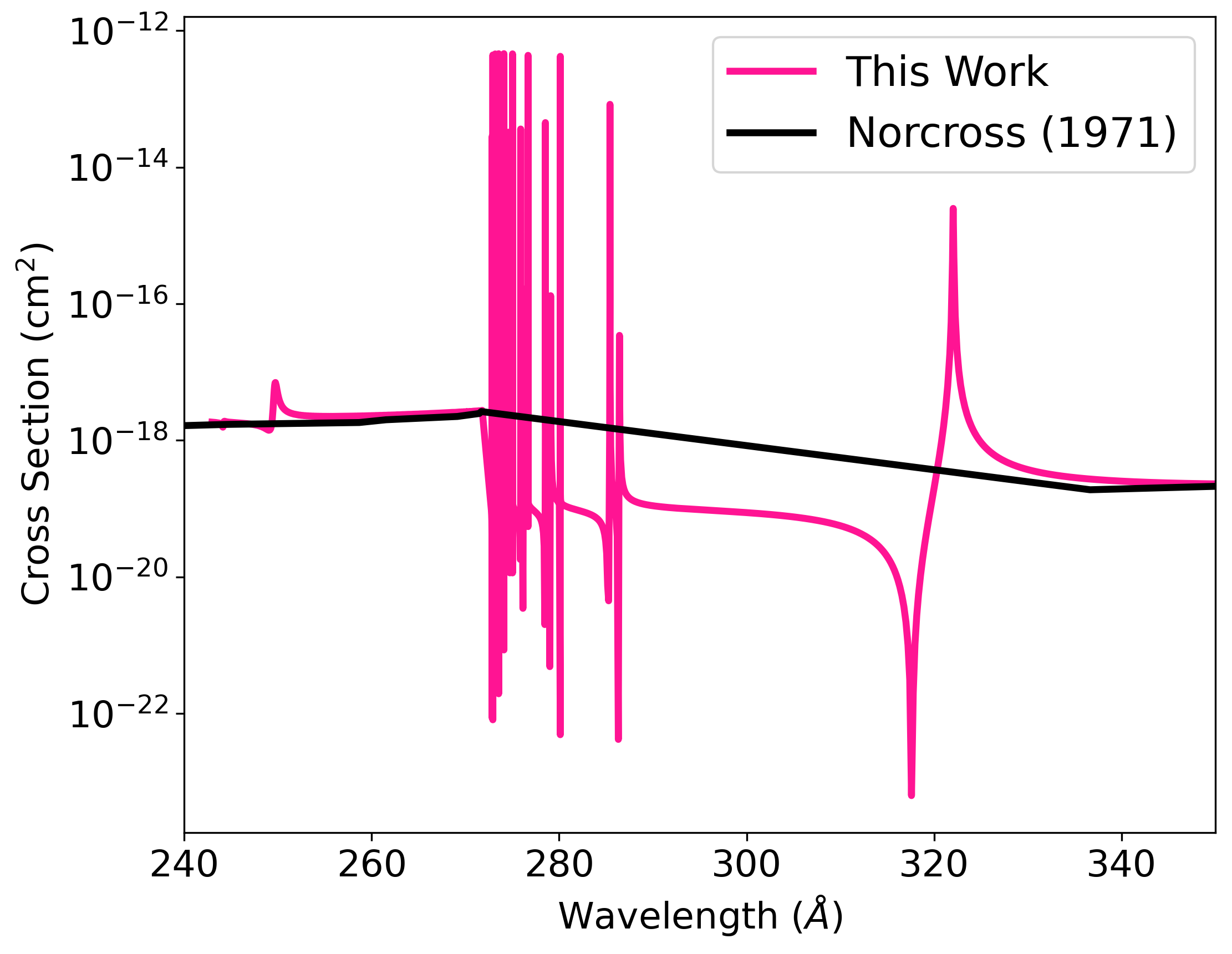}
  \end{minipage}
    \caption{The photoionization cross section for He I (2$^3$S) versus wavelength. The black line indicates what has been used in previous literature \citep{1971JPhB....4..652N}, while the pink lines show high-resolution cross-sections used in our model.}
    \label{fig:xsect}
\end{figure}

Photoionization is an important loss mechanism for the 2$^3$S state of helium, especially at high altitudes in close-in exoplanet atmospheres (see Section \ref{sec:bf}). Previous studies have typically used a low-resolution cross-section from \cite{1971JPhB....4..652N} that does not capture strong resonances in the EUV. We have generated a new high-resolution photoionization cross-section for the 2$^3$S state of helium using the B-spline K-matrix method \citep{Argenti_2006, Argenti_2008, Argenti_2023}. The calculations were performed within a spherical quantization box with a radius of $R_{box}=800$ Bohr radii. The radial components of the orbitals were expanded in B-splines of order $k=8$~\citep{Bachau2001RPPh...64.1815B}, while spherical harmonics represented the angular components. The configuration space comprises two sectors: a close-coupling sector, including all channels of the form $n\ell \epsilon_{\ell’}$ with $n\leq 3$ and angular momenta $\ell,\,\ell’$ compatible with the overall $^3S^e$ and $^3P^o$ symmetries; and a dynamic-correlation sector, composed of configuration state functions of the form $n\ell n’\ell’$ with $n,n’\leq 28$ and $\ell,\ell’\leq 8$, constructed from orbitals localized within 48~Bohr radii of the origin. Cross sections were computed in the velocity gauge~\citep{bransden2003atoms}. We make the cross sections available in a machine-readable format with this paper and on The University of Arizona's Research Data Repository\footnote{\url{https://redata.arizona.edu/}}. Figure \ref{fig:xsect} shows the calculated cross-section compared with the \cite{1971JPhB....4..652N} cross-section. In the right-hand panel of Figure \ref{fig:xsect}, our high-resolution cross-sections at EUV wavelengths show sharp resonance features that are orders of magnitude stronger than those in the previously used low-resolution data, areas where they overlap with strong stellar coronal emission lines.


Penning ionization with hydrogen is another important loss mechanism for He I (2$^3$S), especially in the lower thermosphere and below it (see Section \ref{sec:bf}). Previous studies have used a temperature-independent value for the rate coefficient \citep{oklop2018ApJ...855L..11O, Lampon2020A&A...636A..13L} based on a study on the formation and destruction of HeH$^+$ in astrophysical plasmas \citep{Roberge1982ApJ...255..489R}. In the latter study, they find that the rate coefficient for the sum of associative ionization reaction [He(2$^3$S) + H $\rightarrow$ HeH$^+$ + e$^-$] and the Penning ionization reaction [He(2$^3$S) + H $\rightarrow$ He(1$^1$S) + H$^{+}$ + e$^-$] is $\sim 5 \times 10^{-10} \text{ cm}^3\text{s}^{-1}$. Cross-sections are available for the sum of the associative ionization and the Penning ionization reaction,
which can be used to calculate a temperature-dependent reaction rate by using the Maxwell-Boltzmann distribution. The cross-sections we use for this calculation are from \cite{1979JPhB...12.1805M} and \cite{1971CPL....10..623C}. We fit the resulting temperature-dependent rate to obtain the rate coefficient in Table \ref{table:rxns}. This revised rate coefficient leads to an updated assessment of the relative role of Penning ionization in the depletion of metastable helium (see Appendix for a more detailed discussion of these calculations).

Free electron collisions are interactions between free electrons (those not bound to atoms or molecules) and other particles that can transfer energy between the electron and the particles, resulting in an excited or de-excited target particle. For collisions with thermal electrons, we use the rate expressions that have been referenced in other works (see Table \ref{table:rxns}) \citep{oklop2018ApJ...855L..11O, Lampon2020A&A...636A..13L} with temperature-dependent oscillator strengths from \cite{Bray2000}. Note that the oscillator strengths (denoted by $\Upsilon$ in Table \ref{table:rxns}) have a limited temperature range, with the lowest available temperature tabulations at 6000K. As our temperature profiles range from 3000K -- 9000K, we use only the oscillator strengths at 6000K. The $2^3$S $\rightarrow$ $2^1$S transition due to thermal electron collisions is an important loss mechanism for metastable helium in the thermosphere. The value of the oscillator strength for this transition is $2.389$. We also include the thermal electron collisions that result in the  $1^1$S $\rightarrow$ $2^3$S transition and the $2^3$S $\rightarrow$ $2^1$P transition although these are not significant production or loss mechanisms for metastable helium. The oscillator strengths at 6000 K are $6.198\times 10^{-2}$ for the $1^1$S $\rightarrow$ $2^3$S transition and $7.965 \times 10^{-1}$ for the $2^3$S $\rightarrow$ $2^1$P transition. Note that since the He I (2$^1$S) and (2$^1$P) states are so short-lived, we assume that they decay to ground-state helium instantaneously. 

Photoelectrons are released by photoionization events and their typical energies far exceed those of thermal electrons. Photoelectron impact on helium is an important production mechanism of metastable helium in the Earth's atmosphere \citep{waldrop2005JGRA..110.8304W} but it has been mostly ignored in exoplanet studies thus far. Here, we estimate a rough upper limit on the contribution of photoelectron excitation to the metastable helium population, based mostly on the production of primary photoelectrons. This involves using known photoionization cross-sections and stellar fluxes to calculate the energy-dependent production of primary photoelectrons throughout the upper atmosphere. Cross-sections for electron impact excitation to the 2$^3$S state are taken from laboratory measurements and fitted following the methods of \cite{waldrop2005JGRA..110.8304W} and \cite{1977JPhB...10.3741D} (see Appendix for more details). To more realistically account for the contribution of high-energy photoelectrons, we adopt a simplified scheme for estimating the contribution by secondary electrons. Specifically, for each primary photoelectron with energy $E_p$ exceeding twice the characteristic excitation energy $E_{cs}$ of metastable helium (roughly 36 eV, where the cross-section for metastable excitation peaks), we assume that $\lfloor E_p / E_{cs} \rfloor$ - 1 secondary electrons are produced with energy $E_{cs}$. These secondary electrons are then added to the appropriate energy bin before computing the total excitation rate. This approach ignores the interaction of the primary photoelectrons with other particles and assumes that secondary electron energies coincide with the peak interaction probability for exciting He. We followed this simplified approach because we found that photoelectron excitation does not significantly affect the transit depth for HD209458b, even with this upper limit. 

In addition to the above revisions, we have included proton de-excitation to the ground state of helium. For this process, we use cross sections from \cite{AUGUSTOVICOVA201427}, and calculate a temperature-dependent reaction rate by using the Maxwell-Boltzmann distribution. We fit the resulting rate coefficients with analytical equations to derive the rate expressions shown in Table \ref{table:rxns}. This means that collisional rates to these states are combined into one rate for collisions to the ground state. We note that the proton de-excitation rates do not significantly impact the He I (2$^3$S) transit depth on HD209458b and are included here for completeness. The Appendix provides a more detailed discussion of the relevant calculations.

\begin{table}[h!]
\begin{threeparttable}
\tablenum{2}
\scriptsize
\centering
\caption{He(2$^3$S) chemistry reaction rate coefficients}
\begin{tabular}{l c c} 
 \hline\hline
  Reaction & Rate (cm$^{3}$s$^{-1}$)& Reference \\
 \hline
  H + h$\nu$ $\rightarrow$ H$^+$  + e$^-$ & SC & \cite{10.1093/mnras/125.5.461} \\

 He + h$\nu$ $\rightarrow$ He$^+$  + e$^-$ & SC & \cite{1998ApJ...496.1044Y} \\

  H$^+$ + e$^-$ $\rightarrow$ H  + h$\nu$ & $4.0 \times 10^{-12} (300/T_e)^{0.64}$ & \cite{1995MNRAS.272...41S} \\
 
   He$^+$ + e$^-$ $\rightarrow$ He  + h$\nu$ &  $1.48 \times 10^{-12}(T/300)^{-0.54}$  & \cite{Benjamin1999ApJ...514..307B} \\
 
 H + e$^-$ $\rightarrow$ H$^+$  + e$^-$ + e$^-$ & $2.91 \times 10^{-8} U^{0.39} \frac{\exp(-U)}{0.232 + U}$, U = 13.6/$E_e$(eV) & \cite{1997ADNDT..65....1V} \\
 
  He + e$^-$ $\rightarrow$ He$^+$  + e$^-$ + e$^-$ & $1.75 \times 10^{-8}  U^{0.35} \frac{\exp(-U)}{0.180 + U}$, U = 24.6/$E_e$(eV) & \cite{1997ADNDT..65....1V} \\
 
   H + He$^+$ $\rightarrow$ H$^+$  + He & $1.30\times 10^{-15}(300/T_e)^{-0.25}$ & \cite{Stancil1998ApJ...509....1S} \\
 
 He + H$^+$ $\rightarrow$ He$^+$  + H & $1.75\times 10^{-11}(300/T_e)^{0.75}\exp(-127500/T)$ & \cite{Glover_2007} \\
 
  He$^+$ + e$^-$ $\rightarrow$  He$^{2+}$ + e$^-$ + e$^-$ & $2.05 \times 10^{-7} U^{0.25} \exp(-U) \frac{1 + U^{0.5}}{0.265 + U}$, U = 52.4/$E_e$(eV) & 
\cite{1997ADNDT..65....1V} \\

 He$^{2+}$ + e$^-$ $\rightarrow$ He$^+$ & $1.92 \times 10^{-11}(300/T_e)^{0.61}$ & \cite{1996ApJS..103..467V} \\

 He$^{2+}$ + H $\rightarrow$ He$^+$ + H$^+$& 1 $\times 10^{-14}$& \cite{Kingdon=Ferland1996} \\

 He$^{+}$ + h$\nu$ $\rightarrow$ He$^{2+}$ + e$^-$& SC& \cite{VFCS1996ApJ...465..487V}\\
 
  He(1$^1$S) + e$^{*-}$ $\rightarrow$ He(2$^3$S)  + e$^-$ & SC & This work \\

   He$^+$ + e$^-$ $\rightarrow$ He(2$^3$S)  + $h\nu$ & $2.10 \times 10^{-13}(T_e/10^4)^{-0.778}$  & \cite{Benjamin1999ApJ...514..307B} \\

  He(1$^1$S) + e$^-$ $\rightarrow$ He(2$^3$S)  + e$^-$  & $2.1 \times 10^{-8} \sqrt{\frac{13.6}{kT}}$exp$(\frac{-19.81}{kT}) \Upsilon_{13}$ & \cite{ Bray2000} \\

 He(1$^1$S) $\rightarrow$ He(2$^3$S) & $ 1.272 \times 10^{-4} $ &\cite{DrakePhysRevA.3.908} \\

  He(2$^3$S) + $h\nu$ $\rightarrow$ He$^+$ + e$^-$ &SC & This Work\\
 \multirow{2}{*}{He(2$^3$S) + H $\rightarrow$ He(1$^1$S)  + H} & T $\leq$ 4000 K: $1.9\times10^{-9} (300/T_e)^{0.07}$ & \multirow{2}{*}{This Work} \\

& T $>$ 4000 K: $9.1\times10^{-9} (300/T_e)^{0.50}$ &\\

 He(2$^3$S) + e$^-$ $\rightarrow$ He(2$^1$S)  + e$^-$  & $2.1 \times 10^{-8} \sqrt{\frac{13.6}{kT}}$exp$(\frac{-0.80}{k_B T}) \frac{\Upsilon_{31a}}{3}$ & \cite{ Bray2000} \\

 He(2$^3$S) + e$^-$ $\rightarrow$ He(2$^1$P)  + e$^-$  & $2.1 \times 10^{-8} \sqrt{\frac{13.6}{kT}}$exp$(\frac{-1.40}{k_B T}) \frac{\Upsilon_{31b}}{3}$ & \cite{ Bray2000} \\

 He(2$^3$S) + e$^-$ $\rightarrow$ He  + e$^-$ & $2.60\times 10^{-7}(T_e)^{-0.531}\text{exp}(-5.43\times 10^2/T_e)$ & \cite{Schulik2024arXiv241205258S} \\

He(2$^3$S) +  H$^+$ $\rightarrow$ He + H$^+$ & $5.55 \times 10^{-24} T_e^{-0.5}$ & \cite{AUGUSTOVICOVA201427} \\
 \hline 
\end{tabular}
\label{table:rxns}
\begin{tablenotes}
      \scriptsize
      \item \hspace{1cm} SC: Calculated self-consistently based on the model's ionization cross-sections, stellar flux, and density profiles.
    \end{tablenotes}
\end{threeparttable}
\end{table}

\subsection{H I Level Populations} \label{sec:scat}

\noindent
In addition to the He I 10830~\AA~line, we model the H$\alpha$ absorption line of hydrogen using a previously developed non-LTE model to simulate the level populations of hydrogen \citep{2017ApJ...851..150H, Huang__2023}. As demonstrated by previous studies \citep{2017ApJ...851..150H, 2019ApJ...884L..43G}, the excitation of the H(n=2) state is primarily driven by radiative excitation, with the population of H($2p$) largely excited by the Ly$\alpha$ intensity in the atmosphere. In this study, we employ the Ly$\alpha$ Monte Carlo resonant scattering radiative transfer model developed by \cite{2017ApJ...851..150H} to determine the Ly$\alpha$ mean intensity within a plane-parallel atmosphere. 

Due to the computational intensity of this calculation, it is not feasible to directly integrate the non-LTE model for H with our model of the upper atmosphere. Instead, we calculate the Ly$\alpha$ radiation field separately based on the upper atmosphere model output and then use the results to update the H(n=2) density in the upper atmosphere. Thus, the hydrodynamic model and radiation transfer calculations are performed iteratively until the electron densities and temperature profiles in the upper atmosphere model converge, typically after three iterations.  This iterative process ensures that the level populations are consistent with the atmospheric conditions. We note that ionization of H(n=2) is a significant source of free electrons that contributes to the heating rate of the upper atmosphere. 

\subsection{Simulated Transit Depths} \label{sec:radtran}

\noindent
To compare our atmospheric models to observations of the H I and He I 10830 \AA\ transit depths, we calculate the transit depths based on the modeled pressure, temperature, and density profiles from the upper atmosphere and lower/middle atmosphere models. To compute the relative diminution of the flux from the star caused by the presence of the planet, we use equation (11) from \cite{Brown2001ApJ...553.1006B}: 
\begin{equation}
    \frac{\partial F(k)}{F(k)} = \frac{1}{\pi r_*^2} \int_0^{z_{max}} 2 \pi (r_p + z^*)[1-\exp(-\tau(z^*,k)]dz^*
\end{equation}
where $r_*$ is the radius of the star, $z^*$ is the ray depth, $r_p$ is the radius of the planet, $k$ is wavenumber, and $z_{max}$ is the maximum ray depth. The optical depth $\tau$ is given by 
\begin{equation}
    \tau(z',k) = 2 \int_0^{x_{max}} \kappa(z[z', x])dx 
\end{equation}
where $z'$ is the minimum height of a ray above the 1-bar pressure level, $\kappa$ is the opacity, and $x_{max}$ is the maximum line-of-sight distance in the atmosphere. We use line properties for atomic lines from the NIST database \footnote{\url{https://physics.nist.gov/PhysRefData/ASD/lines_form.html}}. Here we only focus on the He I 10830~\AA~and H$\alpha$ transit depths while the details for the lower/middle atmosphere transit depth calculation and the relevant molecular line lists are given in Section \ref{sec:lowmid}. 

For this study, we calculate center-of-transit transit depths by accounting for thermal and natural broadening using a Voigt profile. We include velocity broadening by incorporating the line-of-sight outflow velocity as an additive term in the Gaussian component of the Voigt profile and add rotational broadening of 2 km/s based on the spin of the planet. We follow this relatively simple approach to include velocity and rotational broadening to facilitate an easier comparison with previous studies and to provide a clear constraint on the required broadening to match observations, which our parameters do (see Section \ref{sec:results}). To model the He I 10830~\AA\ and H$\alpha$ features, we use density and temperature profiles from both the lower/middle and upper atmosphere models. We then combine the He I 10830 \AA\ and H$\alpha$ line profiles and the lower/middle atmosphere transit spectrum, so that the excited hydrogen and helium absorption features appear as enhancements on top of the continuum transit spectrum (see Figure \ref{fig:totaltds}). This approach ensures that the atomic lines are treated consistently with the atmospheric structure and that the resulting transit depths are normalized to the expected continuum level from the lower atmosphere.

We note that transit depths are also influenced by factors such as three-dimensional dynamics, possible planetary magnetic fields, and interactions with stellar winds that are not included in our 1D model \citep{Owen2020SSRv..216..129O}. One-dimensional models can, however, be used to explore the physics and processes in detailed balance models that help to identify key inputs to more complex models. One-dimensional models are also much faster and facilitate population-level surveys of broad theoretical trends.  Here, we also note recent studies showing that a hotter dayside means that transmission spectra are more reflective of the average conditions on the dayside than along the terminator \citep{Jaziri2024A&A...684A..25J}. This is especially true of hot, extended upper atmospheres because their large extent means that radiation on close-in exoplanets can penetrate past the terminator to the night side, suggesting that a significant fraction of the upper atmosphere exhibits dayside-like behavior \citep{Koskinen2007}.

\section{Results} \label{sec:results}
\noindent
Here, we present results from our atmospheric models for HD209458b, focusing on critical parameters such as temperature and velocity distributions, atmospheric composition, metastable helium production and loss processes, and transport mechanisms. We begin by examining a baseline model configured to replicate the framework of the isothermal Parker wind model, characterized by an isothermal temperature profile and the absence of diffusion processes, using this setup to fit existing observational data. We then introduce our upper atmosphere model where the temperature profile is self-consistently calculated from the energy equation rather than assuming an isothermal structure. Building upon this, we integrate the lower and middle atmosphere model to produce a fully coupled atmospheric model, enabling direct comparisons with observational constraints from HST and JWST. Then, we present our best-fit model for the transit depths of He I 10830 \AA\ and H$\alpha$ which constrain the photoelectron heating efficiency and mass loss rate. We also examine how key parameter variations—including the stellar activity level, molecular and eddy diffusion, and tidal forcing—shape the thermal structure, composition, and transit depths predicted by our models. Finally, we incorporate heavy elements into the upper atmosphere to investigate their impact on the atmospheric structure and observable transit signatures.

\subsection{Isothermal Upper Atmosphere Model} \label{sec:upper}


\noindent
To establish a foundation for our analysis, we first use our model to reproduce the approach of previous studies, such as those utilizing the isothermal \texttt{p-winds} framework and the modeling techniques presented by \cite{santos2022A&A...659A..62D, Lampon2020A&A...636A..13L, oklop2018ApJ...855L..11O}. This baseline model, which we call Model A (see Table \ref{tab:modelsummary} for a summary of model configurations), assumes a constant temperature of 8500 K with altitude, bulk transport with no diffusion processes, and the updated excitation and de-excitation rates for He I (2$^3$S) as described in Section \ref{sec:chem}. We model radiative transfer in the atmosphere by using a solar zenith angle of 60$^{\circ}$ for a dayside average and further dividing the incident stellar flux by a factor of two to approximate global redistribution \citep{Smith1972JGR....77.3592S} that may be more suitable for the terminator that is probed in transit than the overhead illumination adopted in several other 1D studies. Unlike \cite{Xing2023ApJ...953..166X}, we do not include the substellar tide by default because it does not apply at the terminator (see Section \ref{sec:param} for further discussion on this). 

\begin{figure}[h!]
    \centering
    \includegraphics[width=.4\textwidth]{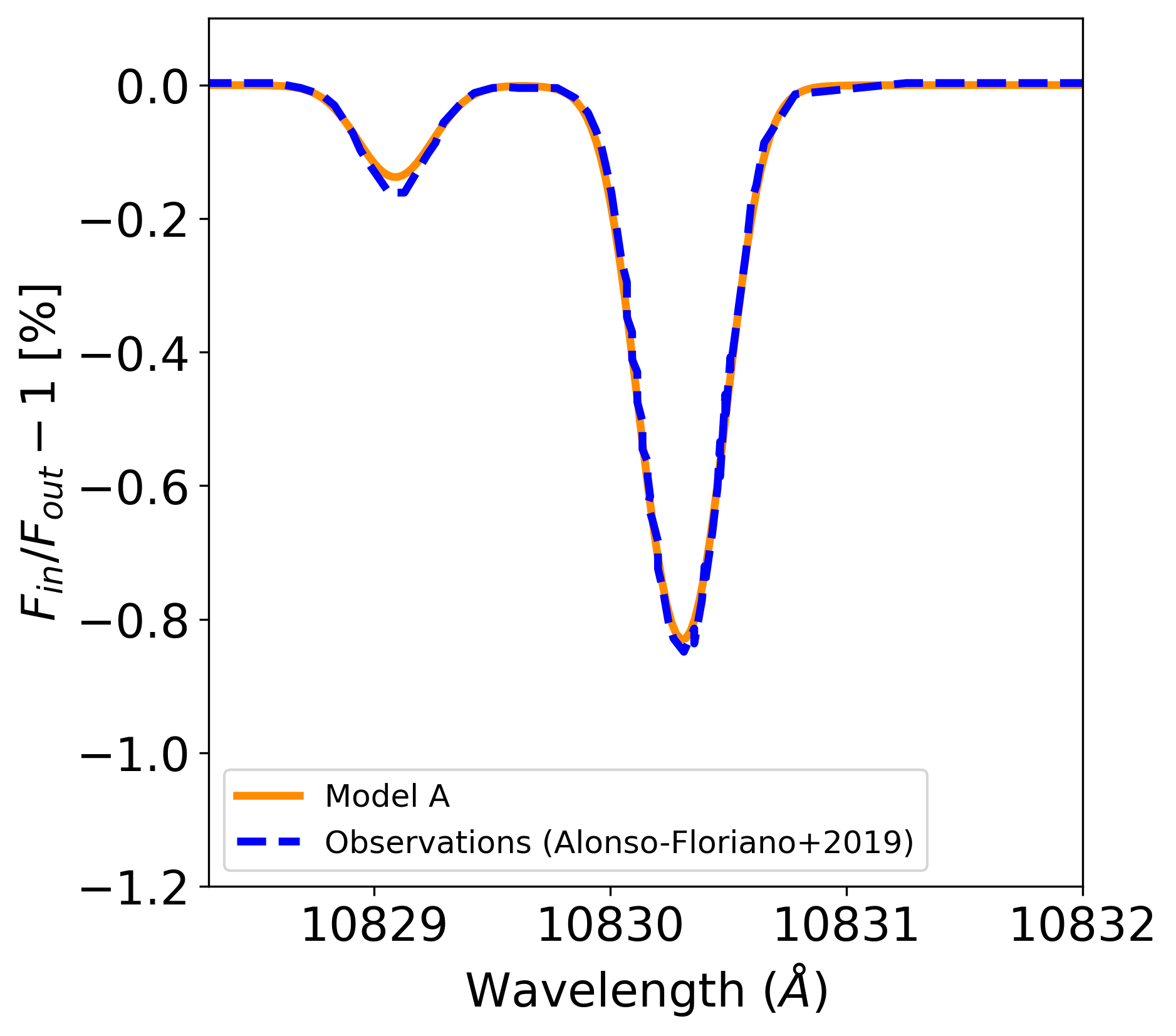}
    \includegraphics[width=.375\textwidth]{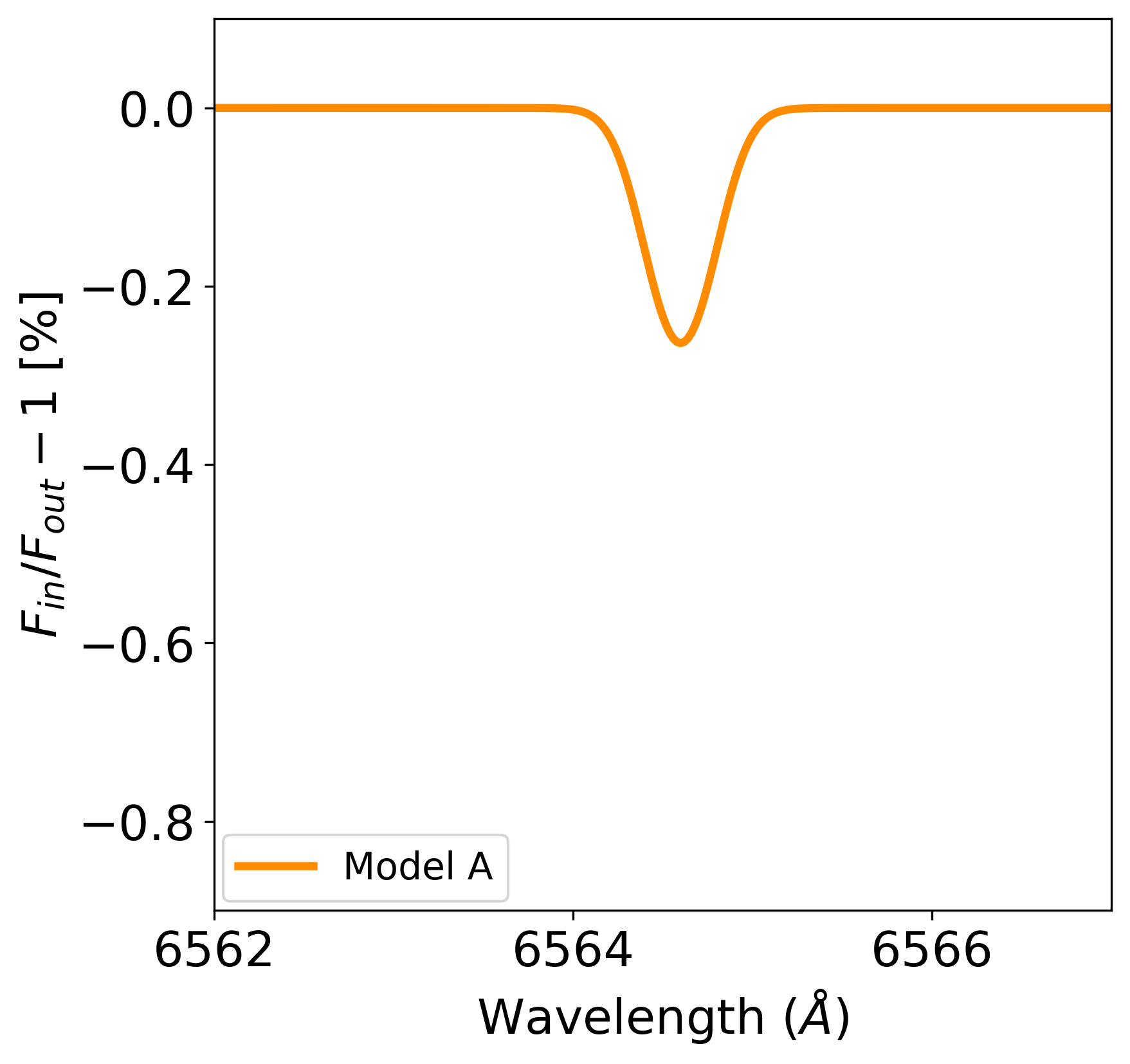}
    \caption{Transit depth results from our baseline model (Model A), comparing the modeled He I 10830 Å (left) and H I Balmer $\alpha$ (right) transit depths from our baseline model (orange solid lines) with the best-fit model to the observed transit depth for the He I 10830 \AA~line from \citet{alonso2019A&A...629A.110A} (blue line), respectively, demonstrating that the baseline model reasonably captures the overall absorption features.}
    \label{fig:pwindstd}
\end{figure}

We set the temperature to 8500 K because this is at the lower end of temperatures used to match the FUV transit observations \citep{2010ApJ...723..116K} and because it is consistent with our best-fit full atmosphere model. Using this temperature to match the He I 10830 \AA\ observations \citep{alonso2019A&A...629A.110A}, we determine a mass-loss rate of 1.86 $\times 10^{10}$ g/s that is consistent with best-fit models from \citet{Lampon2020A&A...636A..13L}. We note that temperature and mass loss rate are degenerate in isothermal fits like this (see below). Figure~\ref{fig:pwinds} presents the temperature and velocity profiles (left panel), and the composition of the atmosphere (right panel). Figure \ref{fig:pwindstd} shows the resulting transit depth compared with observations for the He I 10830 \AA\ line (left panel) and the H$\alpha$ line (right panel). We explored the effect of diffusion processes on Model A, but since the lower boundary pressure of 7 nbar in this model is relatively low and the temperature is high everywhere, the outflow velocities are substantial even from the bottom boundary and we see no effect from diffusive separation as a result.

While Model A provides a reasonable fit to the data, real atmospheres are not isothermal. The assumption of a constant temperature neglects key physical processes such as the dependency of the radiative heating and cooling rates on altitude, adiabatic expansion, and heat conduction that shape the actual thermal structure of the atmosphere. In addition, the choice of the mass loss rate and temperature in the isothermal Parker wind model dictates the lower boundary pressure, which in our best-fit case is 7 nbar. This dependence arises because the Parker wind model enforces a specific relationship between temperature, velocity, and pressure, which constrains the lower boundary conditions in a way that does not necessarily reflect actual atmospheric structure. To understand this constraint, we briefly outline the fundamental equations governing the solution. The basic Parker wind model uses a set of equations to describe the hydrodynamic escape of an atmosphere, focusing on mass and momentum conservation. For an isothermal wind, the momentum equation and mass conservation can be expressed as:
\begin{equation}
    \frac{d\psi}{d\xi} \left(1 - \frac{\tau}{\psi}\right) = -2\xi^2 \frac{d}{d\xi}\left(\frac{\tau}{\xi^2}\right) - \frac{2\lambda}{\xi^2}
\end{equation}
where \( \psi = \frac{1}{2}\frac{m_iv^2}{kT_0}\) is a dimensionless measure of kinetic energy, \( \xi = \frac{r}{r_0} \) is the dimensionless radial distance, \( \tau = \frac{T(r)}{T_0} \) represents the temperature profile, and \( \lambda = \frac{GM_Sm_i}{2r_0kT_0} \) is the thermal escape parameter. Parker's model assumes that the temperature remains constant (isothermal condition) up to a certain radius \( r_b \) beyond which the temperature drops to negligible levels, although this implicitly assumed decrease in temperature is not included in the models. The transonic Parker wind solution, assumed by the above models, includes a critical point \((\xi_c, \psi_c)\) where \(\xi_c = \lambda / 2\) and \(\frac{d\psi}{d\xi}= 0 \rightarrow \psi_c = 1\). As was pointed out by \citet{Chamberlain1960ApJ...131...47C}, for a given escape parameter \(\lambda\) (i.e., temperature), there is only one unique transonic solution (i.e., only one value of the lower boundary kinetic energy parameter $\psi_0$ that will provide the transonic solution). This means that the velocity at the lower boundary depends on the temperature and cannot vary. The lower boundary density and therefore pressure depend on the assumed mass loss rate, given by  
\begin{equation}
    \dot{M} = 4 \pi r_0^2 m_i n_0 v_0.
\end{equation}
This naturally leads to a degeneracy between mass loss rate and temperature when Parker wind models are fitted to transit observations. In other words, setting the mass loss rate \(\dot{M}\) in Parker wind models only changes the density/pressure profile rather than the velocity profile if the temperature is held fixed. This allows for arbitrary changes in density/pressure that may not be consistent with the energy balance of the upper atmosphere. This is because the parameter \(\lambda \) dictates the flow's velocity profile but this parameter has no dependence on density. Thus changing the mass loss rate does not alter the velocity profile that is governed by \(\psi(\xi)\) and the critical point condition. In models such as the isothermal p-winds \citep{santos2022A&A...659A..62D}, the
user therefore selects the temperature and mass loss rate to match observations and this selection fixes the lower boundary pressure. We emphasize that the lower boundary density should be set by the conditions at the actual wind launching radius, which depend on the detailed properties of the atmosphere. 

The lower boundary pressure of 7 nbar in our best-fit isothermal model is located in the middle of the photoionization region for He (where the helium photoionization rate peaks) that is ionized deeper than H due to its higher ionization potential and the majority of He ionization occurs at altitudes near and above the lower boundary rather than higher up in the atmosphere. This artificially enhances the ionized helium, and therefore the metastable helium density at lower radii (see left panel in Figure \ref{fig:he23scomp}) while suppressing it in the extended upper atmosphere, leading to a reduced He~I transit depth compared to a self-consistent thermal model with a more appropriate lower boundary pressure below the ionization peak. The H$\alpha$ differential transit depth is also lower, partly because the transit continuum is set to the 7 nbar level while the true continuum is much deeper.

\begin{figure}[h!]
    \centering
    \includegraphics[width=.43\textwidth]{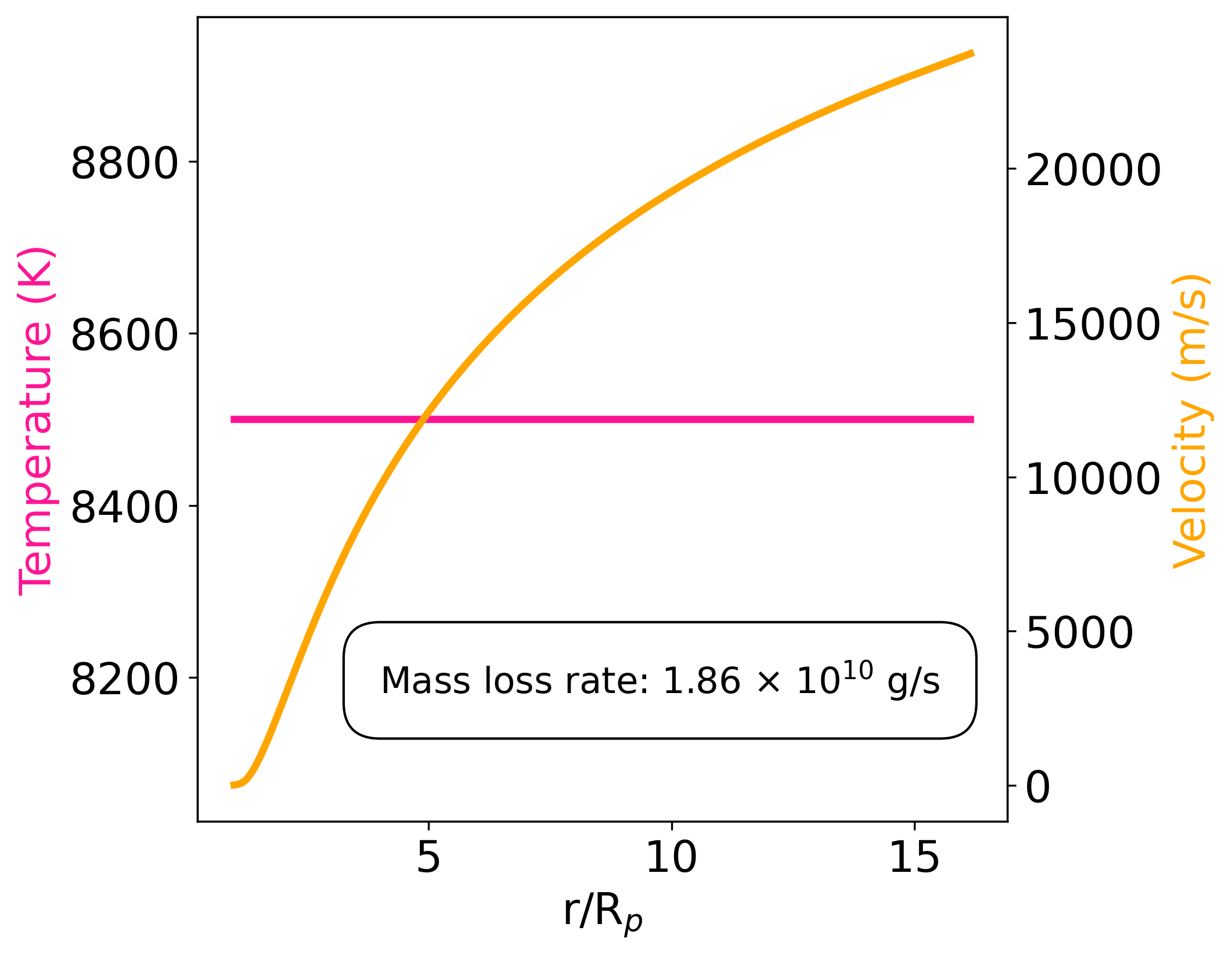}
    \includegraphics[width=.48\textwidth]{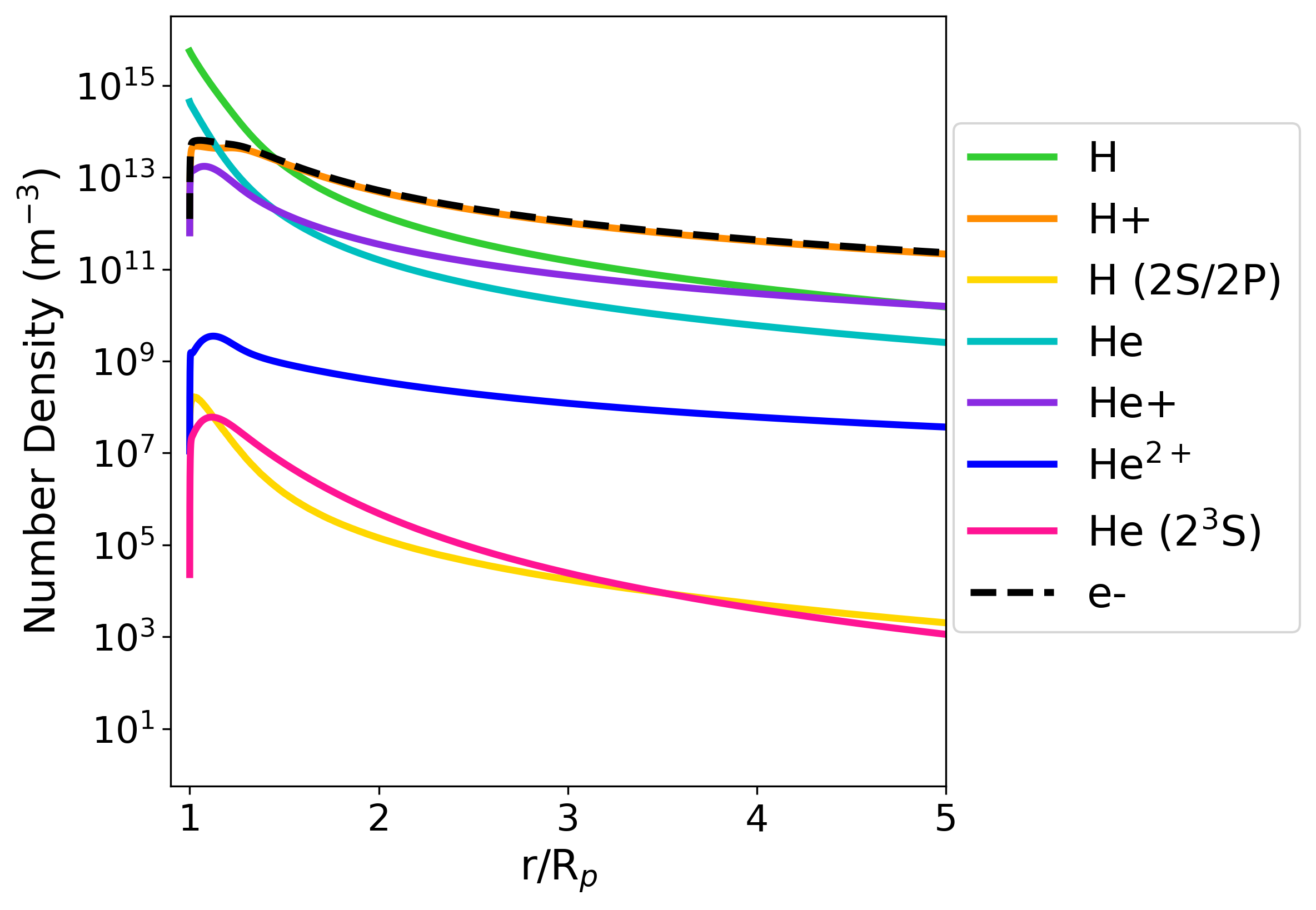}
    \caption{Isothermal model (Model A) results for HD209458b. The left panel shows the temperature (pink) and velocity (orange) profiles as a function of planetary radii, with a mass loss rate of 1.86$\times 10^{10}$ g/s. The right panel presents the number densities of key atmospheric species, including neutral hydrogen (H), ionized hydrogen (H$^+$), neutral helium (He), ionized helium (He$^+$), doubly ionized helium (He$^{2+}$), metastable helium He I (2$^3$S), and free electrons.  }
    \label{fig:pwinds}
\end{figure}

\subsection{Self-consistent Upper Atmosphere Model}

\begin{figure}[!h]
  \centering
    \includegraphics[width=.47\textwidth]{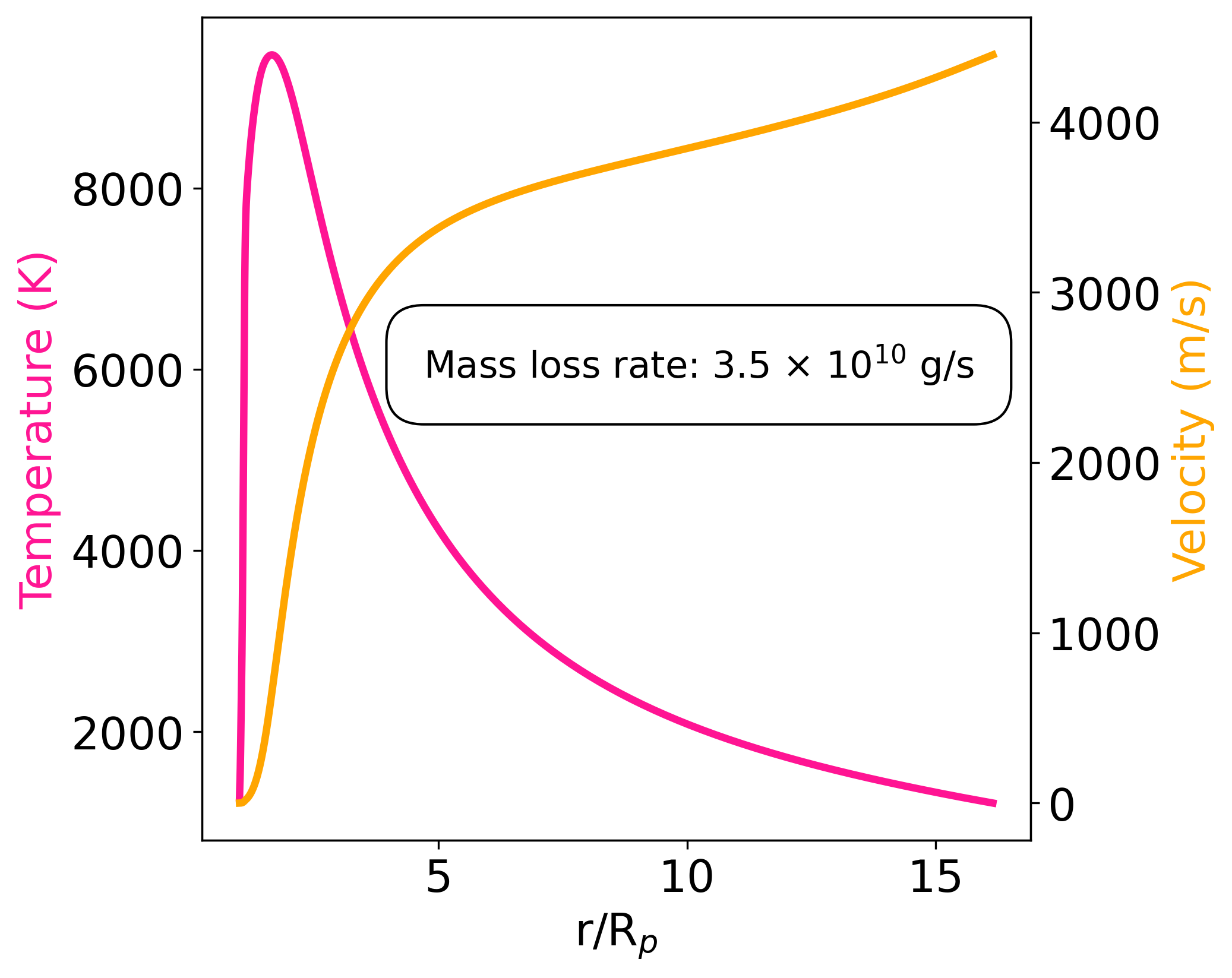}
    \includegraphics[width=.4\textwidth]{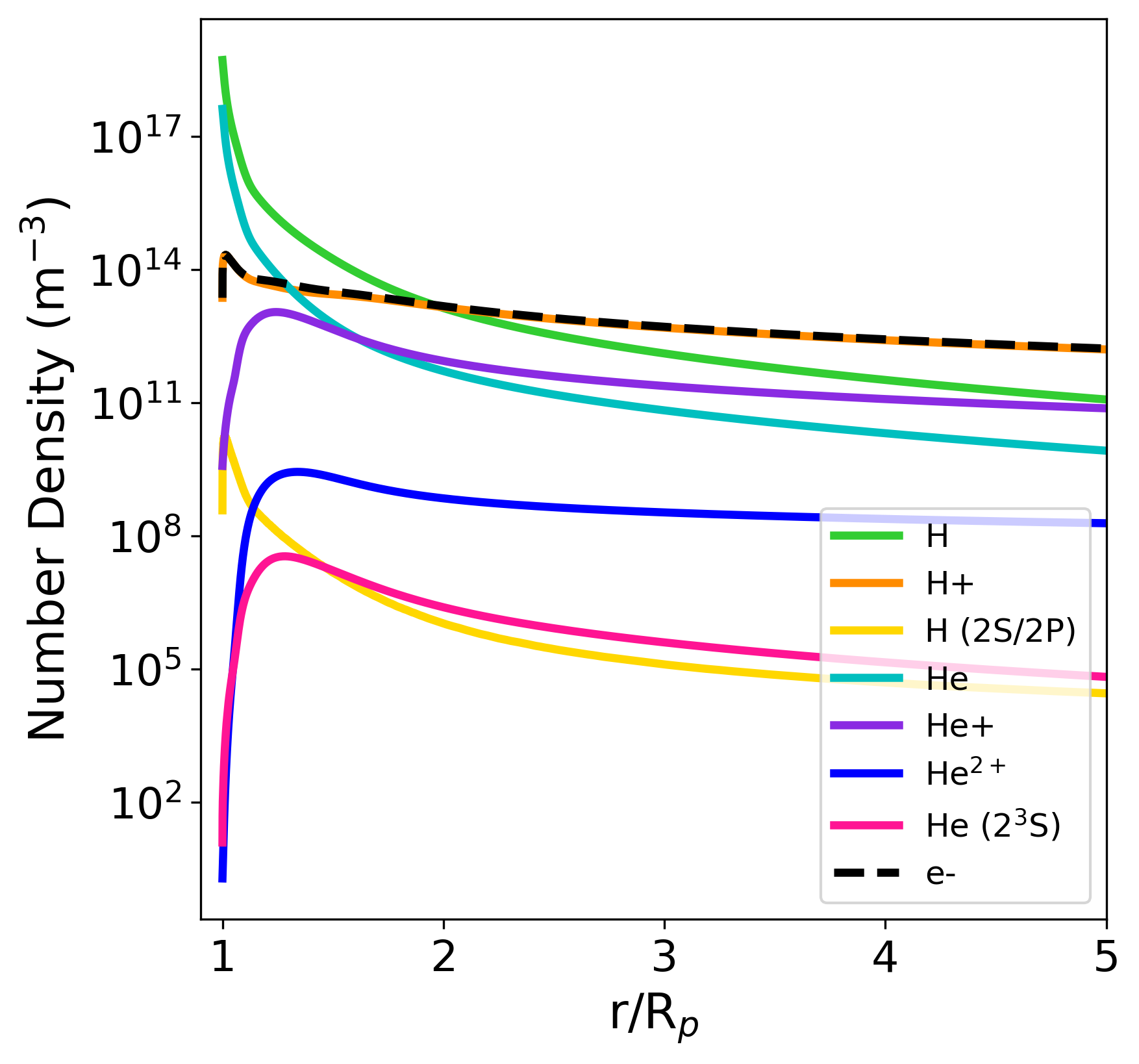}
    \caption{\textit{Left: }Temperature and bulk outflow velocity predicted by our full upper atmosphere model for HD209458b (Model B). \textit{Right: }Composition predicted for HD209458b by Model B. Note that the H+ profile is practically identical to the electron (e$^-$) density profile.  }
    \label{fig:tempv}
\end{figure}

\noindent
Motivated by these considerations, we move beyond the isothermal approximation by solving the full energy equation. We also include multi-species transport with molecular and eddy diffusion. This Model B (see Table \ref{tab:modelsummary} for a summary of model configurations) is initialized with a lower boundary pressure of $10^{-6}$ bar, a more physically motivated choice that coincides with the transition between the middle and upper atmosphere. In addition, previous studies fitting the H Lyman-$\alpha$ transit of HD209458b required a lower boundary pressure near 10$^{-6}$ bar, above which temperature increases with altitude, to achieve a match with observations \citep{2010ApJ...723..116K, Koskinen2013Icar..226.1678K}. Model B does not yet include the lower and middle atmosphere. Instead, the lower boundary altitude is 0 km with a temperature of 1270 K. The model assumes a photoelectron heating efficiency of 100\%, based on some previous studies \citep[e.g.,][]{2009ApJ...693...23M}, and an eddy diffusion coefficient of $10^5$ m$^2$~s$^{-1}$ \citep{Arfaux_2022}. The left panel of Figure \ref{fig:tempv} shows the temperature and bulk outflow profiles for Model B. The temperature peaks at $\sim$9500 K where $r =$~1.65 $R_p$ ($p =$ 1.45 $\times 10^{-10}$ bar). The energy deposition rate peaks at a radius of 1.18 R$_p$ and pressure of $3.94\times 10^{-9}$ bar. The outflow velocity is 3.6 km/s at 4 R$_p$ (the approximate Roche lobe radius). We note that the altitude of the sonic point is strongly dependent on the details of the temperature profile \citep{Koskinen2013Icar..226.1678K}. It also depends on Roche lobe effects that we do not include in our simulations here. If we include tidal forcing and run our model at the substellar point, we find a sonic point of 4.4 R$_p$, consistent with past models \citep{Xing2023ApJ...953..166X}. We find a steady-state mass-loss rate ($\dot{M} = 4\pi v \rho r^2$) of $3.5 \times 10^{10}$ g/s. We calculate a pressure-averaged temperature of the hydrodynamic model above the 7 nbar to the top of the model at $3\times 10^{-4}$ nbar to compare with the isothermal model:
\begin{equation}
    \overline{\mbox{T}}_P = \frac{\int_{P_1}^{P_2} T(P) d(\ln p)}{\ln (P_2/P_1)}
\end{equation}
where $P_1$ and $P_2$ are the pressure bounds for the integration. Using this method we obtain a temperature of 8860 K. The right panel of Figure \ref{fig:tempv} shows the composition of major species in Model B. Neutral helium and hydrogen dominate towards the base of the atmosphere, but as photoionization becomes stronger at higher altitudes, hydrogen and helium ions dominate over their neutral counterparts. The photoionization timescale becomes shorter than the transport timescale at a radius of around 2 $R_p$, allowing the ions to begin to dominate over the neutral atoms. The He I (2$^3$S) density is significantly lower than that of the dominant species and follows the general shape of the He$^+$ profile due to recombination being the dominant production process. 

\begin{figure}[h!]
    \centering
    \includegraphics[width=.3\textwidth]{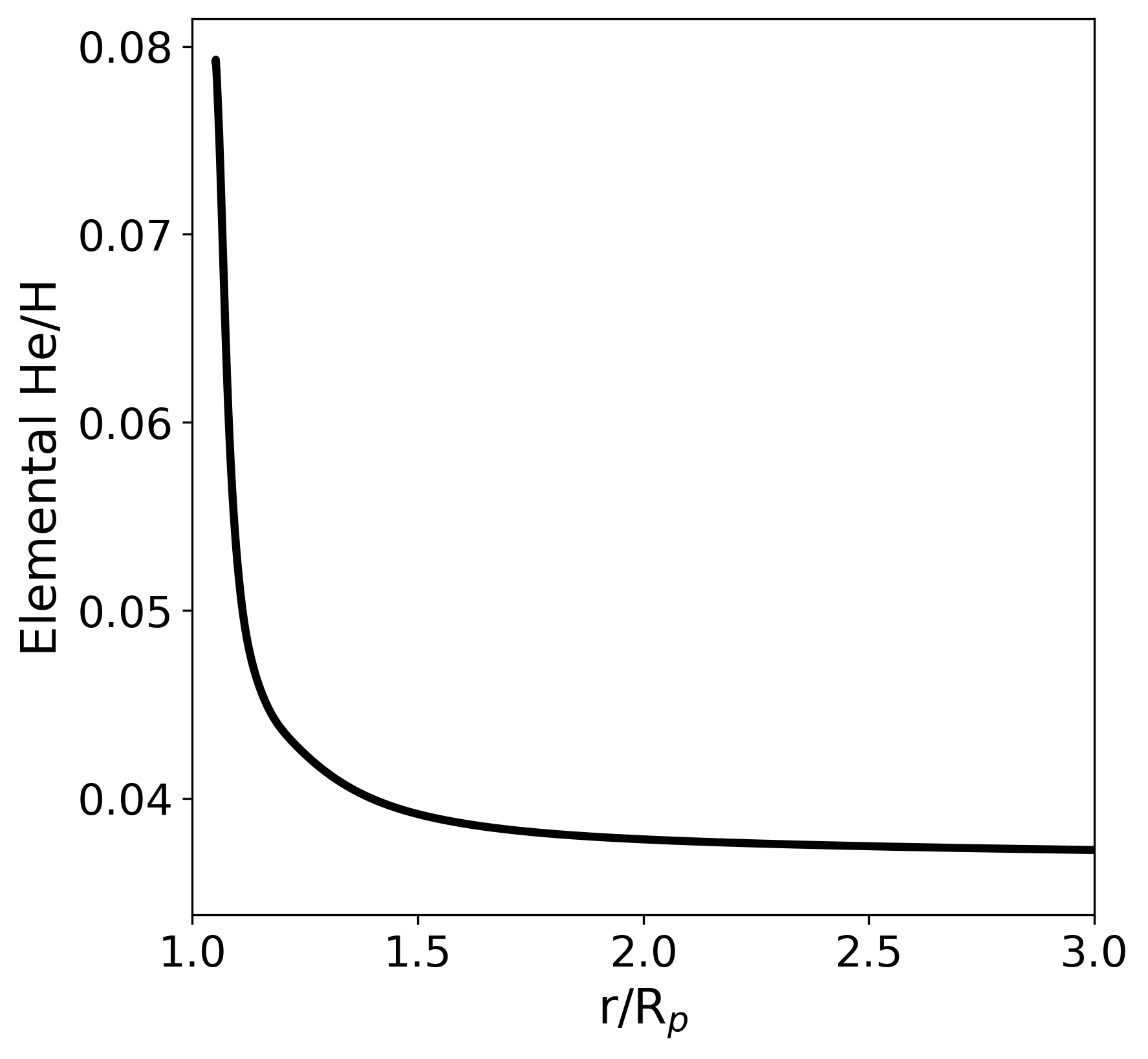}
    \includegraphics[width=.3\textwidth]{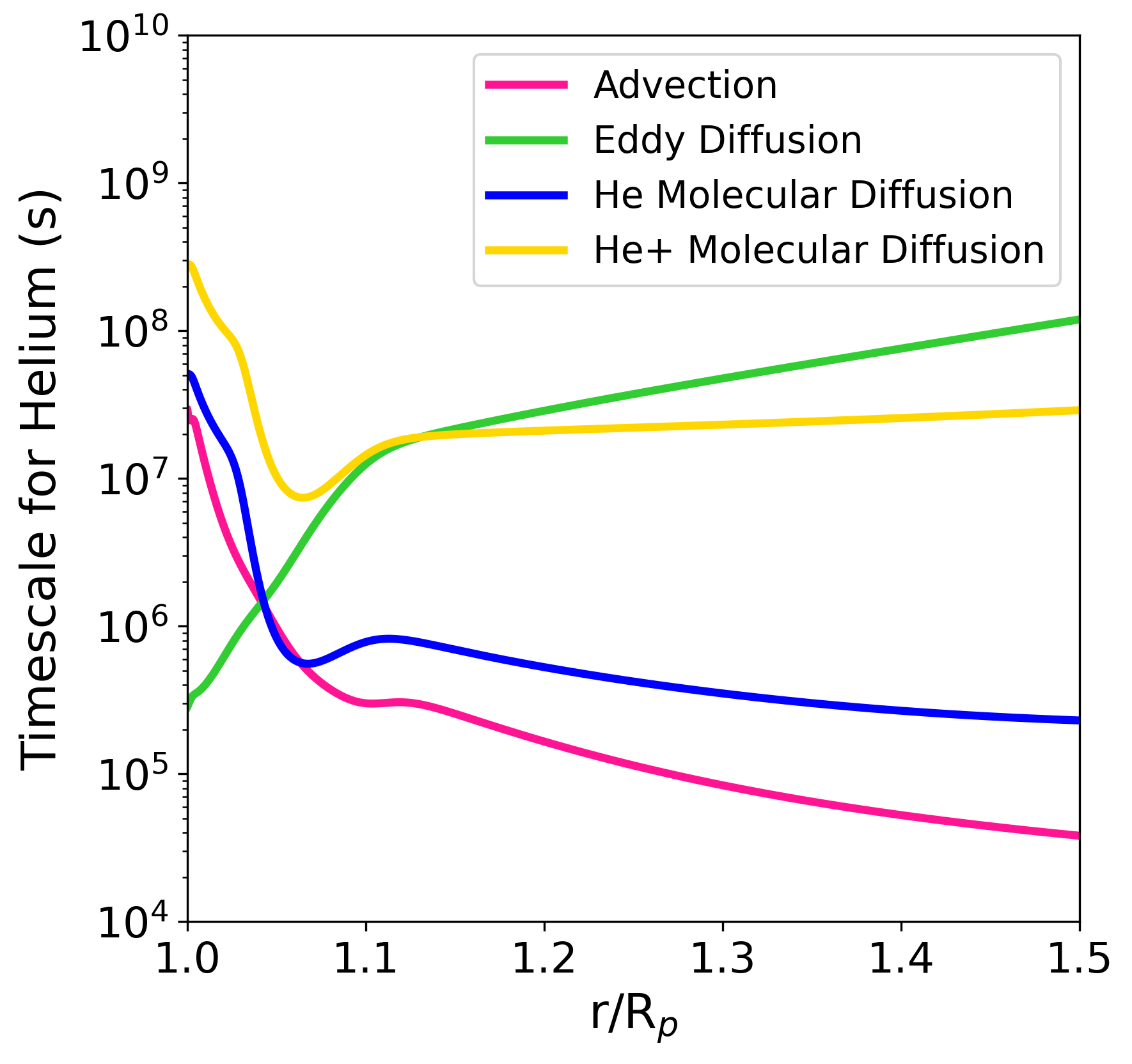}
    \includegraphics[width=.3\textwidth]{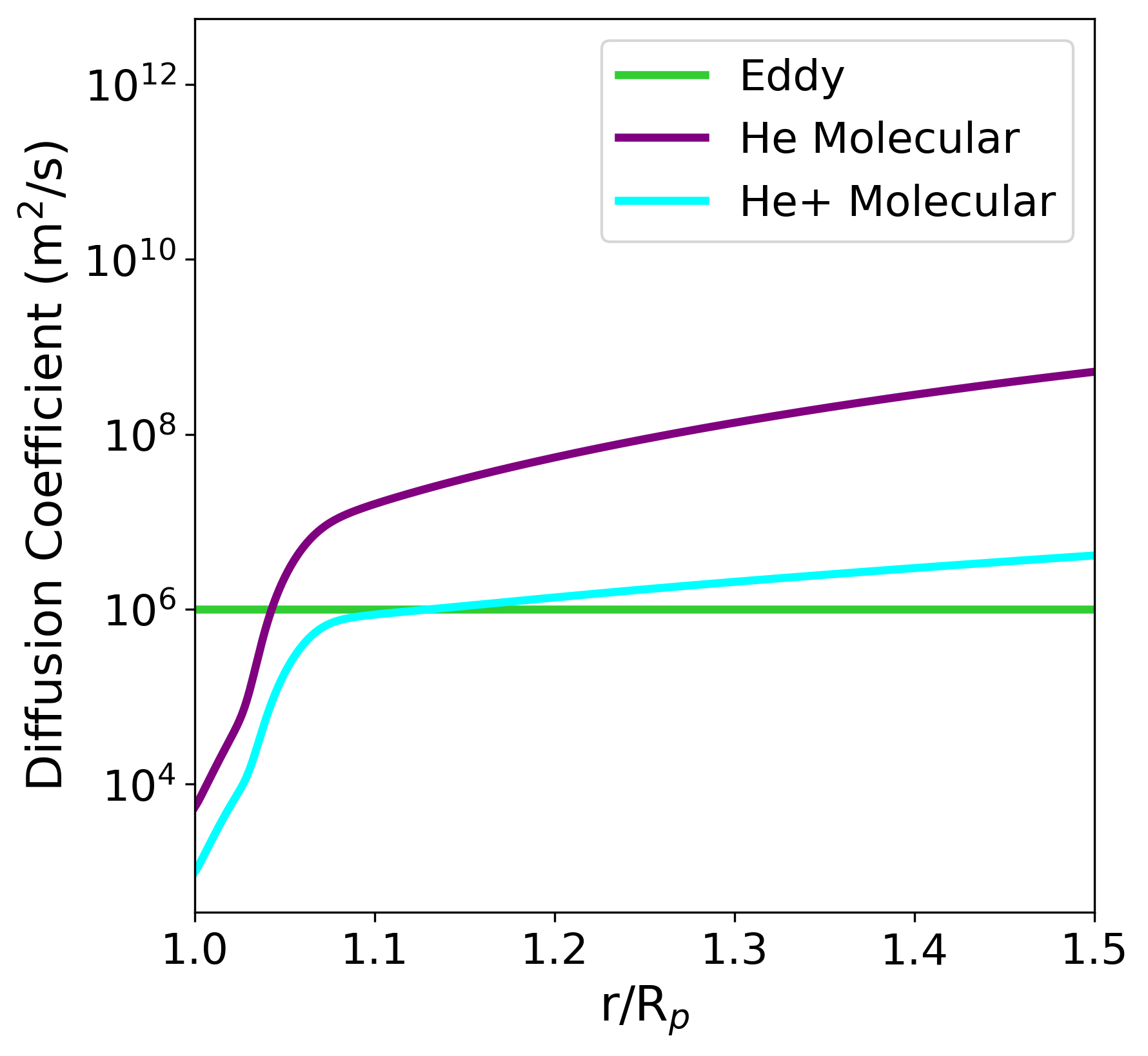}
    \caption{The elemental H/He ratio (\textit{left panel}) and transport and diffusion timescales for helium (\textit{middle panel}) and the eddy versus molecular diffusion coefficients (\textit{right panel}) for our upper atmosphere model (Model B). 
    }
    \label{fig:hhe}
\end{figure}

\citet{Lampon2020A&A...636A..13L} argued for a He/H ratio lower than 90/10 in the upper atmosphere of HD209458b, possibly due to diffusive separation, and we confirm that diffusive separation is possible. Our model produces similar results to those of \citet{Xing2023ApJ...953..166X} and \citet{Schulik2024arXiv241205258S} who solved separate momentum equations for different species, showing that the diffusion approximation to multi-species transport that we use works reasonably well. Figure \ref{fig:hhe} shows the hydrogen-helium ratio in Model B decreasing from 92/8 at the base of the thermosphere to a ratio of about 96/4 at high altitudes. The strongest separation happens around 1.05 R$_p$, which is also where the helium diffusion timescale is shorter than or comparable to the advection timescale and where the helium molecular diffusion coefficient becomes larger than the eddy diffusion coefficient (see the middle and right panel of Figure \ref{fig:hhe}). We note that this result depends on the stellar XUV flux and the assumed eddy mixing coefficient, where the latter was not included by \citet{Xing2023ApJ...953..166X} or \citet{Schulik2024arXiv241205258S}. A higher stellar flux would lead to a higher mass loss rate while a higher eddy diffusion coefficient would lead to more efficient mixing of He to altitudes where it can more easily escape. In both cases, the effect of diffusive separation would become less pronounced. 

\begin{figure}[h!]
    \centering
    \includegraphics[width=0.35\linewidth]{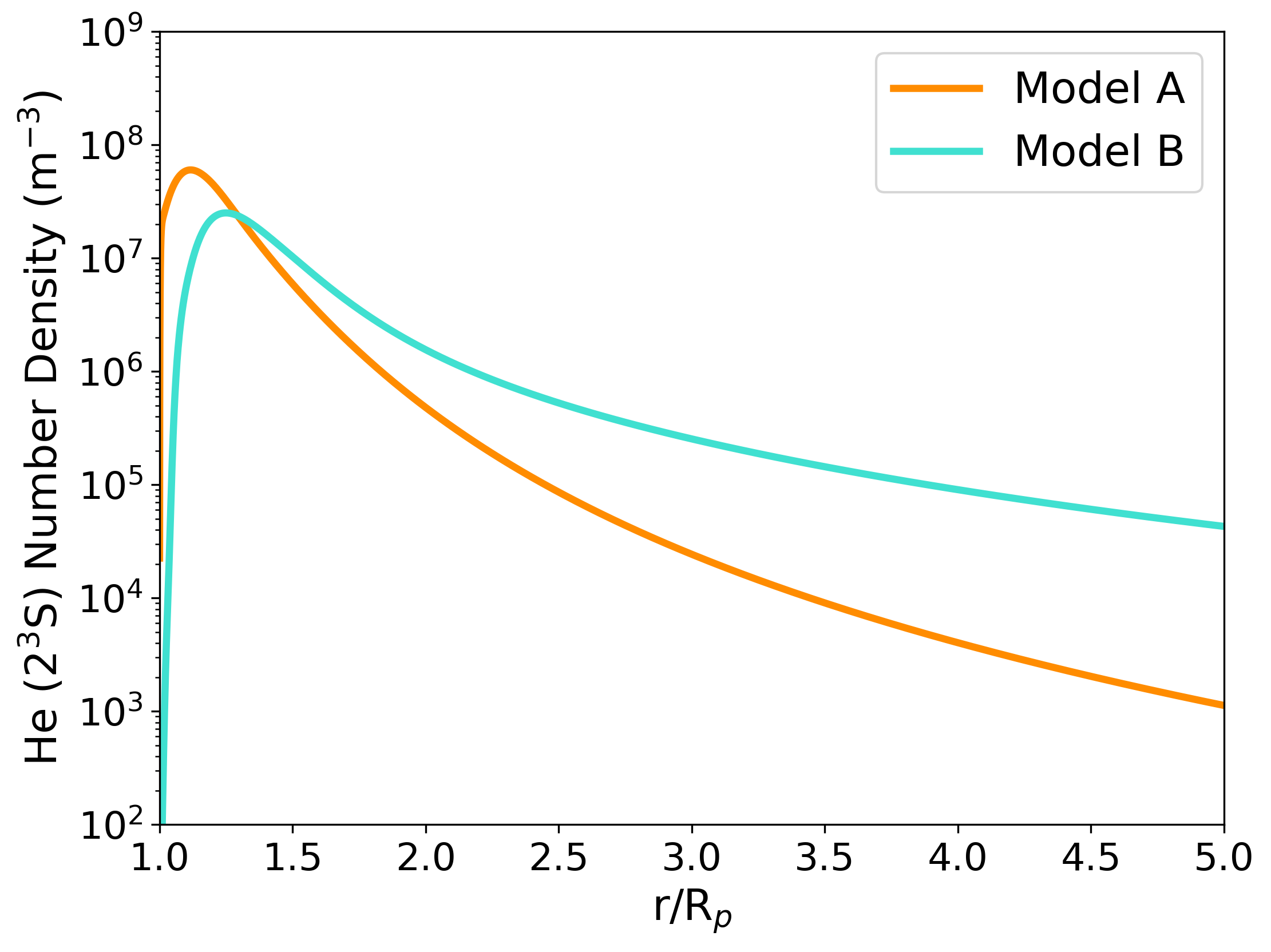}
    \includegraphics[width=0.3\linewidth]{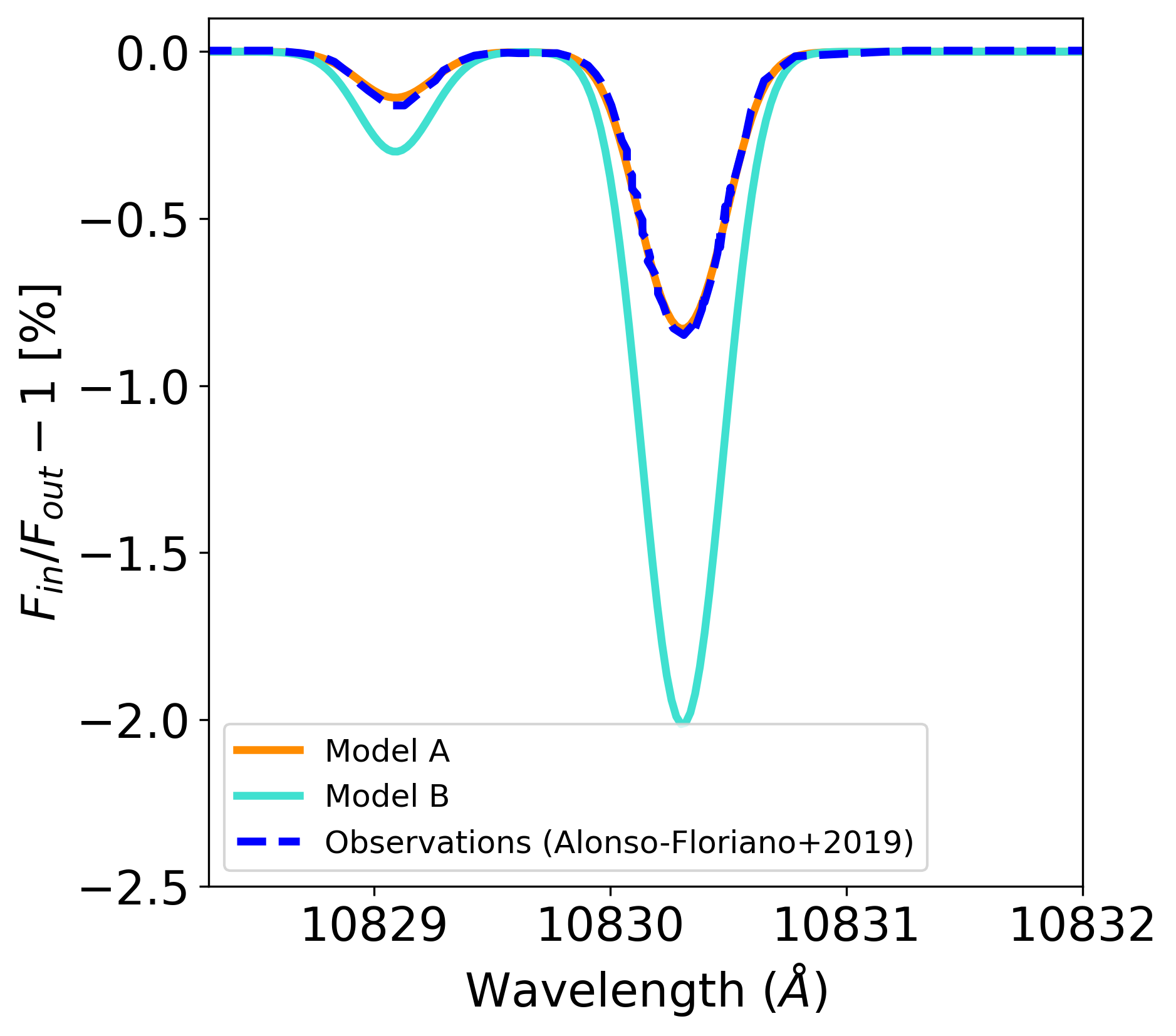}
    \includegraphics[width=0.28\linewidth]{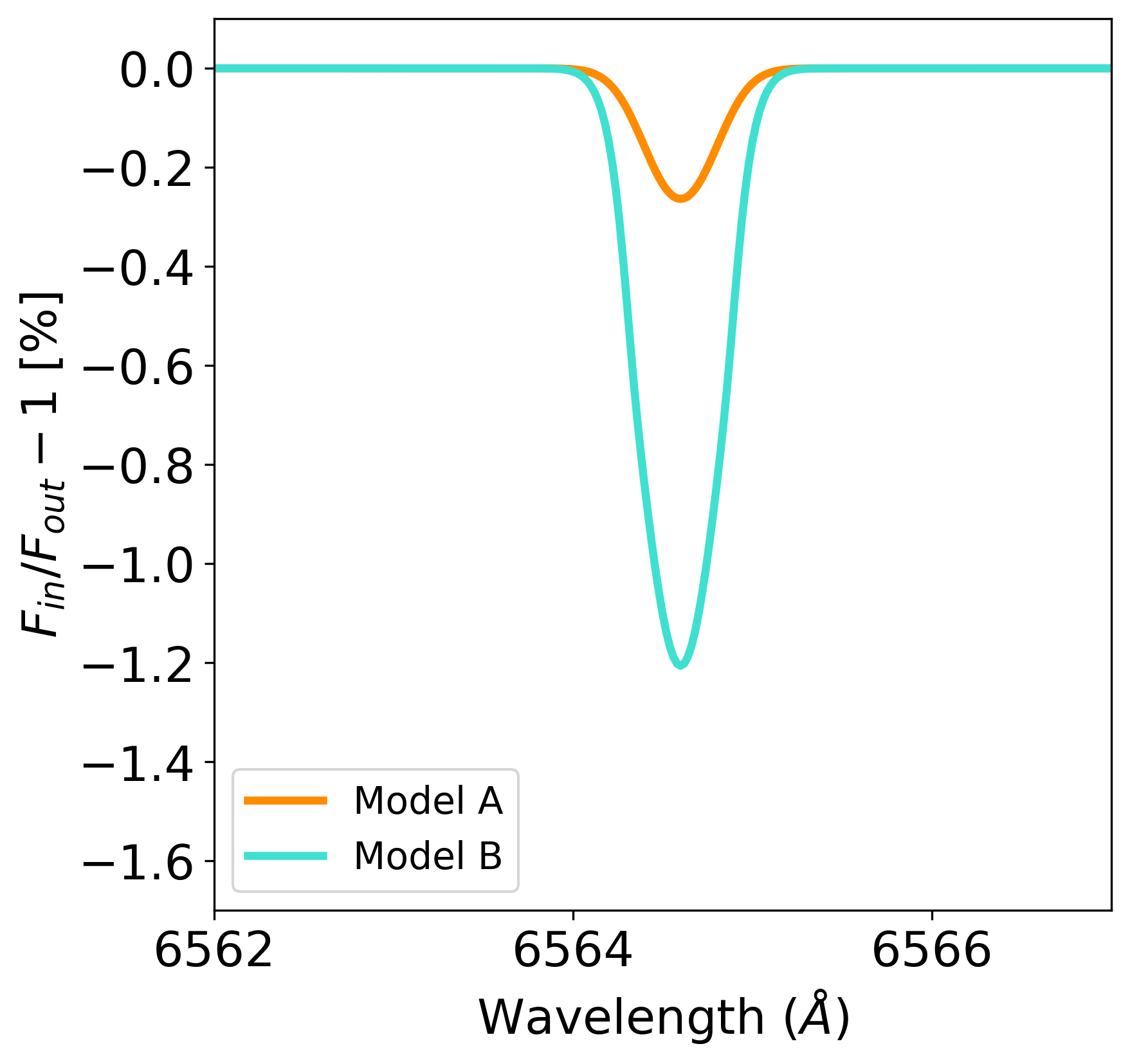}
    \caption{\textit{Left:} Excited helium number density comparison between our isothermal model (Model A) with no multi-species processes in orange and our upper atmosphere model (Model B) in teal. \textit{Middle: } Predicted transit depth for He I (2$^3$S) for our upper atmosphere model (Model B) in teal, compared to the best-fit fit line to the observations \citep{alonso2019A&A...629A.110A} in blue and the isothermal model in orange (Model A). \textit{Right: } Predicted transit depth for H$\alpha$ for our upper atmosphere model (Model B) in teal, compared to the isothermal model (Model A) in orange.}
    \label{fig:he23scomp}
\end{figure}

The differences between the isothermal (Model A) and self-consistent models (Model B) become evident when comparing the He I (2$^3$S) densities and transit depths (Figure \ref{fig:he23scomp}). The peak in excited helium density is lower in altitude but higher in magnitude in Model A compared to Model B. Since atmospheric extent has changed due to the change in lower boundary pressure, the peak in the excited helium density moved to higher altitudes producing a higher transit depth. While the mass-loss rate in Model B is less than a factor of two higher than that of the Model A, the two models produce different transit depths as shown in the middle and right panels of Figure \ref{fig:he23scomp}. Additionally, the pressure-averaged temperature above the 7 nbar lower boundary to the top of the model at $3\times 10^{-4}$ nbar in Model B (8860 K) is comparable to the constant temperature used in the isothermal model (8500 K) reinforcing that differences in atmospheric structure—rather than temperature or mass-loss rate alone—can also drive differences in He I 10830 \AA\ transit depths. Furthermore, the observed He~I and H$\alpha$ transit depths are inconsistent with a photoelectron heating efficiency of 100\%, suggesting that lower efficiencies (20–40\%) are required to reproduce observations.

\subsection{Full Atmosphere Model} \label{sec:fullatmo}

\noindent
Building on the upper atmosphere model, we now incorporate the lower and middle atmosphere to construct a coupled model of HD209458b’s vertical atmospheric structure, which we call Model C (see Table \ref{tab:modelsummary} for a summary of model configurations). 
By including the lower/middle atmosphere, we aim to assess its influence on the He I 10830 \AA\ and H$\alpha$ transit depths and compare the resulting visible to infrared spectrum predicted by the full model with observations from HST \citep{Sing2016Natur.529...59S} and JWST \citep{Xue2024ApJ...963L...5X}, in addition to exploring the He I 10830 \AA\ and H$\alpha$ line observations probing the upper atmosphere. By combining the upper atmosphere transit depths with the full spectrum, we obtain a self-consistent normalization for model differential transit depths that are compared with the observations (e.g., Figure~\ref{fig:he23scomp}), instead of simply normalizing the model transit depths with the assumption that the atmosphere is fully opaque below the lower boundary of the upper atmosphere model with negligible extent. Despite their critical role in regulating energy deposition and escape processes \citep{Chadney2017A&A...608A..75C,Huang__2023}, lower/middle atmosphere models have typically not been included in previous studies of He I 10830 \AA~absorption, potentially leading to incomplete or biased interpretations of exoplanet outflows.

\begin{figure}[h!]
    \centering
    \includegraphics[width=.31\textwidth]{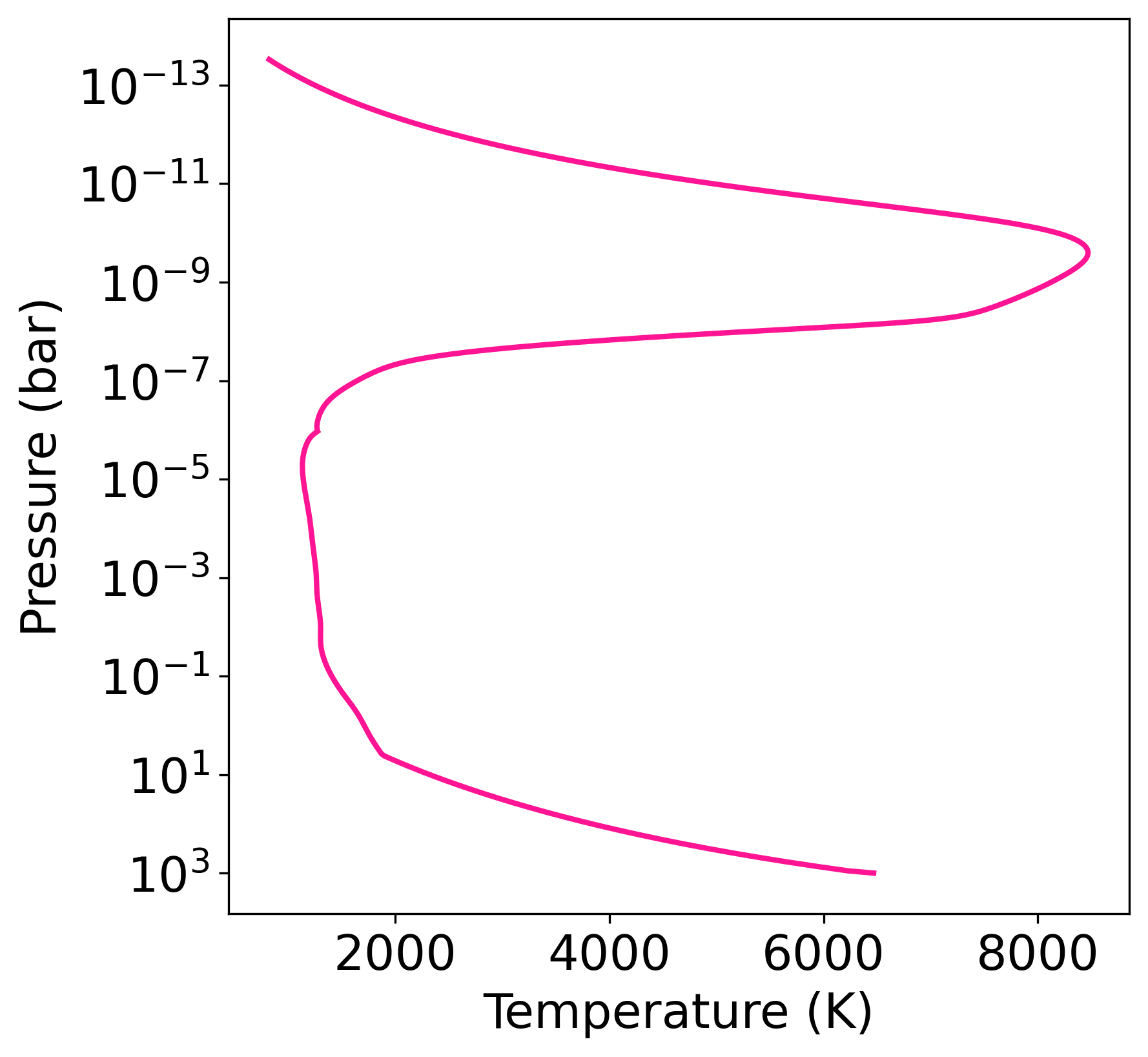}
    \includegraphics[width=.32\textwidth]{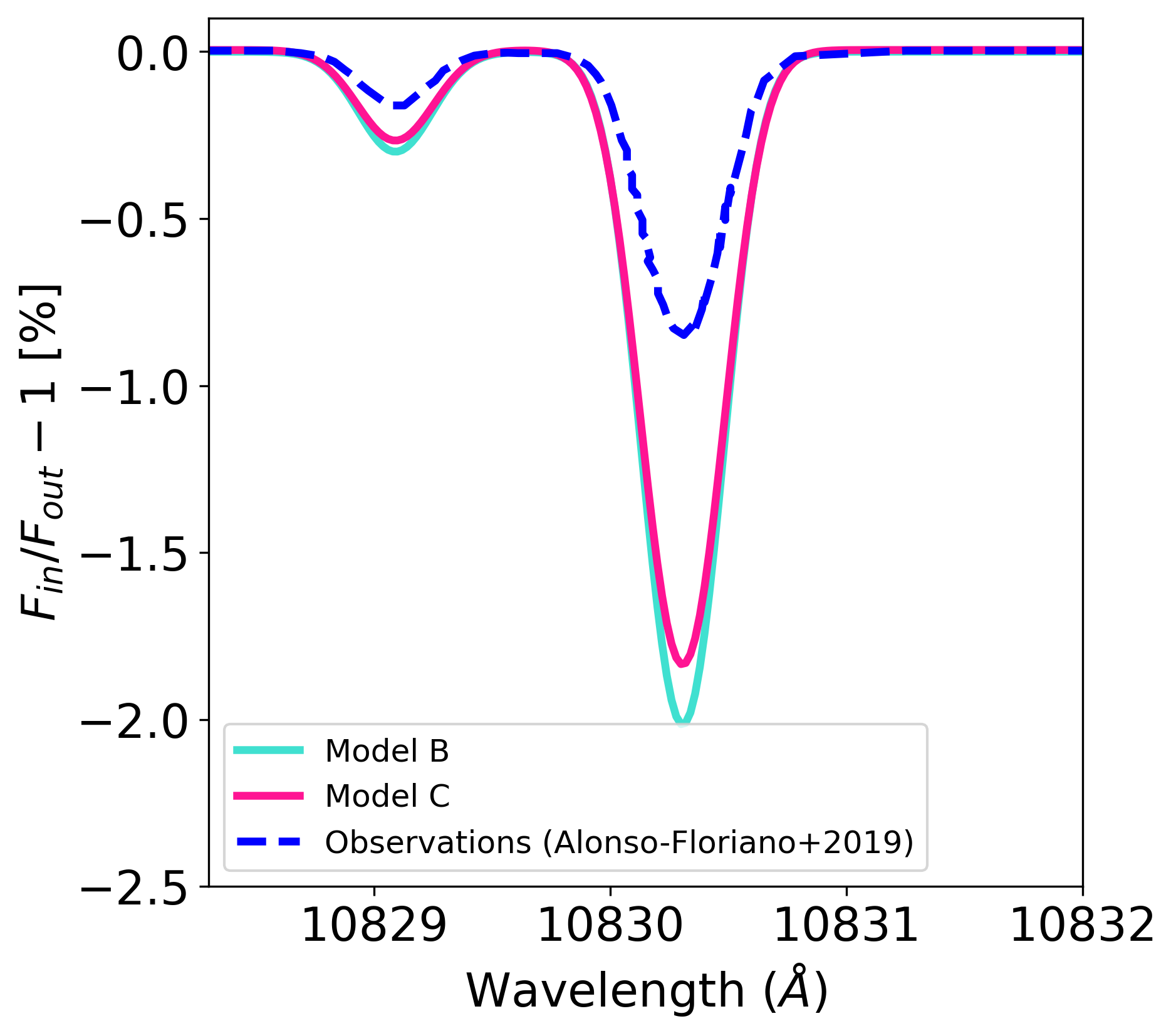}
    \includegraphics[width=.3\textwidth]{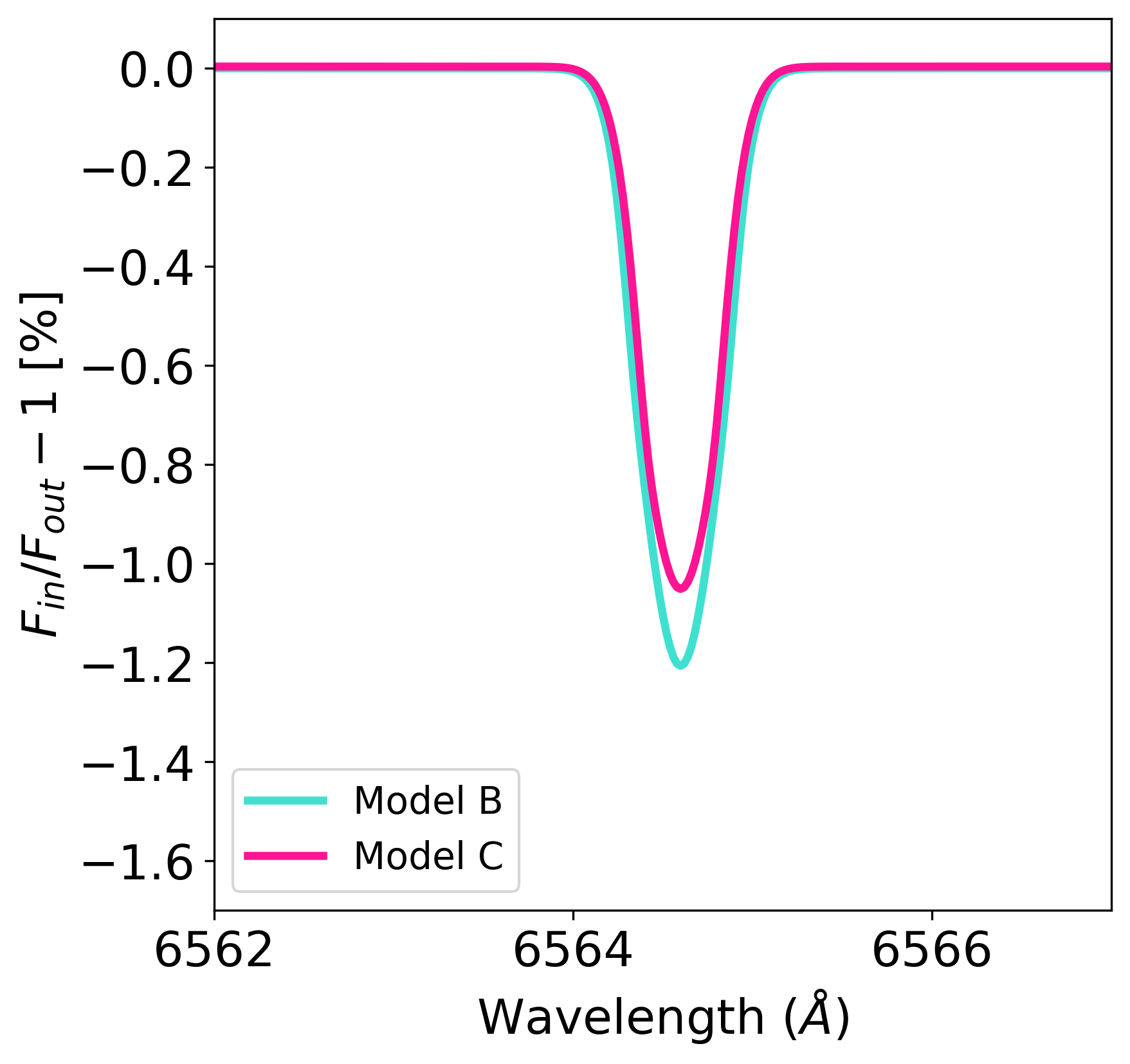}
    \caption{\textit{Left}: P-T profile of our full atmosphere model (Model C). \textit{Middle}: Transmission spectrum around the He I triplet. In teal is the self-consistent upper atmosphere model (Model B), and in pink is the full atmosphere model (Model C). \textit{Right}: Transmission spectrum around the H I Balmer $\alpha$ line. In teal is the self-consistent upper atmosphere model (Model B), and in pink is the full atmosphere model (Model C).}
    \label{fig:lowmid}
\end{figure}

The inclusion of the lower and middle atmosphere alters the overall atmospheric structure, as shown in Figure \ref{fig:lowmid}. The left panel presents the temperature-pressure profile of the full atmosphere model, illustrating a well-defined lower atmosphere that smoothly transitions into the thermosphere. In Model C, the total mass loss rate is found to be $5.5 \times 10^{10}$ g/s, compared to $3.5 \times 10^{10}$ g/s in Model B. Additionally, we compute the pressure-averaged temperature above the 7 nbar level and the 1 microbar level (up to the top of the model at $3\times 10^{-4}$ nbar) for both models. In Model B, the pressure-averaged temperature above 7 nbar is found to be $T = 8860 $ K, while in Model C, it is slightly lower at $T = 8640 $ K. Similarly, above 1 microbar, the pressure-averaged temperature is $T = 6040 $ K for Model B and $T = 5610 $ K for Model C. 
Figure \ref{fig:lowmid} compares the He I 10830 Å transit depth (middle panel) and H I Balmer $\alpha$ transit depth (right panel) for Models B and C. Model B (teal) predicts a slightly stronger He I 10830 \AA\ and H$\alpha$ absorption signal compared to Model C (pink). The decreased He I 10830 \AA\ and H$\alpha$ transit depths result from the change in lower boundary altitude and modified velocity structure, which reduce the excited hydrogen and helium abundances at higher altitudes. While the inclusion of the lower and middle atmosphere modifies the velocity and temperature structure and increases the total mass-loss rate by about 60\%, the predicted He I 10830~\AA\ and H$\alpha$ transit depths remain relatively similar. This suggests that coupling to the lower/middle atmosphere provides improvements in consistency and mass-loss estimates but does not dramatically change the modeled transit signatures of HD209458b. We also emphasize that the conclusions drawn in the previous subsection regarding the Parker wind models are largely independent of assumptions made about the lower and middle atmosphere and apply as long as the upper atmosphere is modeled self-consistently above the 1 $\mu$bar level. 


\begin{figure}[h!]
    \centering
    \includegraphics[scale =0.37]{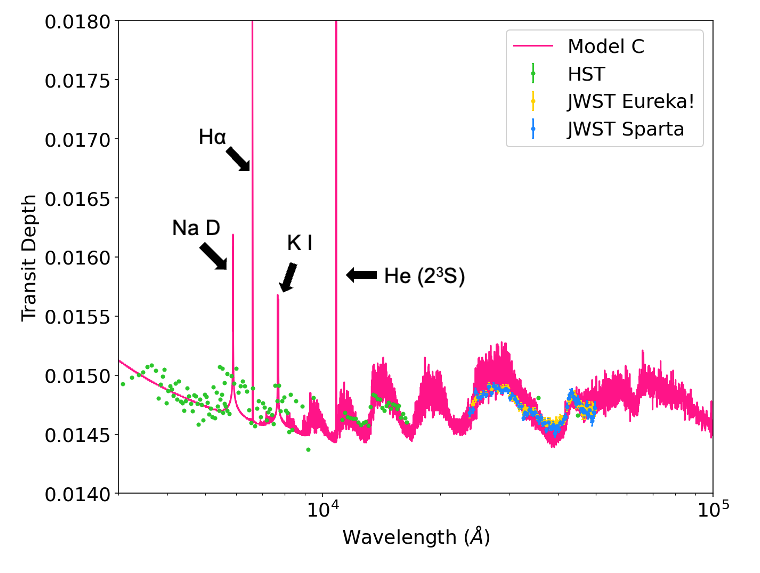}
    \caption{Transit depth versus wavelength for our full atmosphere model of HD209458b (Model C). We zoom in past the extent of the H$\alpha$ and He I 10830 \AA\ line to show the comparison with HST observations \citep{Sing2016Natur.529...59S} and JWST observations (with different data reduction pipelines in different colors) \citep{Xue2024ApJ...963L...5X}. The JWST observations analyzed can be accessed via \dataset[DOI:10.17909/f5j3-jq48]{https://doi.org/10.17909/f5j3-jq48}. Note that the Na D and K I lines are a result of only the lower/middle atmosphere model, and we do not calculate transit depths for these lines based on the upper atmosphere model.}
    \label{fig:totaltds}
\end{figure}

Since our atmospheric model includes both the upper and lower/middle atmosphere, we can generate a complete transmission spectrum for HD209458b. Figure \ref{fig:totaltds} presents the full transmission spectrum derived from the lower/middle atmosphere model, overlaid with observational data from HST and JWST. The lower/middle atmosphere model establishes the transit ‘continuum’ around key atomic absorption lines, providing a reference for differential transit depth calculations. This spectrum captures several prominent spectral features, including the Na D and K I lines originating in the lower/middle atmosphere, as well as the He I 10830 \AA\ triplet and H I Balmer $\alpha$ line, which probe the extended escaping atmosphere. Note that the Na D and K I lines are a result of only the lower/middle atmosphere model, and we do not calculate transit depths for these lines based on the upper atmosphere model. The model successfully reproduces the continuum shape across a broad wavelength range, with the HST and JWST data aligning well with the modeled continuum, particularly in the infrared, further validating the atmospheric structure used in this study. To accurately compare our model predictions with observations from HST and JWST, we accounted for a systematic offset between the transit depths reported in the literature. Previous analyses have noted discrepancies between transit depth measurements, but the origin of these offsets remains unclear. We derived the optimal shift using the Bayesian Information Criterion (BIC) \citep{Kass01061995} to determine the most likely alignment between the datasets. We define the optimal shift \( \Delta d \) as the value that minimizes the BIC when applied to the JWST transit depths. The BIC is given by
\begin{equation}
    \text{BIC} = k \log(n) - 2 \log(\mathcal{L}),
\end{equation}
where \( k = 1 \) is the number of free parameters (the shift), \( n \) is the number of observed data points, and \( \mathcal{L} \) is the likelihood function given by
\begin{equation}
    \log(\mathcal{L}) = - \frac{1}{2} \sum_i \left[ \left( \frac{T_{o,i} - T_{m,i}}{\Delta T_{o,i}}\right) + \log(2\pi \Delta T_{o,i}^2 ) \right]
\end{equation}
where $T_m$ is the model transit depth, $T_o$ is the observed transit depth, and its uncertainty $\Delta T_o$. We performed a minimization of the BIC as a function of the shift parameter, separately for the JWST Eureka! and Sparta datasets. We found optimal shift values of \( \Delta d = 0.000169 \) for the JWST Eureka! data and \( \Delta d = 0.000179 \) for the JWST Sparta data. These values were applied to the JWST transit depths in our analysis to align them with the HST observations and our model predictions.

\begin{table}[h!]
\centering
\caption{Summary of Model Configurations}
\label{tab:modelsummary}
\scriptsize
\begin{tabular}{lccccc}
\hline
\textbf{Feature} & \textbf{Model A} & \textbf{Model B} & \textbf{Model C} & \textbf{Best-fit Model} & \textbf{Metals Model} \\
\hline
Name & Isothermal Model & Self-consistent  & Full Atmosphere Model & Best-Fit Model & Metal Run \\
& & Upper Atmosphere &  & & \\

\hline
Temperature Structure & Fixed (8500 K) & Energy equation & Eenergy equation & Energy equation & Energy equation \\
& & & and lower/middle  & and lower/middle &  and lower/middle \\
& & & atmosphere & atmosphere &  and atmosphere \\
\hline
Photoelectron Heating  & N/A & 100\% & 100\% & 40\% & 23\% \\
Efficiency & & & & &\\ 
\hline
Lower Boundary Pressure & 7 nbar & $10^{-6}$ bar & $10^{-6}$ bar & $10^{-6}$ bar & $10^{-6}$ bar \\
Upper Atmosphere Model & & & & & \\
\hline
Lower Boundary Altitude  & 0 km & 0 km & 5152 km & 0 km & 0 km \\
Upper Atmosphere Model & & & & & \\
\hline
Diffusion Processes & None & Multi-species & Multi-species & Multi-species  & Multi-species  \\
\hline
Revised Rates & Yes & Yes & Yes & Yes & Yes \\
\hline
He/H Ratio & Fixed (92/8) & Evolving  & Evolving  & Evolving  & Evolving  \\
\hline
Lower/Middle Atmosphere & Not included & Not included & Included & Included & Included \\
\hline
Metals Included & No & No & No & No & Yes \\
\hline
Mass-loss Rate (g/s) & $1.9 \times 10^{10}$ & $3.5 \times 10^{10}$ & $5.5 \times 10^{10}$ & $1.9 \times 10^{10}$ & $3.0 \times 10^{10}$ \\
\hline
\end{tabular}
\end{table}
\subsection{Best Fit Model} \label{sec:bf}

\begin{figure}[h!]
    \centering
    \includegraphics[width=.32\textwidth]{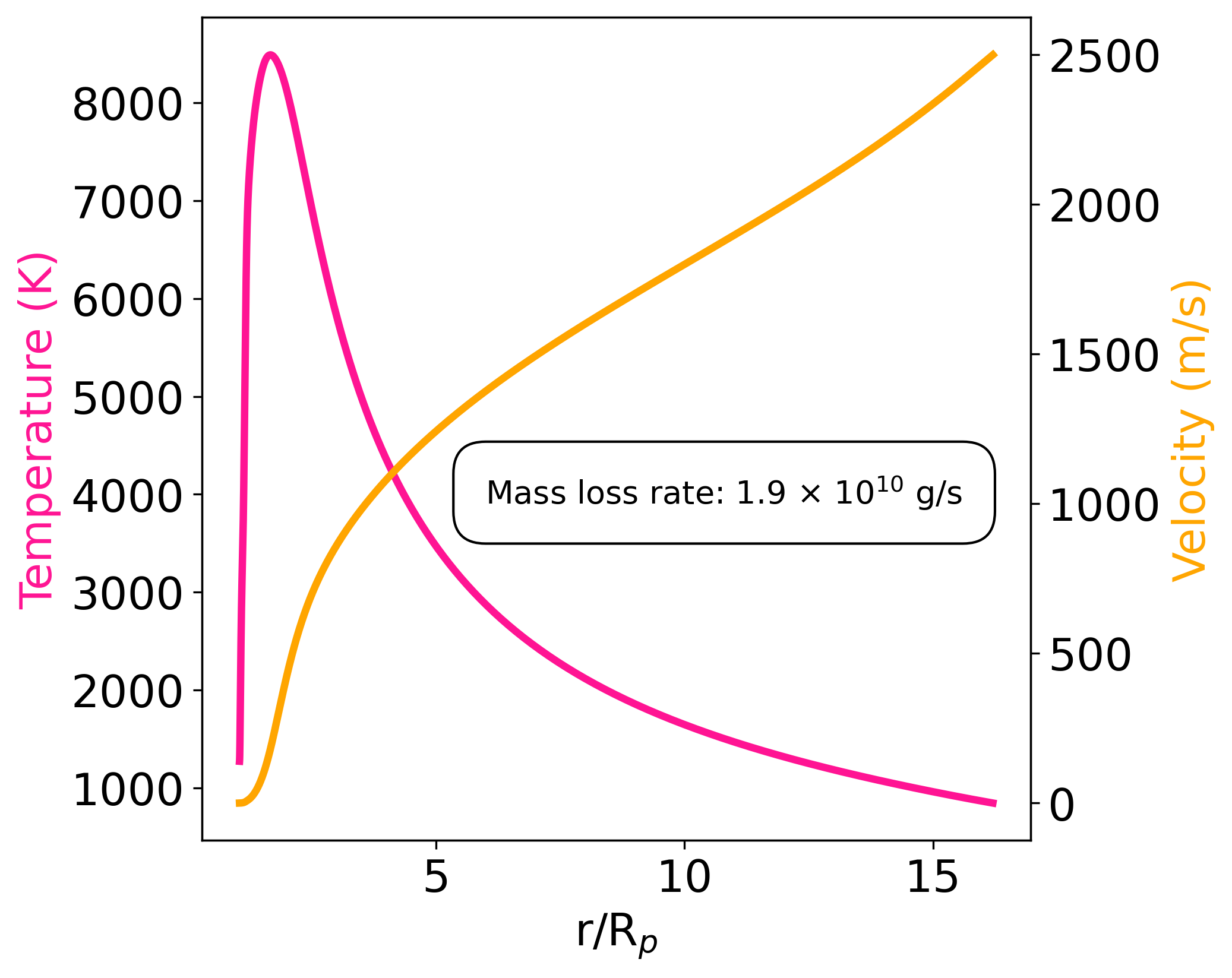}
    \includegraphics[width=.37\textwidth]{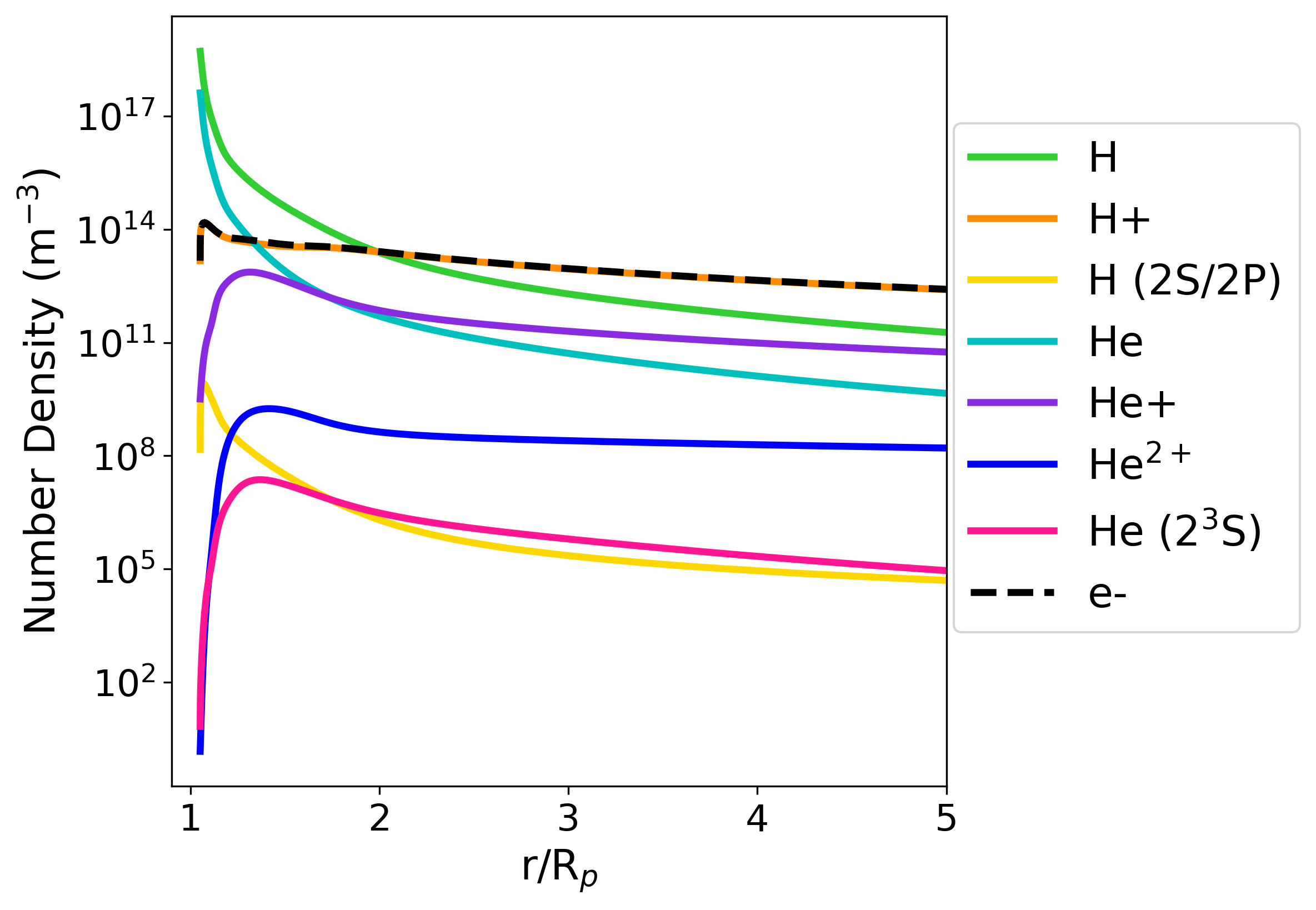}
    \includegraphics[width=.28\textwidth]{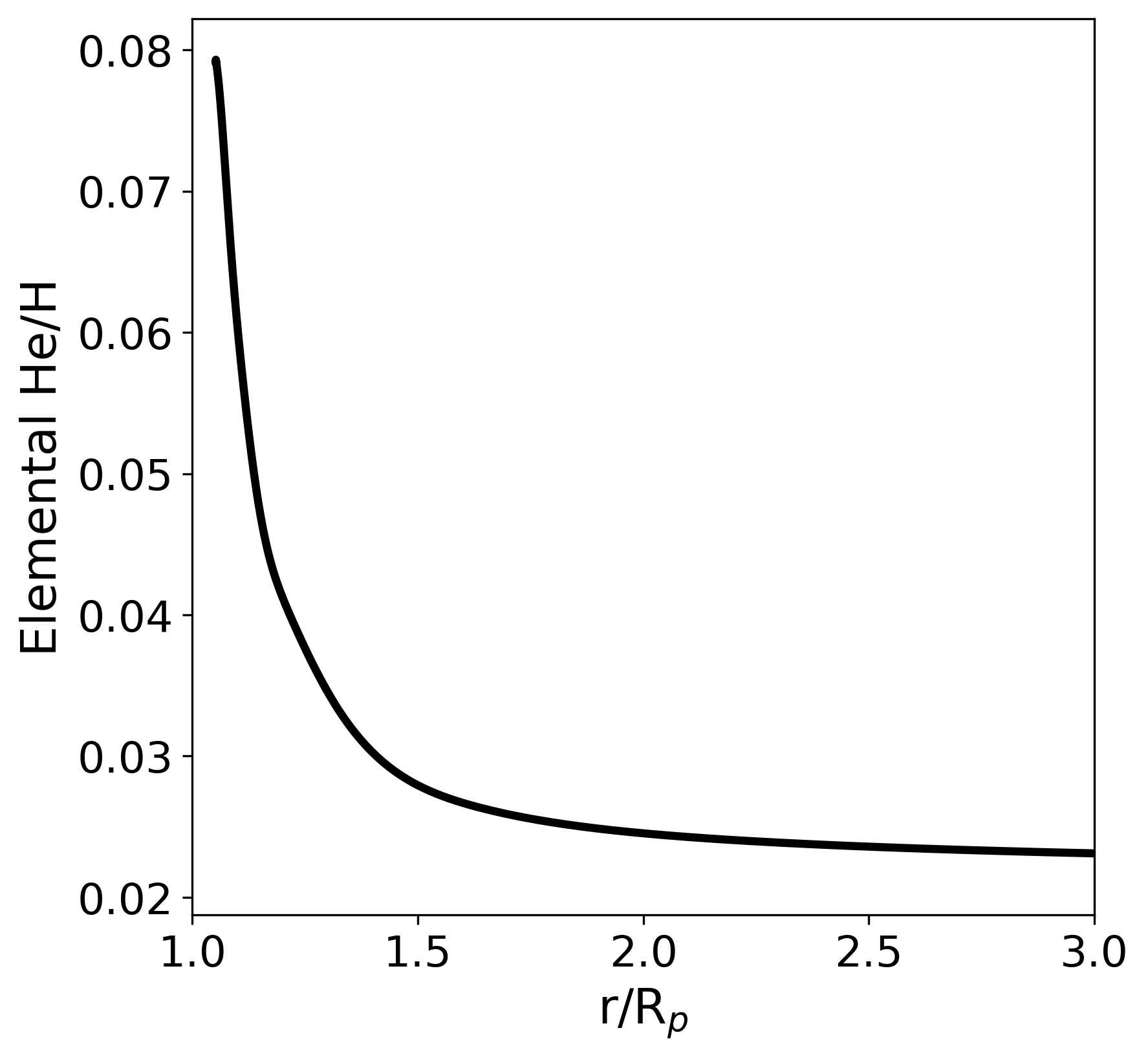}
    \caption{\textit{Left: }Temperature and bulk outflow velocity predicted by our best-fit atmosphere model for HD209458b. \textit{Middle: }Composition predicted for HD209458b. \textit{Right: } The elemental H/He ratio for our best-fit model.
    }
    \label{fig:bfstrc}
\end{figure}

\noindent
After exploring the impact of different assumptions about atmospheric structure, we now present our best-fit H/He model. The free parameters of this model include the photoelectron heating efficiency, the eddy diffusion coefficient, and the activity level of the star, none of which are well constrained for this system. We find that achieving a fit to the He I 10830 \AA\ transit depth that does not violate the upper limit on the H$\alpha$ transit depth set by \citep{jenesen2012ApJ...751...86J} requires a photoelectron heating efficiency of 40\% or lower, leading to a reduced mass-loss rate compared to our previous estimates. The left panel of Figure \ref{fig:bfstrc} shows the temperature and velocity profiles for the best-fit model. The temperature peaks at approximately 8200 K near 1.7 R$_p$, lower than in Model C of Figure \ref{fig:tempv} due to the reduced heating efficiency. The outflow velocity reaches 1.1 km/s at 4 R$_p$, indicating a slower, yet still significant, atmospheric escape compared to the run with 100\% photoelectron heating efficiency. The corresponding mass loss rate is $1.9 \times 10^{10}$ g/s, which is lower than in Model C due to the reduced energy input but still within a physically plausible range for HD209458b. 
The pressure-averaged temperature between 1 $\mu$bar and the top of the model at a pressure of $3\times 10^{-4}$ nbar is 5336 K, and above 7 nbar it is 7970 K. Somewhat by design, the temperature and mass-loss rate in our best-fit model are similar to those derived from the isothermal Parker wind solution (Model A). This indicates that it is possible for the isothermal model to produce a fit that is close to a more self-consistent solution. The isothermal Parker wind framework, however, yields a family of solutions in which mass-loss rate, temperature, and helium abundance can be freely adjusted to match observations, without enforcing a physically consistent atmospheric structure. In contrast, our self-consistent model determines a solution by coupling energy balance, multi-species transport, and radiative processes, ensuring that the temperature, velocity, and density profiles emerge from the underlying physics rather than being arbitrarily prescribed. Additionally, our approach allows us to estimate the photoelectron heating efficiency by matching the observed transit depth, which is particularly valuable given the challenges and uncertainties associated with calculating such heating efficiencies from first principles in exoplanetary atmospheres. The composition of the escaping atmosphere is shown in the middle panel of Figure \ref{fig:bfstrc}, highlighting the number densities of key species. The distribution of neutral and ionized species follows trends consistent with our previous models, with stronger diffusive separation of helium occurring in the right panel of Figure \ref{fig:bfstrc}. The elemental He/H ratio decreases from 8\% at the base of the thermosphere to approximately 2.5\% at high altitudes. This enhanced separation results from the lower mass loss rate allowing molecular diffusion to dominate over a larger region of the atmosphere. 

\begin{figure}[h!]
    \centering
    \includegraphics[width=.4\textwidth]{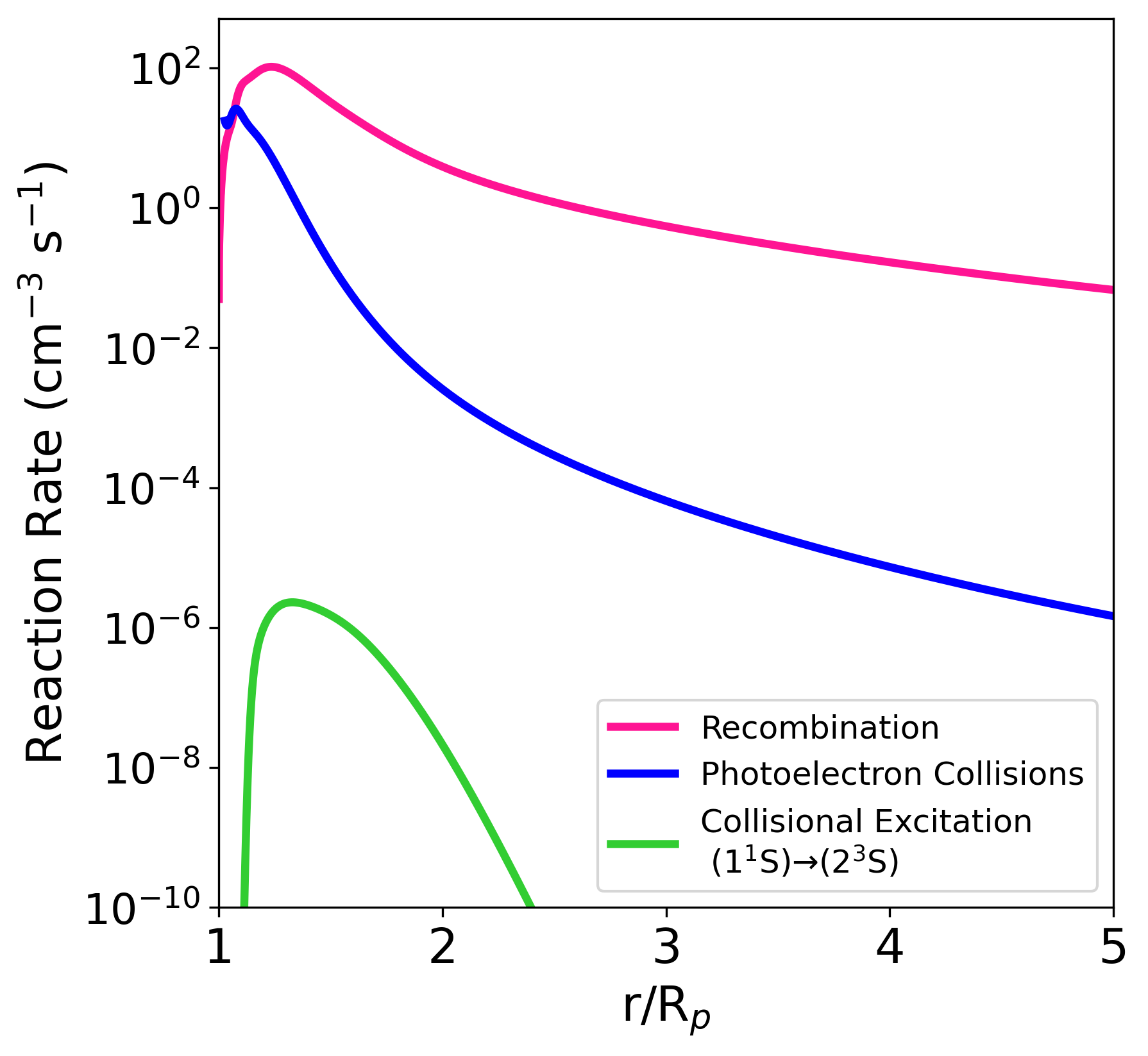}
    \includegraphics[width=.4\textwidth]{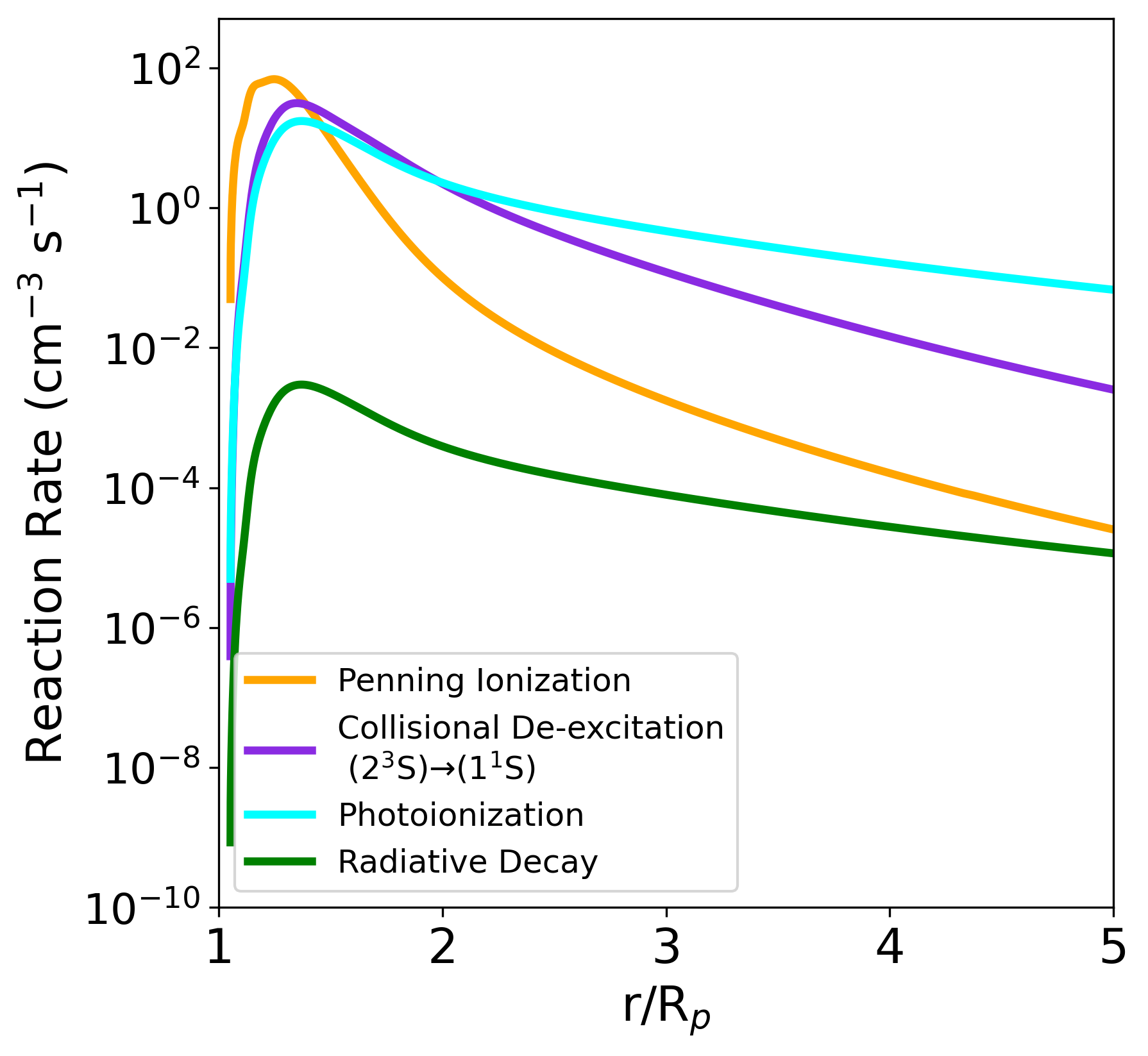}
    \caption{\textit{Left: }Production mechanisms relevant for the formation of He I (2$^3$S). \textit{Right: }Loss mechanisms relevant for the formation of He I (2$^3$S).
    }
    \label{fig:bfenergy}
\end{figure}

\begin{figure}[h!]
    \centering
    \includegraphics[width=.4\textwidth]{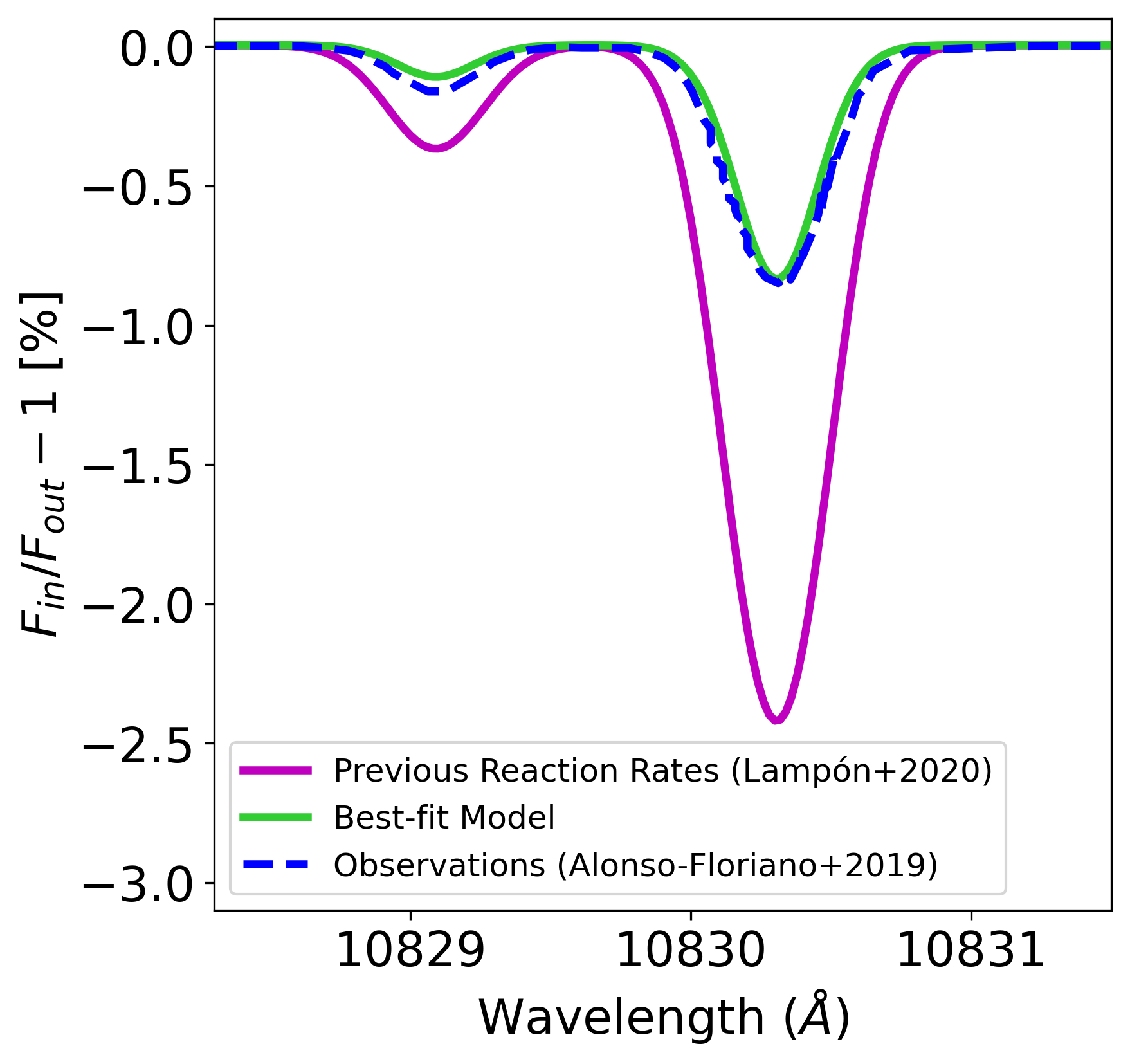}
    \includegraphics[width=.42\textwidth]{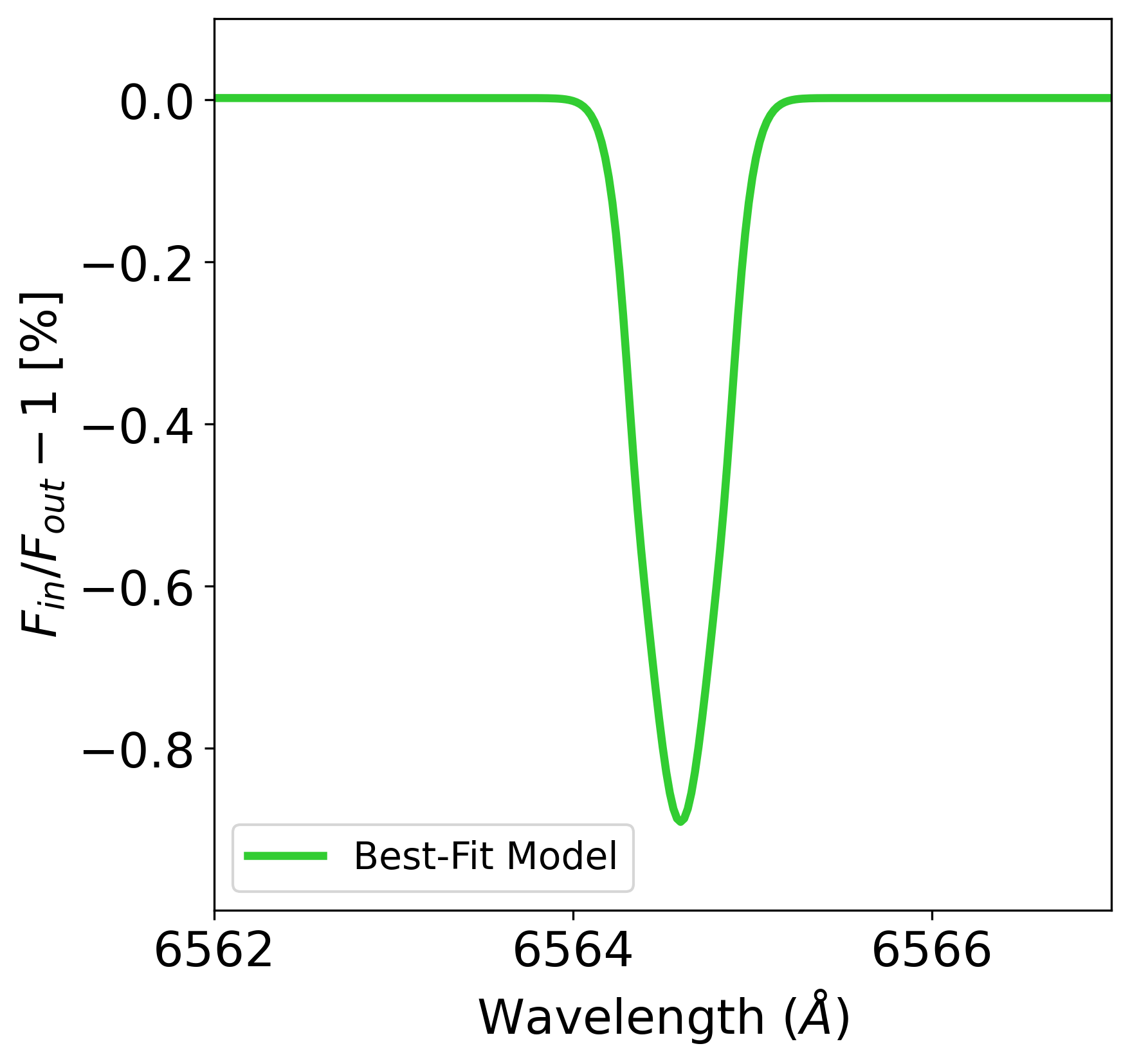}
    \caption{Transit depth results from our best-fit model, comparing the modeled He I 10830 Å (left) and H I Balmer $\alpha$ (right) transit depths from our best-fit model (green solid lines) with the best-fit to the observed transit depth for the He I 10830 \AA~line from \citet{alonso2019A&A...629A.110A} (blue line). He I 10830 Å transit depth results using previously used reaction rates shown in magenta \citep{Lampon2020A&A...636A..13L}.}
    \label{fig:bftds}
\end{figure}

The loss and production rates relevant to He I (2$^3$S) are presented in Figure \ref{fig:bfenergy}, respectively. Recombination remains the dominant production mechanism at all altitudes. Although photoelectron collisions can be important near the lower boundary of our model, they do not significantly affect the He I 10830 \AA\ transit depth on HD209458b. Penning ionization dominates loss at the lowest altitudes (r $<$ 1.5 R$_p$), with collisional de-excitation briefly dominating from (1.5 R$_p$ $<$ r $<$ 2 R$_p$), and finally photoionization dominating past 2 R$_p$. While our best-fit model uses updated reaction rates (outlined in Section \ref{sec:chem}), uncertainties in these rates can significantly impact the predicted He I 10830 \AA\ transit depth.  To test the sensitivity of our results to reaction rates, we run our best-fit model with reaction rates outlined in \citet{Lampon2020A&A...636A..13L}. As shown in the left panel of Figure \ref{fig:bftds}, differences in these rates lead to non-trivial changes in the He I 10830 \AA\ transit depth. The changes in the He I 10830 \AA\ transit depth are mainly due to the new photoionization cross-section and the ground state helium recombination rate. In \cite{Lampon2020A&A...636A..13L}, they use a lower ground state recombination rate than our model, allowing there to be more ionized helium and electrons in the atmosphere than in our model, leading to more recombination to the metastable state. This comparison is limited to the rates listed by \cite{Lampon2020A&A...636A..13L} and comparisons with other works are expected to differ.

The best-fit model successfully reproduces the observed He I 10830 Å absorption feature \citep{alonso2019A&A...629A.110A}. Our results are also consistent with the upper limit of about -1\% on H$\alpha$ absorption at line center set by the observations of \citet{jenesen2012ApJ...751...86J}, with our model predicting an H$\alpha$ excess absorption of -0.9\% at line center. Our H$\alpha$ transit depth is also consistent with the broad upper limit from \citet{Winn2004}. Using the same observations as \citet{Winn2004}, however, \cite{Astudillo-Defru2013A&A...557A..56A} inferred a transit depth of -0.123$\pm$0.012\% detection with a 1.125 \AA\ bandpass while the transit predicted by our model for this bandpass is higher at -0.4\%. Based on the observations of \citet{Casasayas-Barris2021A&A...647A..26C}, we may infer a rough upper limit of -0.3\% for a 0.5 \AA\
bandpass while our model predicts a transit depth of -0.73\% for this bandpass. We note that the H$\alpha$ observations are not simultaneous with the He I 10830 \AA\ transit observations. Future high-resolution observations could help further constrain the H$\alpha$ absorption and its variability to provide additional insight into the hydrogen ionization structure in escaping atmospheres.

\subsection{Sensitivity of Transit Depths to Stellar Activity and Other Factors} \label{sec:param}

\noindent
To assess the key physical mechanisms driving the transit depths, we investigate the effects of stellar activity, diffusion processes, and tidal forcing. Stellar activity variations influence the high-energy flux incident on the atmosphere, altering the heating rate, mass-loss rate, and photoionization rates. Diffusion processes, including molecular and eddy diffusion, regulate the vertical transport of species and determine the degree of diffusive separation between hydrogen and helium. Tidal forcing modifies the gravitational potential and velocity structure, affecting the efficiency of mass loss. By systematically varying these processes, we aim to identify their individual and combined effects on atmospheric structure and the observed transit depths.

\subsubsection{Stellar Activity Level}

\noindent
To investigate the effect of changing stellar activity levels on the upper atmosphere and the He I 10830~\AA~and H$\alpha$ transit depths, we run our best-fit model using stellar spectra corresponding to solar minimum, solar average, and solar maximum conditions (as described in Section \ref{sec:stellar}). These spectra differ in both their overall XUV flux and overall spectral energy distribution (SED) and are all scaled to the orbital distance and stellar surface radius of HD209458b. We find that increasing stellar activity leads to higher mass-loss rates. Specifically, the mass-loss rate rises from $1.9 \times 10^{10}$ g/s in the solar minimum case to $3.2 \times 10^{10}$ g/s in the solar average case, and to $4.2 \times 10^{10}$ g/s in the solar maximum case. This enhancement is primarily due to increases in both the outflow velocity and density (for example, taken here at 4 R$_p$, the approximate Roche Lobe radius). In the solar minimum case the velocity and density at 4 R$_p$ are 1.1 km/s and 8.78$\times 10^{-15}$ kg/m$^3$, respectively, whereas in the solar average case, they increase to 1.5 km/s and 1.10$\times 10^{-14}$ kg/m$^3$, respectively, and in the solar maximum case to 1.7 km/s and 1.25$\times 10^{-14}$ kg/m$^3$, respectively. The pressure-averaged temperature between 1 $\mu$bar and the top of the model at $3\times 10^{-4}$ nbar is 5720 K in the average case and 5906 K in the maximum case, compared to 5336 K in the minimum case. The pressure-averaged temperature between 7 nbar and the top of the model at $3\times 10^{-4}$ nbar is 8690 K in the average activity case and 9021 K in the maximum case, compared to 7970 K in the minimum case. 

\begin{figure}[h!]
    \centering
    \includegraphics[width=.4\textwidth]{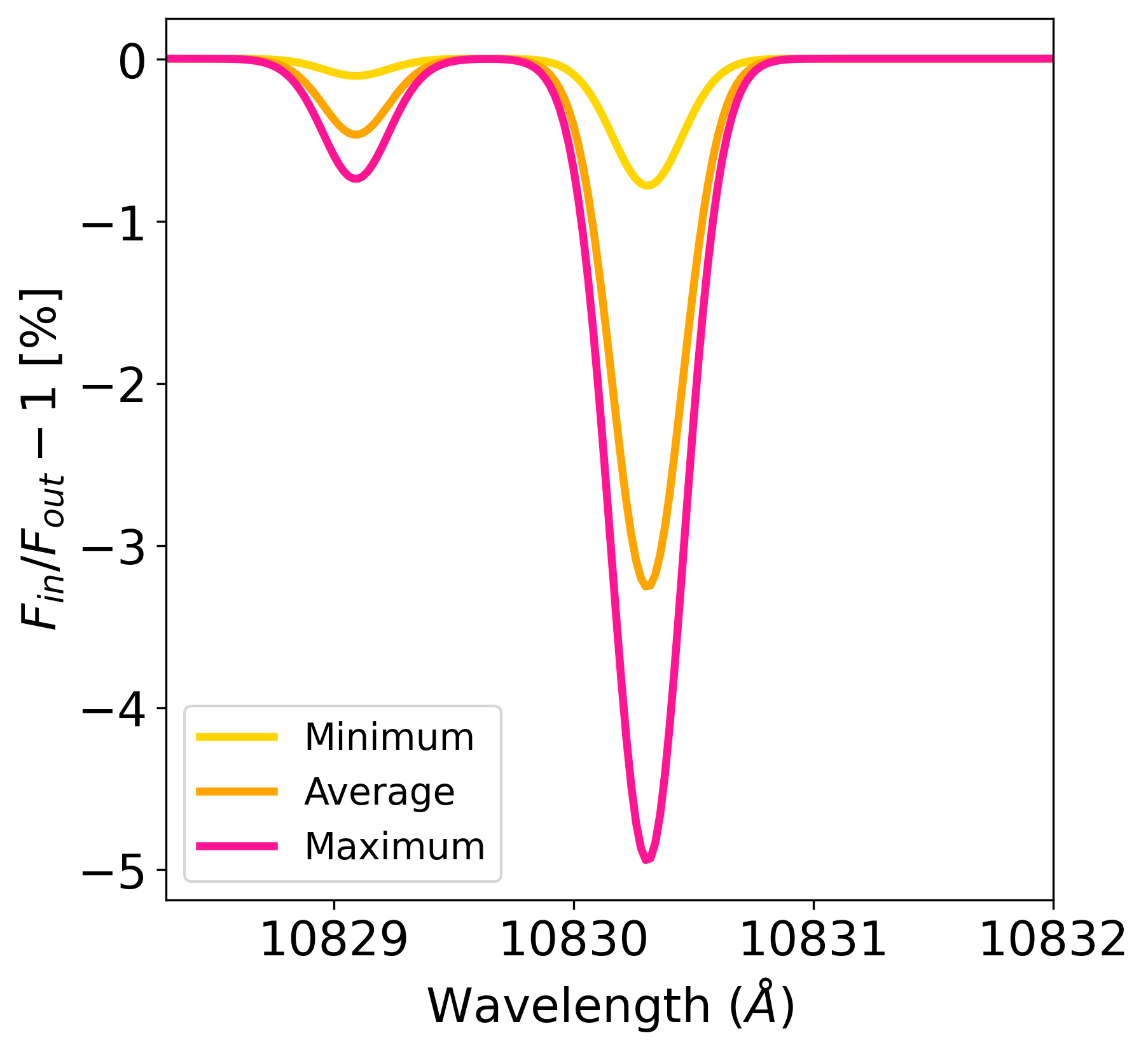}
    \includegraphics[width=.42\textwidth]{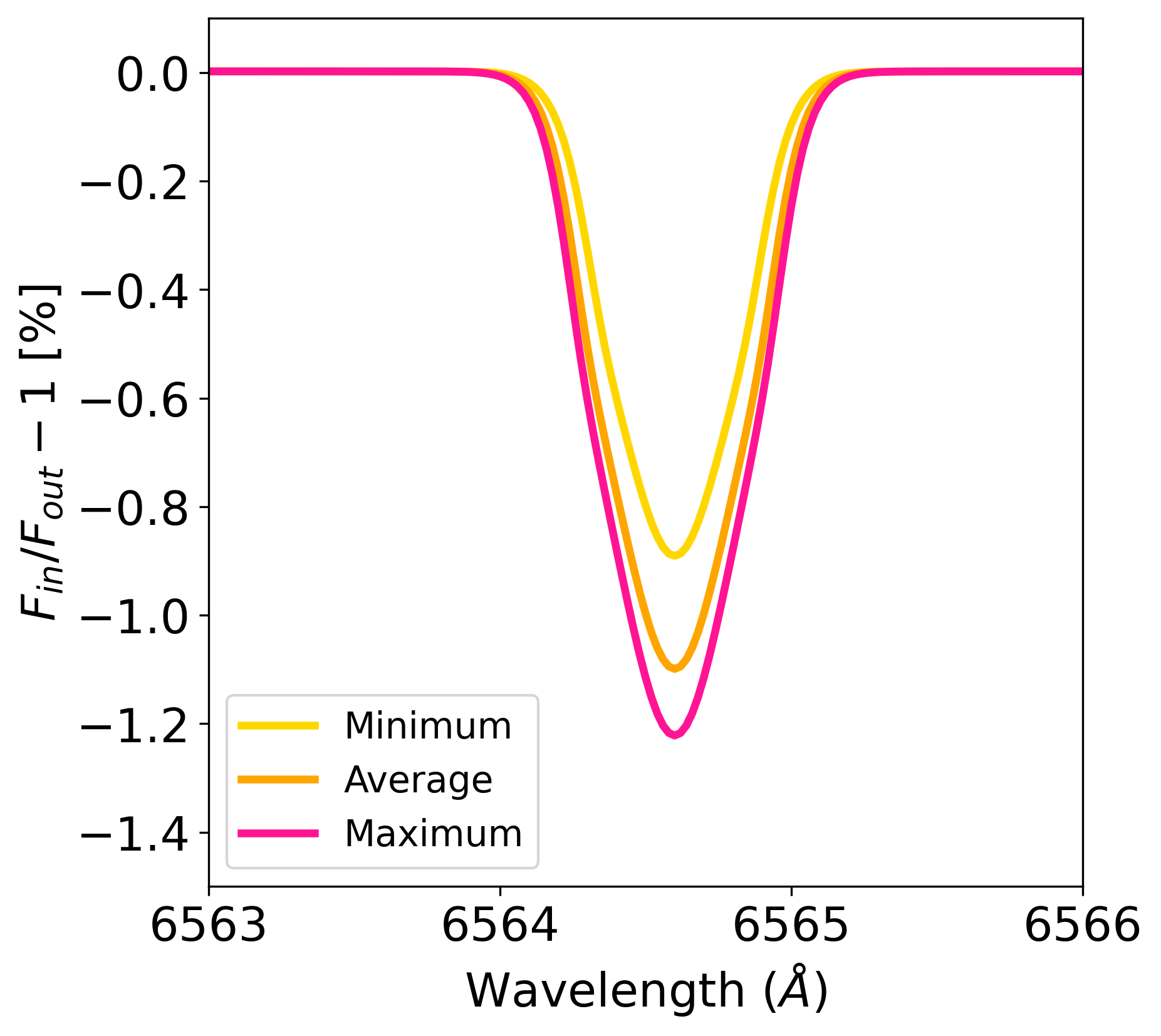}
    \caption{Transmission spectrum around the He I triplet (left) and the H-$\alpha$ Balmer line (right). The transit depth produced using a maximum, average, and minimum solar spectrum is in pink, orange, and yellow respectively.}
    \label{fig:stellaratv}
\end{figure}

The degree of diffusive separation decreases with higher stellar activity, with the elemental He/H ratio at the top of the atmosphere increasing from 0.025 in the solar minimum case to 0.04 and 0.055 in the solar average and maximum cases, respectively. As predicted by previous studies \citep{oklop2019ApJ...881..133O, Krishnamurthy_2024}, the He I 10830~\AA~transit depth is highly sensitive to the stellar XUV flux. Interestingly, the He I 10830~\AA~transit depth is much more sensitive to the stellar XUV flux than the  H$\alpha$ transit depth, as seen in Figure \ref{fig:stellaratv}. These results demonstrate that stellar activity levels significantly influence the He I 10830~\AA~transit depth, while the effect on the H$\alpha$ transit depth is more limited. If HD209458b experiences an activity cycle similar to that of the Sun, variations in XUV flux over time could lead to periods with significantly larger transit depths than those observed. This suggests that future observations of HD209458b, taken at different phases of its host star’s activity cycle, may reveal substantial variability in its atmospheric escape signature. Additionally, the difference in sensitivity to stellar activity between the He I 10830 \AA\ line and the H$\alpha$ line presents an opportunity to better understand the interplay between stellar radiation and planetary escape mechanisms. By simultaneously observing both He I 10830 \AA\ and H$\alpha$ over multiple epochs, we can isolate the role of high-energy stellar radiation in driving escape. 

\subsubsection{Molecular and eddy diffusion}

\begin{figure}[h!]
    \centering
    \includegraphics[width=.4\textwidth]{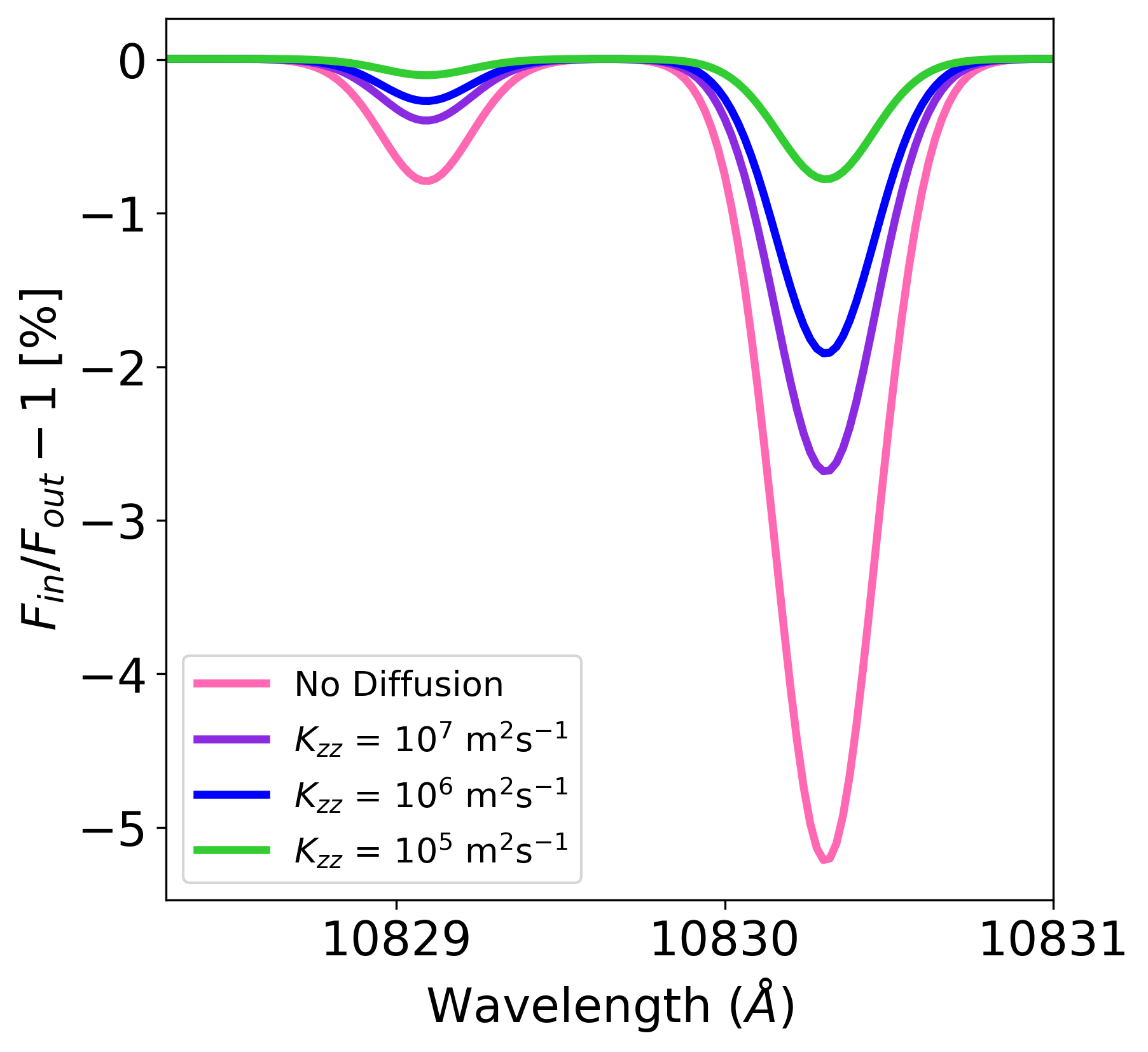}
    \includegraphics[width=.4\textwidth]{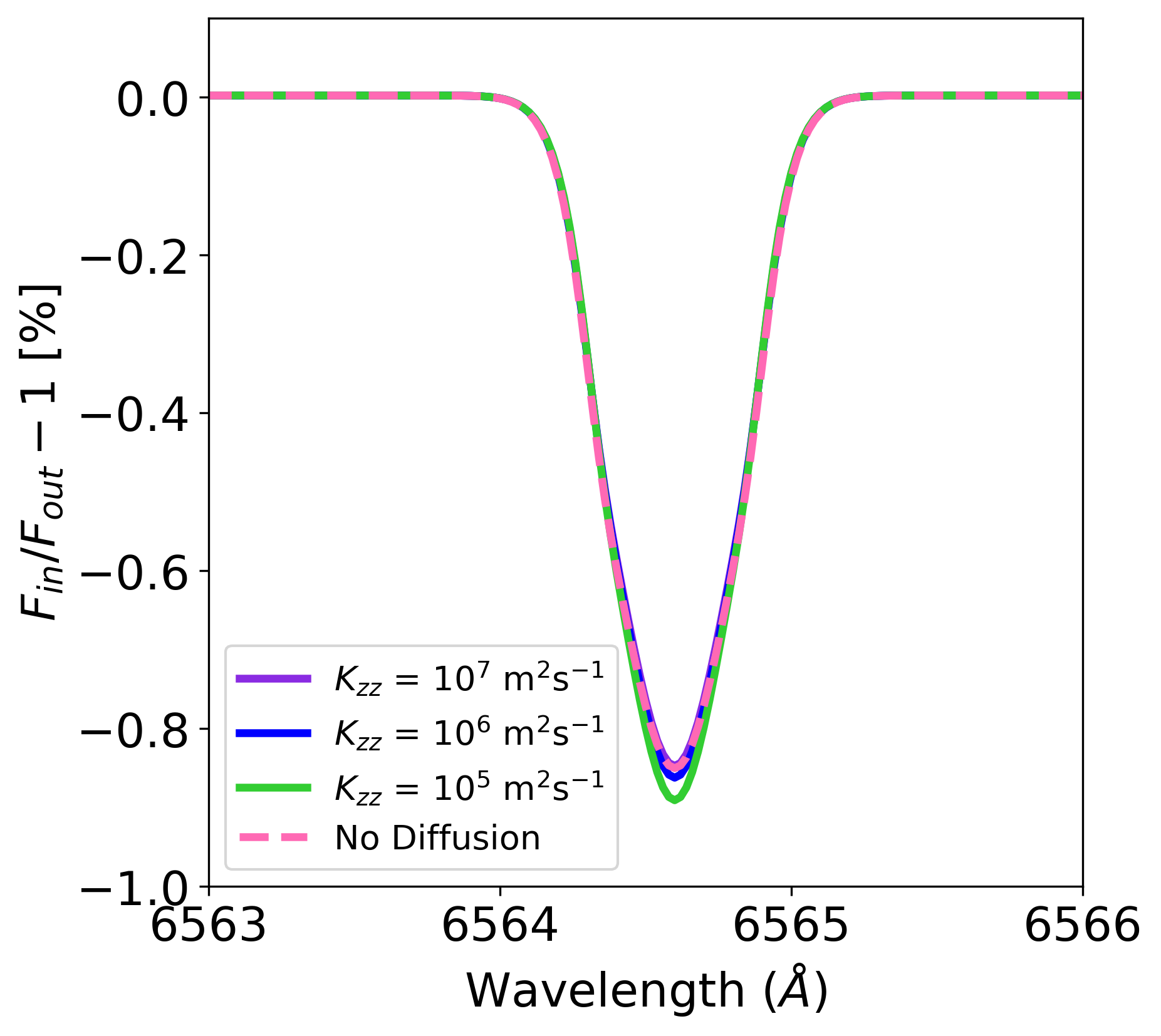}
    \caption{Transmission spectrum around the He I triplet (left) and the H-$\alpha$ Balmer line (right). The model in purple, blue, and green have eddy diffusion coefficients of $10^7$ m$^2$s$^{-1}$,  $10^6$ m$^2$s$^{-1}$,  and $10^5$ m$^2$s$^{-1}$ respectively. The model with no vertical diffusion is shown in pink. }
    \label{fig:kzz}
\end{figure}

\noindent
Studies have suggested that diffusive separation plays a key role in shaping He I 10830 \AA\ transit depths \citep{Lampon2020A&A...636A..13L}. To assess the importance of molecular and eddy diffusion, we investigate how they affect the structure of the upper atmosphere and the resulting He I 10830 \AA\ and H$\alpha$ transit depths. Running our model without any vertical diffusion predicts a He I 10830~\AA~transit depth nearly five times larger than observed, as seen in the left panel of Figure~\ref{fig:kzz}. Eddy diffusion ($K_{zz}$) is an essential component of upper atmosphere models, representing the aggregate effects of global circulation, turbulence, and breaking gravity waves. Without eddy diffusion, species mixing is driven only by bulk outflow and molecular diffusion, which may not accurately represent real atmospheric dynamics. To assess the impact of eddy diffusion, we run our best-fit model with three different values of  $K_{zz} $—$10^5$ m$^2$s$^{-1}$ (best-fit model), $10^6$ m$^2$s$^{-1}$, and $10^7$ m$^2$s$^{-1}$—and compare these to the case with no diffusion. The choice of these values is motivated by previous studies that have estimated a wide range of eddy diffusion coefficients for exoplanet atmospheres. \cite{Moses2011ApJ...737...15M} suggest values between $10^8$ and $10^9$ m$^2$s$^{-1}$ for hot Jupiter atmospheres while \cite{Parmentier2013A&A...558A..91P} propose lower values around $10^6$ to $10^7$ m$^2$s$^{-1}$. For our best-fit model, we adopt a value of $10^5$ m$^2$s$^{-1}$,  consistent with the lower and middle atmosphere model used in this study and based on a recent assessment of eddy diffusion in exoplanet atmospheres \citep{Arfaux2023MNRAS.522.2525A}. Increasing the eddy diffusion coefficient by a factor of 10 reduces the mass loss rate by a factor of 1.1 and decreases the degree of diffusive separation. The elemental He/H ratio at the upper boundary increases from 0.025 for  $K_{zz} = 10^5$  m$^2$s$^{-1}$ to 0.035 for  $K_{zz} = 10^6 $ m$^2$s$^{-1}$ and 0.04 for $ K_{zz} = 10^7$  m$^2$s$^{-1}$. Consequently, higher eddy diffusion enhances the He I 10830 \AA\ transit depth, as shown in the left panel of Figure \ref{fig:kzz}. In contrast, the effect of diffusion is much weaker for the H$\alpha$ transit depth (right panel of Figure \ref{fig:kzz}). Since hydrogen constitutes the bulk of the atmosphere, the H$\alpha$ transit depth remains largely unaffected even when diffusion is absent or with changing eddy diffusion. Comparing both lines—alongside independent constraints on stellar activity—can provide valuable insights into the role of molecular and eddy diffusion in shaping exoplanetary upper atmospheres. This analysis also points to a degeneracy between the eddy diffusion coefficient and the photoelectron heating efficiency: a lower $K_{\mathrm{zz}}$ could, in principle, be compensated by a higher heating efficiency, and vice versa. The adopted value of $10^5$~m$^2$~s$^{-1}$ represents a conservative choice that balances diffusive separation and observational constraints.

\subsubsection{Tidal Forcing} \label{sec:tides}

\noindent
Tidal forces (i.e., Roche lobe effects) play an important role in shaping the atmospheric dynamics of close-in exoplanets by modifying the pressure structure and accelerating mass loss. Some previous studies have incorporated tidal effects in models that are appropriate for the sub-stellar line but not necessarily a good assumption for a global average \citep[e.g.,][]{Xing2023ApJ...953..166X}. To investigate this, we implement tidal forcing in our upper atmosphere model by including the Roche potential in the momentum equation and running simulations at two different locations:
\begin{itemize}
    \item Substellar Case with Tide - At the substellar point, where tidal effects are strongest. Here, we divide the incident stellar flux by four to approximate global heat redistribution.
    \item 60$^\circ$ Stellar Zenith Angle Case with Substellar Tide -  At a stellar zenith angle of 60°, dividing stellar radiation by two (matching the best-fit model) to directly compare to the best-fit model and isolate tidal effects. 
\end{itemize}


\noindent We find that introducing tidal forcing with a stellar zenith angle of 60$^\circ$ increases the mass-loss rate from 1.9 $\times 10^{10}$ in the best-fit model to 3.7 $\times 10^{10}$ g/s, then moving to the substellar point lowers this mass-loss rate to 2.5 $\times 10^{10}$ g/s. 
This occurs due to a combination of differences in outflow velocity, density structure, and the altitude of peak heating. The outflow velocity at 4 R$_p$ increases significantly in the 60$^\circ$ stellar zenith angle case, reaching 6 km/s compared to 1.1 km/s in the best-fit model and 3.8 km/s in the substellar case. However, the mass-loss rate depends on both velocity and density. In the 60$^\circ$ stellar zenith angle case, the expansion of the atmosphere leads to a lower density at 4~Rp, while in the substellar case, the density remains slightly higher. The key factor driving these differences is the location of the heating peak. In the substellar case, the heating peak shifts to lower altitudes, decreasing the efficiency of the atmospheric expansion and dampening escape. Additionally, the sonic point is at 4.4 R$_p$ in the substellar case and 4.2 R$_p$ in the 60$^\circ$ stellar zenith angle case.

\begin{figure}[h!]
    \centering
    \includegraphics[width=.4\textwidth]{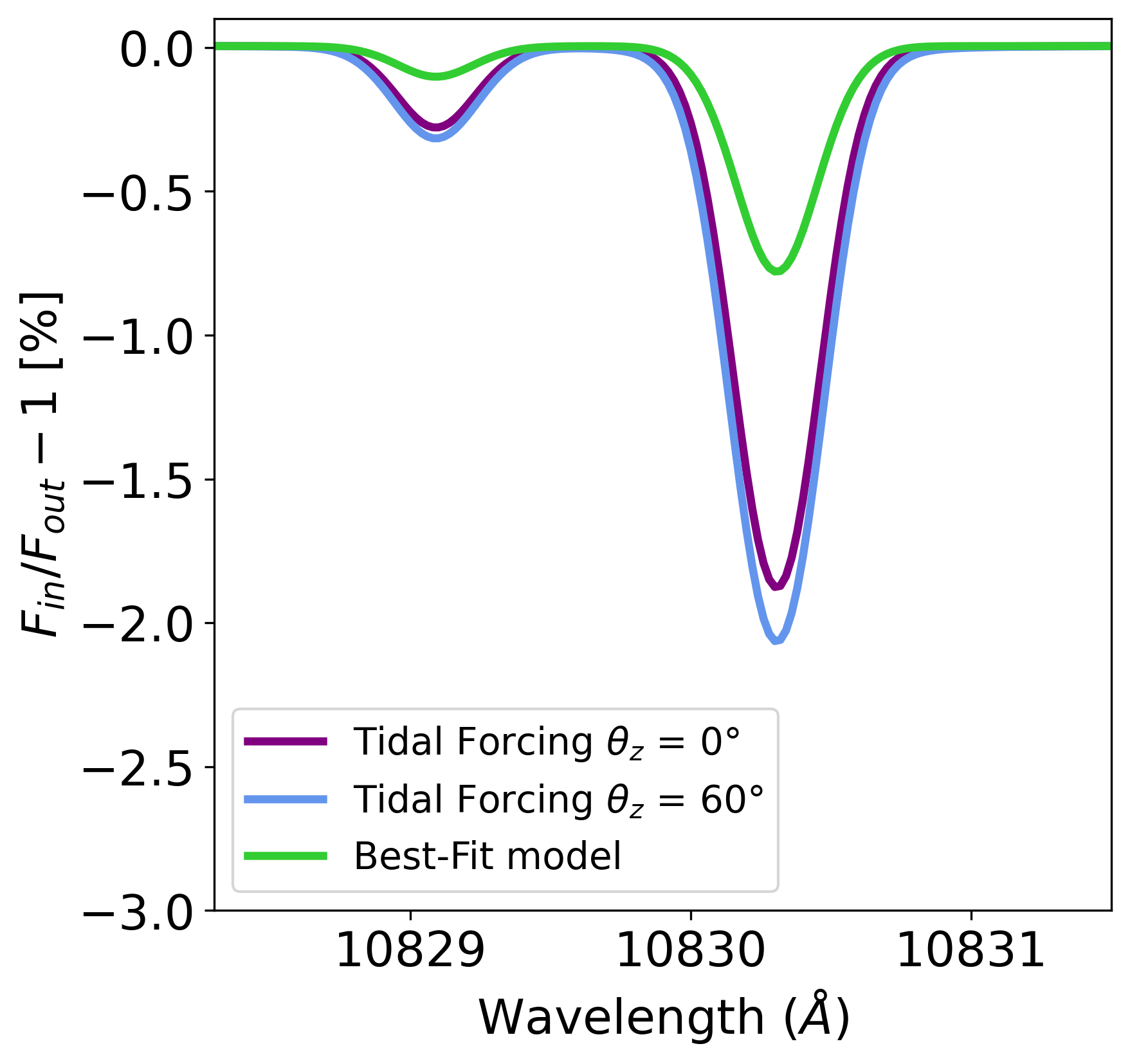}
    \includegraphics[width=.425\textwidth]{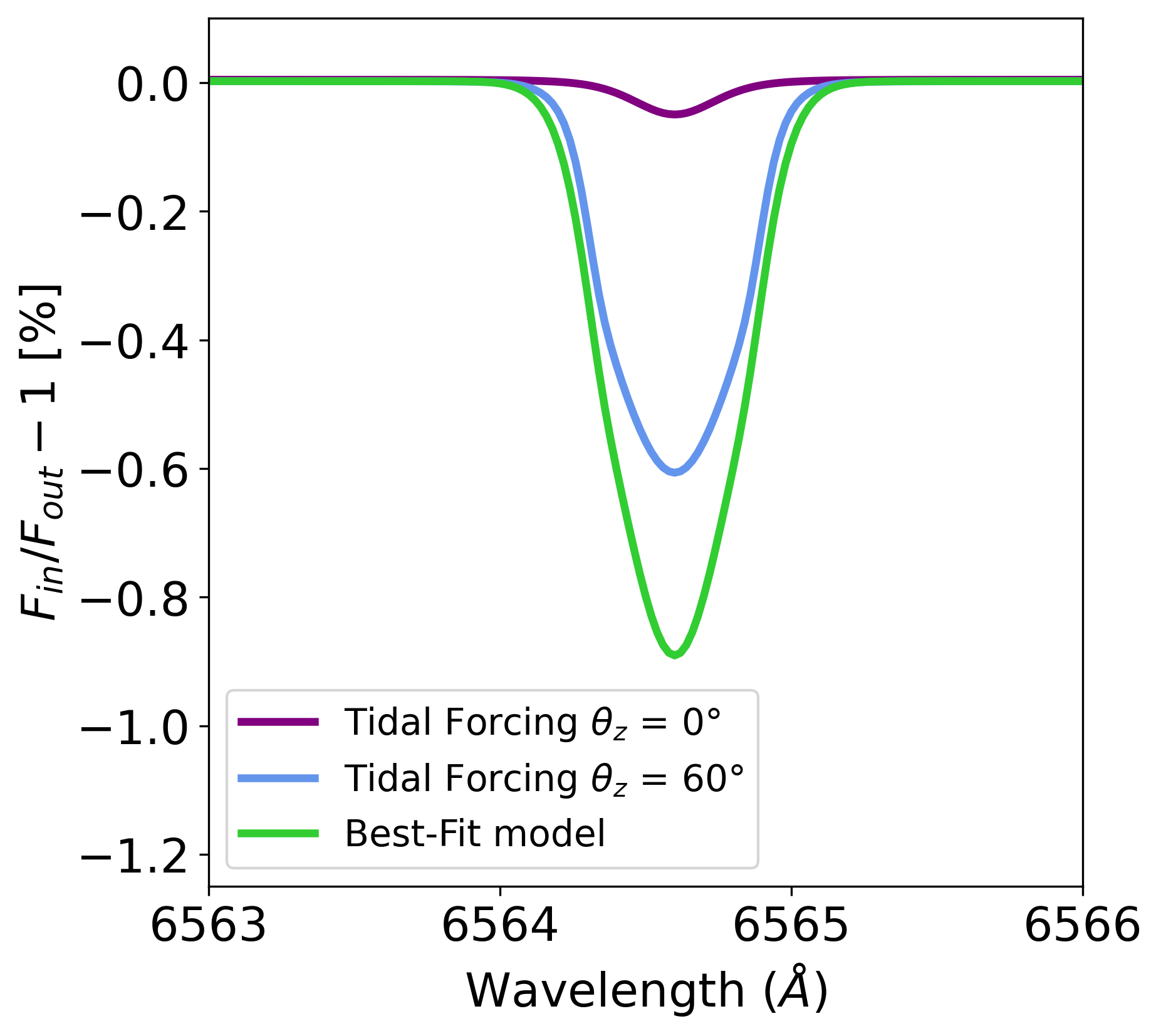}
    \caption{Transmission spectrum around the He I triplet (left) and the H-$\alpha$ Balmer line (right). The model in green is our best-fit model, in blue is the zenith angle of 60$^\circ$ with tidal forcing, and in purple is the model run at the substellar line with tidal forcing.}
    \label{fig:tides}
\end{figure}

Interestingly, we show that tidal effects increase the He I 10830 \AA\ transit depth, as shown in Figure \ref{fig:tides}. This arises from an increase in the metastable helium population at radii of $<$ 2.5 R$_p$ in the model with Roche lobe overflow that is driven by an increase in the He$^+$ population at these altitudes, producing a higher recombination rate to the metastable state. In the global average model with tidal forcing and a solar zenith angle of 60°, the He I 10830 \AA\ transit depth is enhanced compared to the best-fit case without Roche lobe overflow, showing a slightly larger transit depth than the substellar model with Roche lobe overflow. This suggests that the increased outflow velocity and density changes from tidal effects enhance metastable helium absorption, but the exact magnitude depends on the incident stellar flux distribution. In contrast, the H$\alpha$ transit depth substantially decreases in both tidal runs. The reduction in H$\alpha$ transit depth in the tidal forcing models is primarily driven by a decrease in the Lyman-$\alpha$ mean intensity in the atmosphere, which reduces radiative excitation of hydrogen to the n=2 state. This occurs because tidal expansion lowers the overall density of the atmosphere, including H and electrons, decreasing collisional excitation rates and the scattering optical depth to Lyman-$\alpha$ photons. As a result, fewer Lyman-$\alpha$ photons are available to populate the H(n=2) state, leading to weaker H$\alpha$ absorption. The effect is most pronounced in the substellar case, where the density drops the most, causing the strongest suppression of neutral H density and thus the lowest H$\alpha$ transit depth. 

While tidal effects enhance the outflow velocity and alter the density structure, it is important to recognize that transit observations primarily probe the terminator region rather than the substellar line. Previous 1D models applying tidal effects along the substellar line do not necessarily reflect the geometry of transit observations. Our results demonstrate that models using substellar conditions produce transit depths that differ from those using globally averaged conditions. Since globally averaged conditions may better approximate the terminator region, while substellar conditions clearly do not, using substellar conditions in fits to transit observations can be problematic. In the end, 3D models will be required to address these differences \citep[e.g.,][]{Khodachenko2021MNRAS.507.3626K}.

\subsection{Metals} \label{sec:metals}

\noindent Heavy elements and metal species can play a key role in regulating the thermal structure and dynamics of the escaping atmosphere \citep{Huang__2023}, yet they are often omitted from models of giant planet atmospheric escape. Observations of HD209458b suggest the presence of metals such as Fe, O, C, Mg, and Si in the upper atmosphere \citep{Cubillos2020AJ....159..111C,Linksy2010ApJ...717.1291L, Vidal2004ApJ...604L..69V, Vidal-Madjar2013A&A...560A..54V}. While the analysis and interpretation of these observations have been subject to debate and could be significantly affected by changes in stellar activity, it is reasonable to explore the impact of metals on atmospheric structure, atmospheric escape processes, and the He I 10830 \AA\ and H$\alpha$ transit depths. Therefore, we run our best-fit model incorporating Mg, Fe, Si, O, C, N, Na, K, S, Ca, and relevant ions. This model assumes that none of these species are removed from the upper atmosphere by condensation processes in the lower and middle atmosphere. The relevant details of this version of the model are given in full by \citet{Huang__2023}.

\begin{figure}[h!]
    \centering
    \includegraphics[width=.3\textwidth]{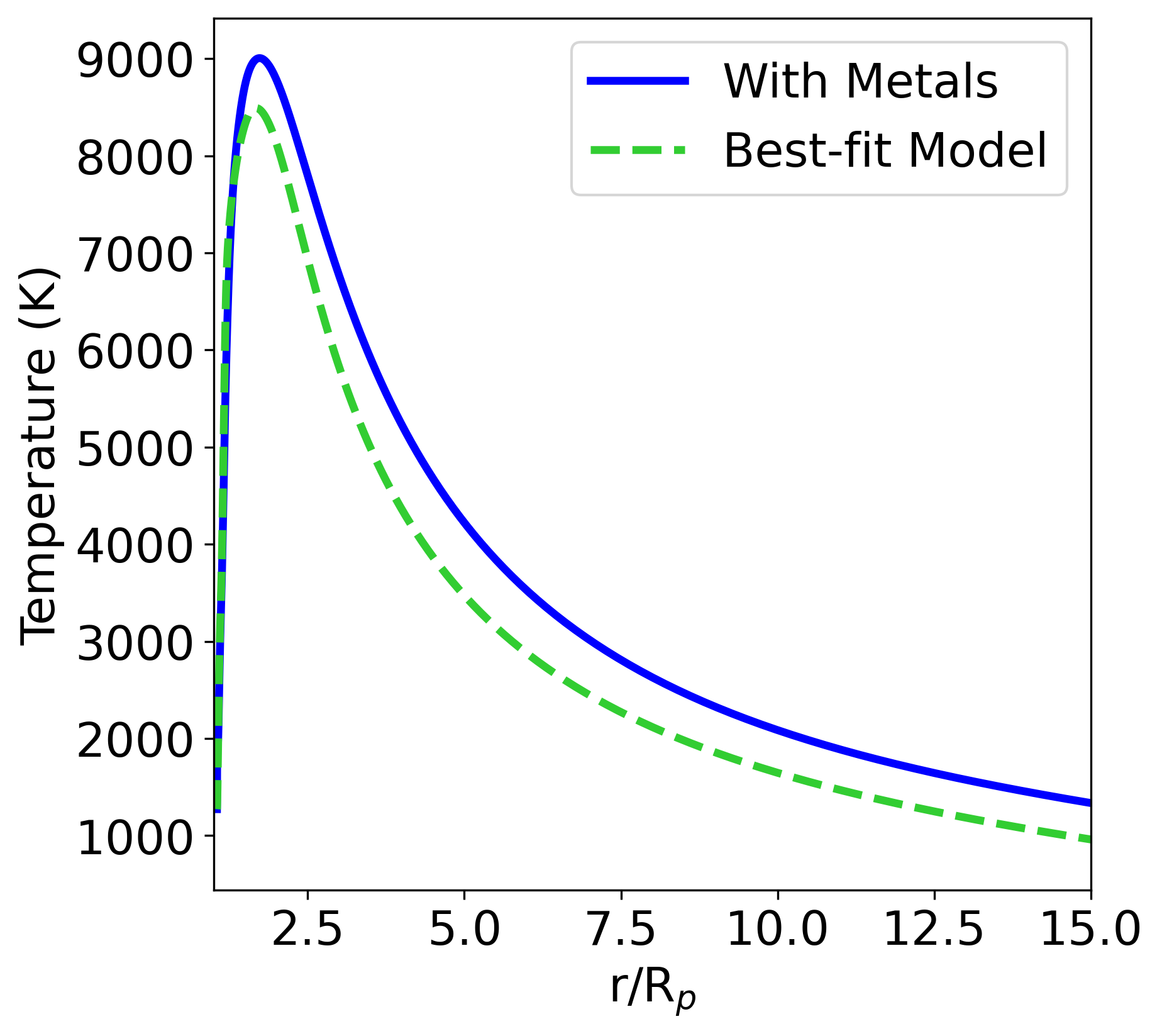}
    \includegraphics[width=.29\textwidth]{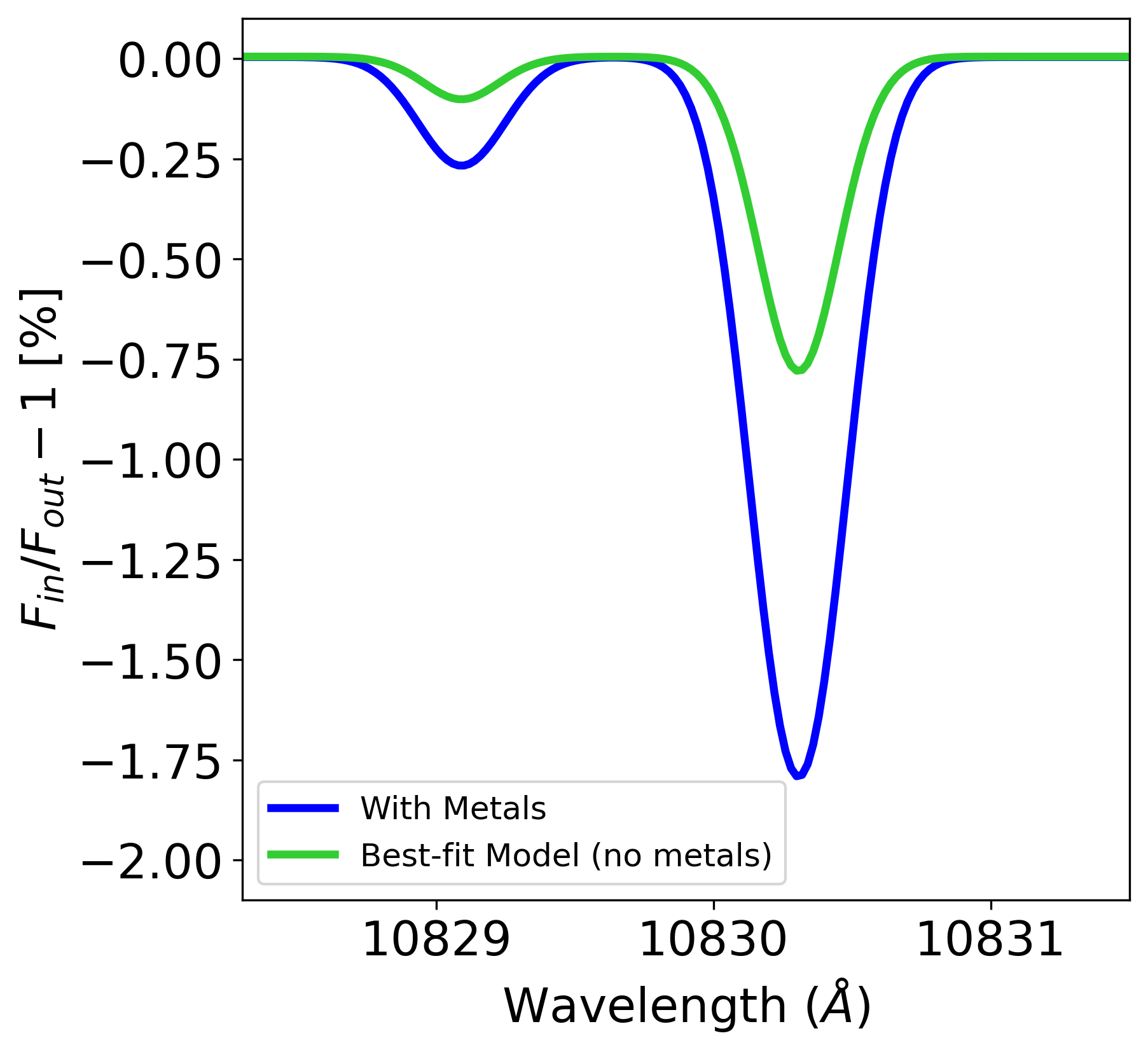}
    \includegraphics[width=.3\textwidth]{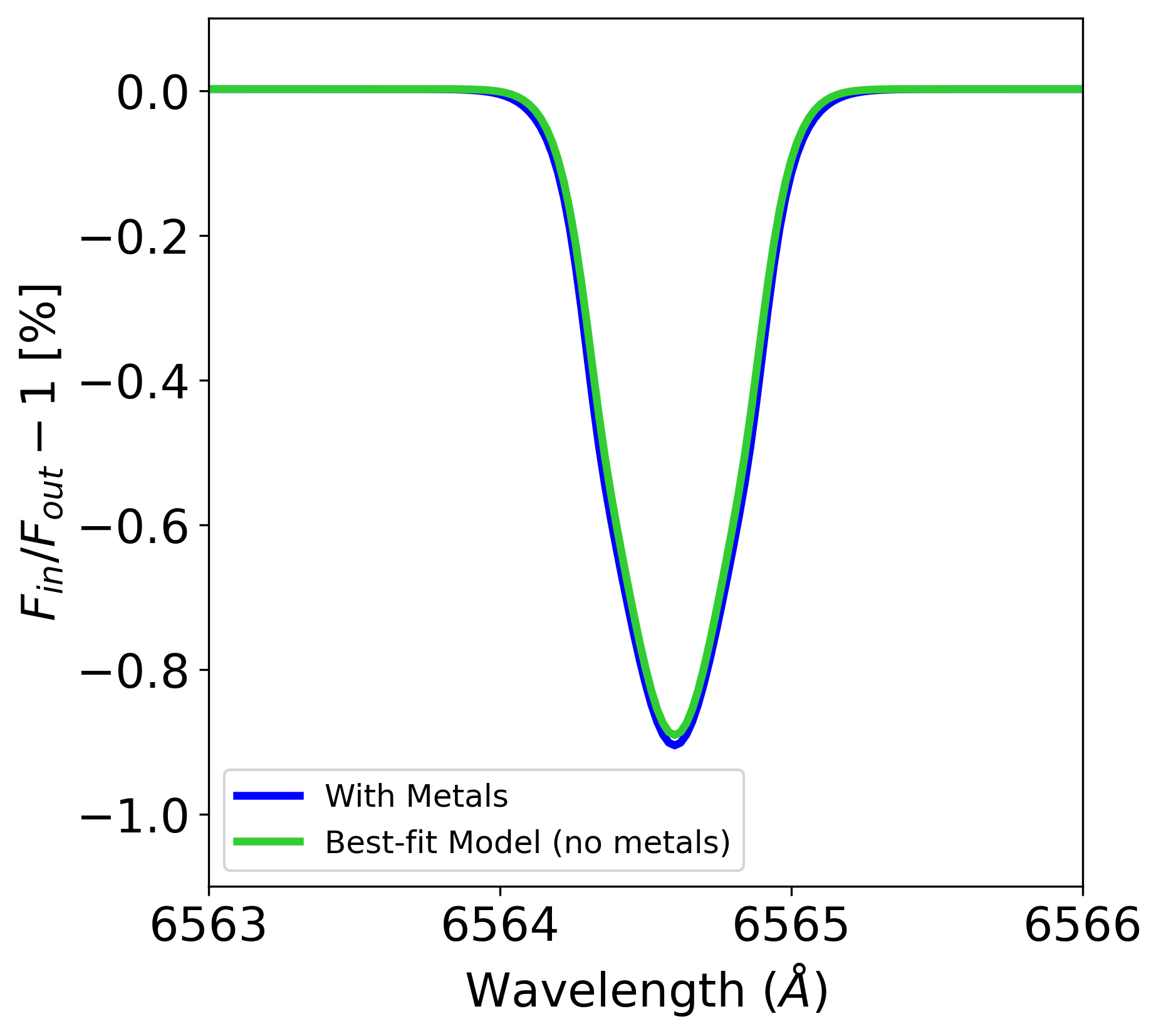}
    \caption{Left: Temperature and bulk outflow velocity predicted by our atmosphere model for HD209458b including heavy elements (blue) compared to our best-fit model (green). Middle and Right: Transmission spectrum around the He I triplet (middle) and the H-$\alpha$ Balmer line (right). The transit depth produced by our best-fit model is in green compared to the model with heavy elements in blue.}
    \label{fig:metaltd}
\end{figure}

As seen in the temperature profiles in the left panel of Figure \ref{fig:metaltd}, the inclusion of metals results in a higher peak temperature in the upper atmosphere, as additional heating and changes in radiative cooling processes alter the thermal balance. The temperature peaks at approximately 9000 K near 1.7 R$_p$, higher than in the best-fit model which peaks at 8200 K at 1.7 R$_p$. The corresponding mass loss rate is $3 \times 10^{10}$ g/s, slightly higher than in the best-fit model, which is $1.9 \times 10^{10}$ g/s, but still within a physically plausible range for HD209458b. The pressure-averaged temperature above 1 $\mu$bar to the top of the model at $3\times 10^{-4}$ nbar is 5724 K, and above 7 nbar to the top of the model is 8614 K. The elemental He/H ratio decreases from 8\% at the base of the thermosphere to approximately 5.5\% at high altitudes, which is a less enhanced diffusive separation than in the best-fit model. As expected based on the diffusive separation of helium in the model, we also see diffusive separation between hydrogen and all of the heavier elements. 

The most direct observational impact of metals in the atmosphere is reflected in the He I 10830 Å transit depths (middle panel Figure \ref{fig:metaltd}). The He I 10830 \AA\ absorption depth in the metal-rich model is notably enhanced compared to the best-fit model. This increase is predominantly driven by increased ionization caused by the presence of metals that elevate the electron density and recombination to the metastable He I (2$^3$S) state. In order to match the He I 10830 Å observations with this model, we now require a photoelectron heating efficiency of only 23\%. On the other hand, the H$\alpha$ transit depth (right panel Figure \ref{fig:metaltd}) remains essentially unchanged between models. This is because H$\alpha$ line formation primarily depends on Lyman-$\alpha$ radiative excitation rate and the neutral hydrogen density profile, both of which remain relatively stable even in the presence of additional metals.

\section{Discussion} \label{sec:dis}

\noindent
Our results highlight the importance of comprehensive atmospheric modeling in interpreting the escaping atmosphere of HD209458b. Assumptions about temperature structure, composition, and transport processes can significantly influence the inferred mass-loss rates and predicted transit depths. Parker wind models can provide a useful analytical framework for constraining the parameter space of the solution while self-consistent models help to narrow the solution and offer more detailed insights into upper atmospheric structure and processes. Including the lower and middle atmosphere increases the mass-loss rate by roughly 60\%, while the He I 10830 \AA\ and H$\alpha$ transit depths remain largely unchanged. Although this suggests limited sensitivity to the properties of the lower and middle atmosphere of HD209458b, the structure and composition of the lower and middle atmosphere may play a more important role in other planets—particularly those with more extended lower and middle atmospheres or those with less extended lower and middle atmospheres but enhanced metallicities that feed higher abundances of heavy elements to the upper atmosphere. Finally, we also explore how different excitation and de-excitation rates, the inclusion of metals, and stellar variability affect the predicted line profiles.


Parker wind models have been widely used to account for hydrodynamic atmospheric escape in forward model fits to observations of escaping atmospheres. This approach has led to mass-loss estimates for HD209458b of $8 \times 10^{10}$ g~s$^{-1}$ \citep{oklop2018ApJ...855L..11O, santos2022A&A...659A..62D}, 0.42--1 $\times 10^{11}$ g~s$^{-1}$ \citep{Lampon2020A&A...636A..13L}, and a high upper limit of $< 2.4 \times 10^{11}$ g~s$^{-1}$ \citep{masson2024probing}. Our best-fit models require a photoelectron heat efficiency of 20-40\% and predicts a mass-loss rate of $1.9-3 \times 10^{10}$ g~s$^{-1}$, depending on whether metals are included or not in the upper atmosphere. We note that photoelectron heating efficiency here refers specifically to the fraction of photoelectron energy deposited as heat, rather than the net heating efficiency of the atmosphere. The overall heating efficiency, which accounts for all energy deposition processes relative to the incoming XUV flux and determines the energy-limited mass loss rate, is 15\% for our best-fit model. Our best-fit models also show diffusive separation between hydrogen and helium. Our results underscore the importance of self-consistently modeling the upper atmosphere at least above the 1 $\mu$bar level. This is because temperature gradients and dynamics shape the structure and composition of the upper atmosphere’s, influencing He I 10830 Å and H$\alpha$ absorption. The processes that govern them must be treated carefully. While the lower and middle atmosphere help establish the continuum level for differential transit depth comparisons, the structure of the upper atmosphere remains the dominant factor in shaping the He I 10830 Å and H$\alpha$ line cores on HD209458b. Our best-fit models are also consistent with the upper limit on H$\alpha$ transit depth set by \cite{jenesen2012ApJ...751...86J}. However, as noted in Section~\ref{sec:results}, our model slightly exceeds other published constraints on H$\alpha$ absorption in this system \citep{Astudillo-Defru2013A&A...557A..56A, Casasayas-Barris2021A&A...647A..26C}. This modest discrepancy may arise from geometric effects—combined with tidal forcing—which we show can strongly influence the H$\alpha$ line, or from variations in stellar activity, to which the measured H$\alpha$ transit depth is somewhat sensitive. That said, we refrain from drawing strong conclusions until longer-baseline observations of this line become available. 

As shown in Sections \ref{sec:fullatmo} and \ref{sec:bf}, significant diffusive separation was observed in our models, and the degree of diffusive separation increased in our best-fit model. This demonstrates that reducing the heating efficiency enhances diffusive separation due to the decrease in mass-loss rate. We believe, however, that the degree of diffusive separation is surprising, given theoretical expectations based on the crossover mass equation \citep{HUNTEN1987532}. The crossover mass equation provides a rough estimate of the limiting mass-loss rate, $\dot{M}_{\rm lim}$, required for a species of mass $M_c$ (in units of $m_H$) to escape along with hydrogen via neutral-neutral collisions \citep{Koskinen2014}:
\begin{equation}
    \dot{M}_{\rm lim} = 4\pi m_H^2 G M_p (M_c - 1) \frac{n D_c}{kT}.
\end{equation}
Here, $D_c$ is the mutual diffusion coefficient for species c with H, given by \cite{Marrero1972JPCRD...1....3M}: 
\begin{equation}
  \log(D_c p) = \log(A) + s \log(T)
\end{equation}
where A and s are constants given by $14.2 \times 10^{-5}$ $\frac{\text{atm}\cdot \text{cm}^2}{\text{s(K)}^\text{s}}$ and 1.721, respectively for He with H. At a temperature of 5336 K (the pressure-averaged temperature of our best fit model), the mass loss rate required to mix He with H is approximately $6.4 \times 10^9$ g/s. Given that our best-fit model estimates a mass-loss rate of $1.9 \times 10^{10}$ g/s, which is well above the limit, diffusive separation should be suppressed, yet we still observe significant separation between H and He in our model. This is despite the fact that the above equation is based on neutral-neutral collisions while the upper atmosphere of HD209458b is strongly ionized and this should further reduce the importance of diffusive separation. 

The discrepancy between our results and the predictions of the crossover mass equation highlights the need to reconsider key assumptions in its derivation. The standard formulation assumes (1) no eddy diffusion, (2) a constant temperature, and (3) a volume mixing ratio for helium that remains constant with altitude. However, none of these assumptions hold in our self-consistent model. We find that the atmosphere exhibits a strong temperature gradient rather than a uniform temperature structure. Additionally, our simulations include non-zero eddy diffusion, which enhances vertical mixing and alters the competition between advection and molecular diffusion. Furthermore, there exists a region in our model where the molecular diffusion timescale is shorter than or comparable to both vertical advection and eddy diffusion, indicating that helium cannot maintain a constant mixing ratio throughout the atmosphere. These deviations from the idealized conditions of the crossover mass equation explain why we observe significant diffusive separation of helium even at mass-loss rates that exceed the theoretical limit for full mixing. Previous studies have also found that diffusive separation could be happening on HD209458b, but they did not include eddy diffusion \citep{Xing2023ApJ...953..166X, Schulik2024arXiv241205258S}. The choice of the eddy diffusion coefficient remains an important uncertainty that is degenerate with the mass loss rate in controlling the degree of diffusive separation. 



Excitation/de-excitation rates and photoionization cross-sections obviously play an important role in determining metastable helium abundances and, consequently, He I 10830 \AA\ transit depths.  In this work, we present updates to the excitation/de-excitation rates and cross-sections relevant to the formation and destruction of the metastable 2$^3$S state of He I 10830 \AA\ in exoplanet atmospheres. The most notable revisions that we present are new photoionization cross-sections and a new Penning ionization rate, described in Section \ref{sec:chem}. 
We also consider excitation/de-excitation mechanisms such as photoelectron collisions, which are important in Earth's atmosphere, and proton de-excitation. We provided an upper limit on photoelectron excitation for HD209458b and included proton de-excitation rates but did not find that they made a significant difference to the results (see Figure \ref{fig:bfenergy}). 
For demonstration, we compare our best-fit model using the (de-)excitation rate scheme from \cite{Lampon2020A&A...636A..13L} to our updated model that includes the new excitation scheme. The combined effect of these updates lowers the He I 10830 \AA~transit depth by a factor of $\sim$2.5. The dominant factor in this reduction is the new high-resolution photoionization cross-section, which includes the previously neglected EUV resonances, and the update to the ground state recombination rate, which controls the population of ionized helium. These changes reduce the overall metastable helium population and, consequently, the predicted transit depth. We note, however, that this difference is strongly model and rate-dependent. In particular, in isothermal models where temperatures are fixed, the reduction in He I 10830 \AA\ transit depth is expected to be smaller.

Metals are expected to influence the thermodynamics and composition of the escaping atmosphere, yet they are often omitted from models of atmospheric escape. Our model that includes heavy elements shows that the addition of metals has a relatively small effect on the temperature profile and mass loss rate. The presence of metals that have lower ionization potentials than H and He, however, increases the electron density and boosts the recombination rate to the He I (2$^3$S) state. This leads to deeper He I 10830 Å transit depths, requiring us to lower the photoelectron heating efficiency to 23\% in order to maintain the match to the He I 10830~\AA~observations while the H$\alpha$ line shows little change from the best-fit model. In these models, we see significant diffusive separation of metals in the lower thermosphere, as expected based on our results for the He/H ratio in the upper atmosphere. 


One-dimensional models are commonly used to interpret observations of exoplanet upper atmospheres, even though they do not account for the multi-dimensional nature of Roche lobe overflow or temperature and compositional differences between the day and night sides that affect global mass loss rates. In addition to these limitations, other multi-dimensional phenomena—such as leading and trailing tails, asymmetries induced by stellar winds, or magnetic confinement—can introduce further complexities that are obviously not captured by 1D models \citep{WANG2021ApJ...914...99W, wANG2021ApJ...914...98W, Khodachenko2021MNRAS.507.3626K}. These effects are especially important in systems where extended or directional escape signatures are observed. Multi-dimensional simulations are essential for fully characterizing such cases. Many 1D models assume substellar illumination and Roche lobe overflow, which we show in Section \ref{sec:tides}, overestimates the mass-loss rate, and misrepresents the spatial distribution of helium and hydrogen at the transit terminator, impacting the modeled transit depths. Assumptions about the radiation field in 1D models are also important. We run our best-fit model using a dayside-averaged radiation field with global redistribution, achieved by assuming a solar zenith angle of 60$^\circ$ and dividing the incident stellar flux by a factor of two. These are standard assumptions for globally averaged 1D simulations and may be closer to the conditions around the terminator, which is probed in transit. Even without the impact of Roche lobe overflow, the resulting transit depths differ from those in substellar geometry. Therefore, 1D forward models of the transit observations should always include an additional uncertainty in the derived parameters due to the lack of an accurate representation of multi-dimensional effects. In addition to these multi-dimensional considerations, interactions with the stellar wind could also influence mass loss rates and the escape of different species. However, we do not expect stellar wind interactions to be a dominant factor in shaping the observed He I 10830 \AA\ and H$\alpha$ transit depths. The relatively low transit depths for both species are consistent with a transit obstacle that is significantly smaller than the size of the Roche lobe, reducing the likelihood of strong star-planet interactions \citep[see also][]{Khodachenko2021MNRAS.507.3626K}. 

\begin{figure}[h!]
    \centering
    \includegraphics[width=0.5\linewidth]{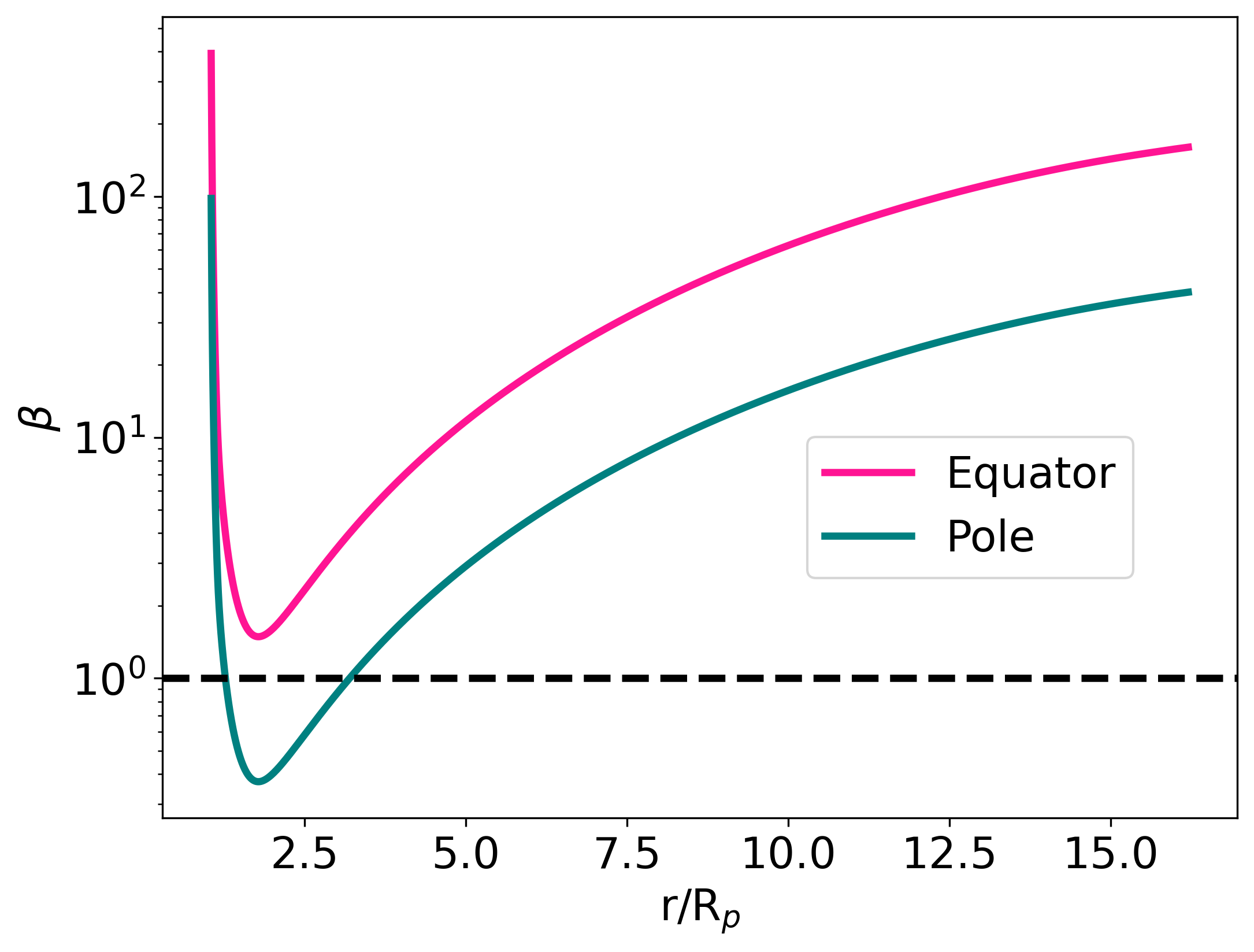}
    \caption{Plasma beta ($\beta$) as a function of radius (r/$R_p$) at the magnetic equator (pink) and pole (teal). The high values of $\beta$ indicate that thermal pressure dominates over magnetic pressure at all altitudes, leading to an extended and dynamic magnetosphere influenced by planetary outflows.}
    \label{fig:pb}
\end{figure}

The presence of a planetary magnetic field, however, can influence atmospheric escape by modifying the structure of the outflow and dictating the regions where mass loss is permitted. The strength of this influence depends on the balance between thermal, magnetic, and dynamic pressures in the upper atmosphere. To better understand this interaction, we estimate the size of the magnetosphere to constrain the regions where the stellar wind and planetary outflow interact. While more complex models of the magnetosphere exist \citep{Khodachenko2021MNRAS.507.3626K}, it is also useful to provide simple order-of-magnitude estimates that are more readily applicable to other planetary systems. One way to estimate the size of the magnetosphere is through the Chapman-Ferraro (C-F) boundary, which represents the distance at which the planetary magnetic pressure balances the incoming stellar wind pressure. The C-F standoff distance is given by:
\begin{equation}
    R_{CF} \approx \zeta \left( \frac{m_B^2}{2 \mu_0 \rho_{sw} v_{sw}^2}\right)^{1/6}
\end{equation}
where $\zeta$ is a scaling factor ($\sim1.26$), $m_B$ is the planet's magnetic moment, $\rho_{sw}$ and $v_{sw}$ are the density and velocity of the stellar wind for which we use  $v_{sw}$ = 220 km/s, $\rho_{sw}$ = 4000 g/cm$^3$ \citep{Khodachenko2021MNRAS.507.3626K}.  We estimate the Chapman-Ferraro stagnation distance to be 4.03 r/$R_p$, with the polar standoff distance at 8.06 r/$R_p$, marking the location where the planet’s magnetosphere balances the incoming stellar wind pressure. However, the Chapman-Ferraro model assumes a static balance between magnetic and dynamic pressures and does not account for the contribution of internal plasma pressure or the additional effects of planetary outflows. To address this limitation, we calculate the plasma beta ($\beta$), the ratio of thermal to magnetic pressure, which determines how much the escaping atmosphere influences the magnetosphere’s structure. The plasma beta is given by:
\begin{equation}
    \beta = \frac{2 \mu_0 p}{B^2}.
\end{equation}
where $\mu_0$ is the permeability of free space, p is thermal pressure, and B is the magnetic field strength, estimated by a dipole given by 
\begin{equation}
    B(r, \theta) = \frac{m_B}{r^3} (1 + 3 \cos^2{\theta})^{1/2}.
\end{equation}
$m_B$ is the magnetic moment, which we estimate to be 10\% of Jupiter's magnetic moment \citep{Kislyakova2014}, giving us a value of $2.83 \times 10^{19}$ T m$^3$ \citep{Connerney2017}, r is radius, and $\theta$ is co-latitude. The magnetic moment estimate used here is based on fitting Lyman-$\alpha$ transit observations with a model of the escaping atmosphere \citep{Khodachenko2021MNRAS.507.3626K}. However, the interpretation of these observations is not unique \citep[e.g.,][]{2010ApJ...723..116K}, so the inferred magnetic moment should be considered uncertain. Based on calculations at the equator, $\beta$ significantly exceeds unity at all radii, as shown in Figure \ref{fig:pb}. 
This implies that the thermal pressure of the plasma dominates over the magnetic pressure, leading to an extended and dynamic magnetosphere. This result is consistent with solar system giant planet magnetospheres that extend beyond the Chapman-Ferraro standoff distance when significant internal plasma sources are present. The dynamic effects of the escaping planetary atmosphere must therefore be considered when estimating the true size of the magnetospheric boundary. To incorporate the effects of planetary outflow, we estimate the location of the magnetic ‘hydropause,’ which is the location where the combined thermal, dynamic, and magnetic pressures of the stellar wind balance the combined thermal, dynamic, and magnetic pressures of the escaping planetary atmosphere. Unlike in solar system studies, where this boundary is often simplified to a static balance of magnetic pressures, exoplanetary systems require a full pressure balance equation that includes all relevant forces. This balance is described by:
\begin{equation}
    \rho_{sw}v_{sw}^2 + p_{sw} + \frac{B_{sw}^2}{2 \mu_0} = \rho_{pl}v_{pl}^2 + p_{pl} + \frac{B_{pl}^2}{2 \mu_0}
\end{equation}
where the subscript $sw$ refers to the stellar wind, and the subscript $pl$ refers to the planetary magnetosphere. For the  and a stellar wind magnetic field strength, we use $5 \times 10^{-9}$ T \citep{Khodachenko2021MNRAS.507.3626K}.Balancing these terms in our model, we find that the hydropause is located at 4.7 r/$R_p$, extending slightly beyond the Chapman-Ferraro boundary. In addition, a stagnation distance of 4–5 R$_p$ suggests a polar cap boundary near a magnetic latitude of 60–63°, indicating that approximately 27\% of the planetary surface feeds open magnetic field lines along which escape is not suppressed. This suggests that the planetary outflow exerts enough pressure to expand beyond the traditional magnetospheric interaction region, reinforcing the notion of a dynamic magnetosphere shaped by both stellar wind interactions and atmospheric escape. 



The upper atmosphere of HD209458b has also been probed by various FUV lines, including H Ly$\alpha$, O I, C II, and Si III \citep{v-m2003Natur.422..143V, Ben-Jaffel2010ApJ...709.1284B, Vidal2004ApJ...604L..69V, Linksy2010ApJ...717.1291L}, although the reality of the detection of Si III has been questioned \citep{Ballester2015ApJ...804..116B}. These observations provided some of the first detections of atmospheric escape on exoplanets and inspired many of the earlier modeling studies. Our best-fit model predicts a mass loss rate $1.9 \times 10^{10}$ g~s$^{-1}$, which is within the mass loss range of about 1-10 $\times 10^{10}$ in other modeling studies studies of HD209458b \citep[e.g.,][]{v-m2003Natur.422..143V, Yelle2004Icar..170..167Y, Tian2005ApJ...621.1049T, GarciaMunoz2007, 2010ApJ...723..116K, Linksy2010ApJ...717.1291L}, albeit at the lower end of this range. Our best-fit model predicts a pressure-averaged temperature of 7180 K when integrated from 1 $\mu$bar to 3 R$_p$, which is comparable to the temperatures required to match the FUV observations. The mass loss rates and net heating efficiencies required to match the He I 10830~\AA~and H$\alpha$ observations, however, are generally lower than the values fitted to the FUV transit observations \citep{2013Icar..226.1695K}. Our best-fit model also predicts diffusive separation between He and H, which presents a possible challenge for matching the FUV observations of escaping heavy elements. Observations of O I and C II in the upper atmosphere suggest that these heavier species are not significantly depleted relative to hydrogen. If diffusive separation of He is occurring, we would expect these species to be less abundant at high altitudes, yet they are observed in the FUV with relatively large transit depths.

The He I 10830~\AA~transit depth is influenced not only by atmospheric properties but also by variations in the stellar XUV flux, as demonstrated in this (see Figure \ref{fig:stellaratv}) and previous studies \citep{oklop2018ApJ...855L..11O, oklop2019ApJ...881..133O, Krishnamurthy_2024, Sanz-Forcada2025A&A...693A.285S, Ballabio2025MNRAS.537.1305B}. Our results suggest that repeated observations of HD209458b should reveal significant changes in the He I 10830 \AA\ absorption unless the star is exceptionally quiet, with activity variations far lower than those of the Sun. This behavior contrasts with the relative stability of the H$\alpha$ transit depth, which remains less sensitive to stellar activity cycles. This is because the H(n=2) population is predominantly excited by Lyman-$\alpha$ radiation and the solar Lyman-$\alpha$ flux exhibits much smaller variability compared to the short-wavelength EUV flux over the solar cycle. We believe that the possible range of variability prevents direct comparisons between He I 10830~\AA~and the existing H Ly$\alpha$ or other FUV transit depths, as attempted in previous studies \citep[e.g.,][]{oklop2018ApJ...855L..11O}. This is because the observations were taken at different times and, for example, H Ly$\alpha$ measurements are themselves sensitive to changes in stellar activity, in addition to the changes in the planet's atmosphere due to the changes in the incident XUV flux \citep[e.g.,][]{2010ApJ...723..116K,Koskinen2013Icar..226.1678K}. Instead, our model here is intended to be consistent with the He I 10830~\AA~transit depth and roughly consistent with the upper limit on the H$\alpha$ Balmer line depth, both of which are relatively low, possibly reflecting a period of lower stellar activity. Generally, monitoring exoplanet He I 10830 \AA\ and H$\alpha$ transit depths across multiple epochs, together with host star activity indicators, provides valuable constraints on the host star’s activity level and its temporal influence on atmospheric escape. Having said that, observers have struggled to detect the H$\alpha$ line on HD209458b and we have never observed He I 10830 \AA\ transit depth consistent with moderate or higher stellar activity that we simulated based on the XUV flux variations of the sun, a similarly quiet G-type star. 


Finally, our results provide insights into why helium absorption detections do not always follow clear trends across exoplanet populations. While previous studies have suggested correlations between He I 10830~\AA~transit depths and planetary or stellar parameters such as irradiation levels, the likely extent of the planetary atmosphere, and mass-loss rates, significant outliers exist \citep{Sanz-Forcada2025A&A...693A.285S, Ballabio2025MNRAS.537.1305B, Linssen2024arXiv241003228L, Allan2025arXiv250402578A}. The intermittency of helium detections can likely be attributed to a combination of planetary atmospheric conditions and stellar variability. One key factor is the impact of stellar activity variations. As our results also show, the He I 10830 \AA\ transit depth is highly sensitive to changes in the stellar XUV flux. This means that repeated observations of the same planet could yield varying transit depths, depending on the phase of stellar activity at the time of observation. This variability complicates direct comparisons between different observations, as a single snapshot may not be representative of the true long-term behavior of the atmosphere. Unless HD209458 is significantly less active than the Sun, we would expect transit depths to vary across multiple epochs, reinforcing the need for continued monitoring. In addition to stellar activity, atmospheric properties such as diffusive separation and vertical mixing processes remain uncertain and vary between planets, as well as with stellar activity, even for the same planet. A study of these processes in different environments would provide  broader context for understanding the role of vertical mixing, atmospheric composition, and mass-loss rates in shaping transit observations. However, improved models of stellar XUV fluxes and spectra are also necessary to make such a study more valuable, as uncertainties in the stellar input strongly affect He I 10830 \AA\ predictions. Observationally, simultaneous He I 10830~\AA~and H$\alpha$ transit measurements, along with possible other species, combined with stellar activity indicators would provide necessary insights into the interplay between planetary mass loss and host star variability. These measurements would help distinguish between intrinsic planetary properties and external stellar influences on transit depth variations. Additionally, repeated observations of HD209458b and other planets would be useful in building a time-averaged picture of mass loss, mitigating the effects of stellar activity variability and enabling more robust comparisons across exoplanet systems. Since we are aware of the time limitations of the observatories likely required for this purpose, we advocate for this goal to be achieved for a few well selected systems to start with.

\section{Conclusion}

In this study, we develop a self-consistent model of HD209458b’s full atmosphere, combining results from a multi-species hydrodynamic escape code with results from a photochemical model of the lower and middle atmosphere. By updating key excitation and de-excitation rates for metastable helium, and incorporating non-isothermal structure, stellar activity variations, diffusion, and metals, we identify key processes that shape He I 10830~\AA\ and H$\alpha$ transit depths. Our model reproduces the observed transit signals with a photoelectron heating efficiency of 20–40\% corresponding to a mass-loss rate of $1.9-3\times10^{10}$ g/s and confirms that diffusive separation of helium occurs in the upper atmosphere. These results highlight the fact that interpreting helium absorption trends requires accounting for atmospheric structure and mixing, in addition to stellar variability. Simultaneous or near-simultaneous observations of He I, H$\alpha$, other species, and stellar activity indicators such as X-rays or Ca II H\&K emission will be essential to fully constrain mass loss and atmospheric escape processes across different exoplanet systems. While detailed modeling is required to fully interpret He I 10830~\AA\ transit observations and simultaneous observations of multiple absorbers help to refine models of the upper atmosphere, observations of metastable helium alone still provide valuable insights. They are sensitive to the presence of an extended, escaping atmosphere that has consequences on the temperature and escape rate of the upper atmosphere. Even without other complementary constraints, observations of the He I 10830 \AA\ line can help identify whether escape is likely and guide the plausible range of physical conditions, making them a potentially powerful diagnostic of the state of the upper atmosphere. Simple Parker wind models may offer a basic framework for estimating the parameter space for temperature, mass-loss rate and the H/He ratio, but their assumptions are often too limiting. Self-consistent models are required to meaningfully narrow the solutions and rule out unphysical scenarios. Finally, we emphasize that planet-specific properties—such as gravity, irradiation, and composition—can lead to diverse atmospheric outcomes. Although the inclusion of metals had a limited effect on HD209458b, metal-rich atmospheres may exhibit significantly different thermal and compositional structures, including higher mean molecular weights and altered energy balance in the upper atmosphere.

\vspace{0.5cm}

\noindent A.R.T. and T.T.K acknowledge support by NASA/XRP 80NSSC23K0260. N.L. and L.A. acknowledge support by NSF Grant No. PHY-1912507. C.H. is sponsored by Shanghai Pujiang Program (grant NO. 23PJ1414900). P.L. acknowledges support from the Programme National de Planétologie of CNRS/INSU under the project ``Temperate Exoplanets''.

\software{Astropy \citep{2018AJ....156..123A, 2022ApJ...935..167A}, Matplotlib \citep{Hunter2007}, NumPy and SciPy \citep{2020NatMe..17..261V}, Pandas \citep{mckinney-proc-scipy-2010}}

\appendix

\label{appendix:rate}

\begin{figure}[h!]
    \centering
    \includegraphics[scale=0.4]{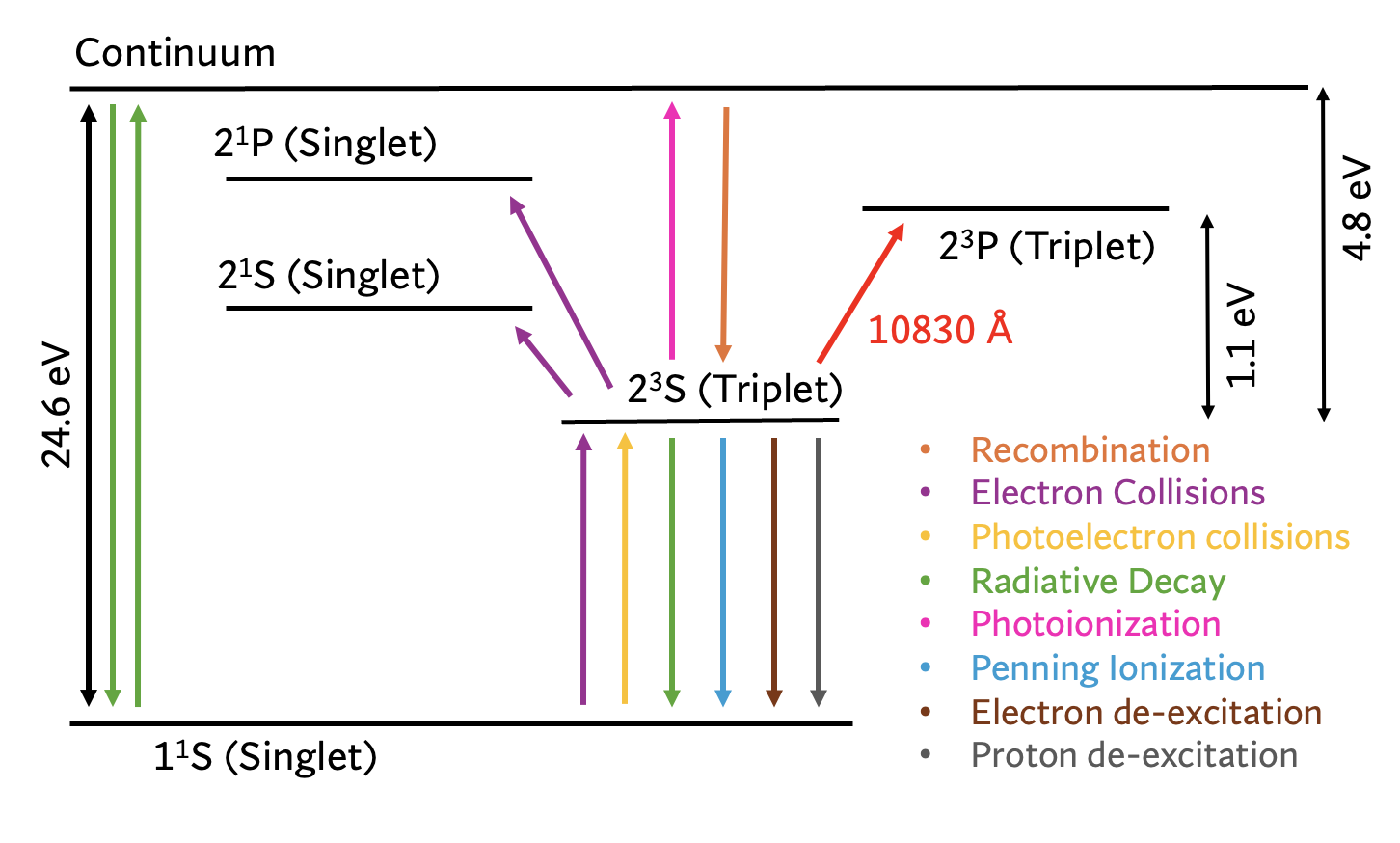}
    \caption{Atomic structure of the helium atom, indicating the radiative and collisional transitions included in our analysis. Adapted from \cite{oklop2018ApJ...855L..11O}.}
    \label{fig:balance}
\end{figure}

Here, we elaborate on how we obtained some of the revised de/excitation rates for metastable helium in this work. Figure \ref{fig:balance} illustrates a schematic overview of all the radiative and collisional transitions considered in our calculation. The radiative transition between the metastable state and the 2$^3$P state, responsible for the 10830 \AA\ line, is not included in the modeled reaction network. This omission is due to the fact that this process does not significantly depopulate the metastable state; the atom in the 2$^3$P state simply decays back into the 2$^3$S state, preserving the triplet configuration. The resulting emission is isotropic and, therefore, does not significantly affect the observed transit depth.

\begin{figure}[h!]
\centering
\includegraphics[width=.4\textwidth]{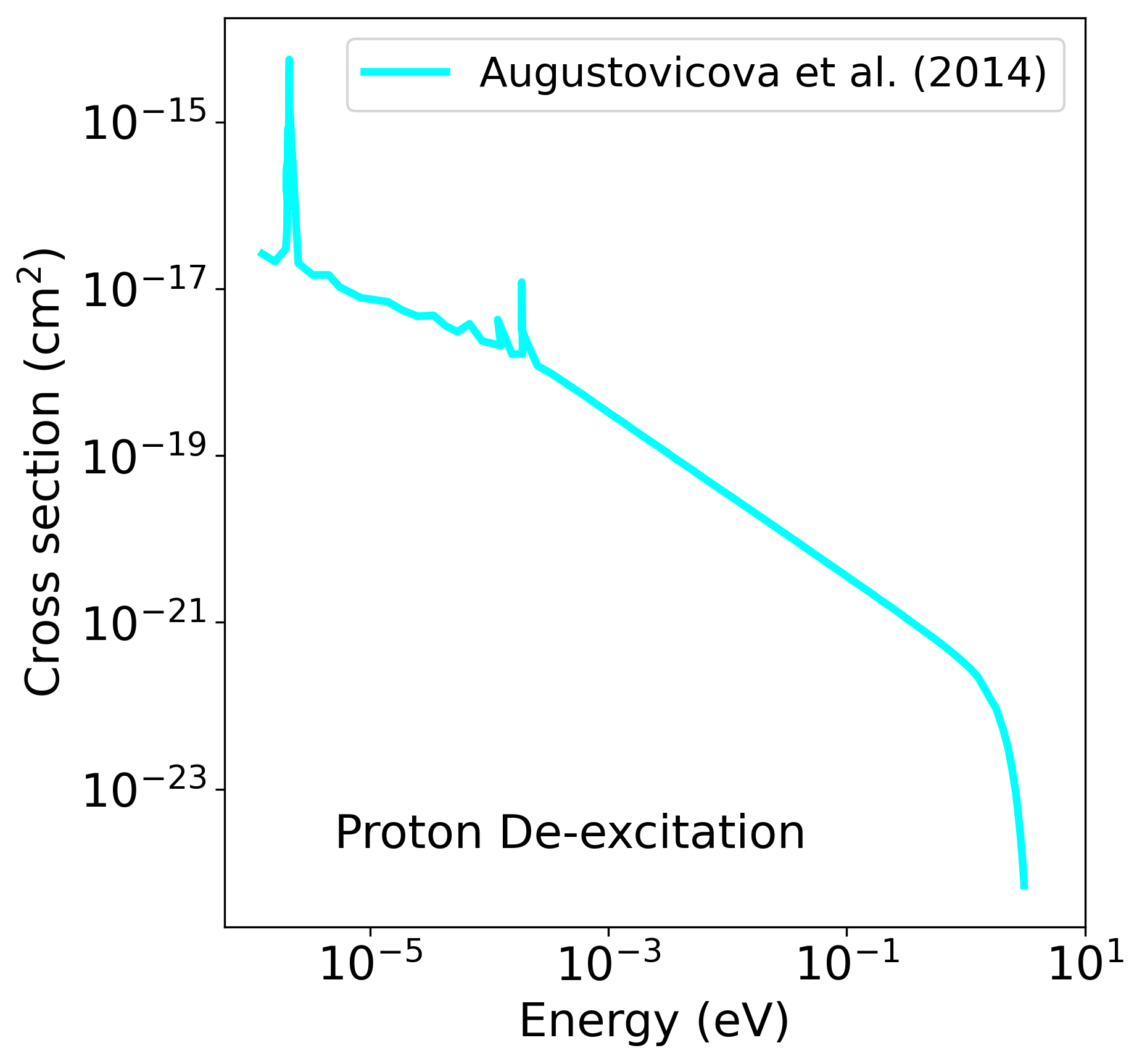}
\includegraphics[width=.35\textwidth]{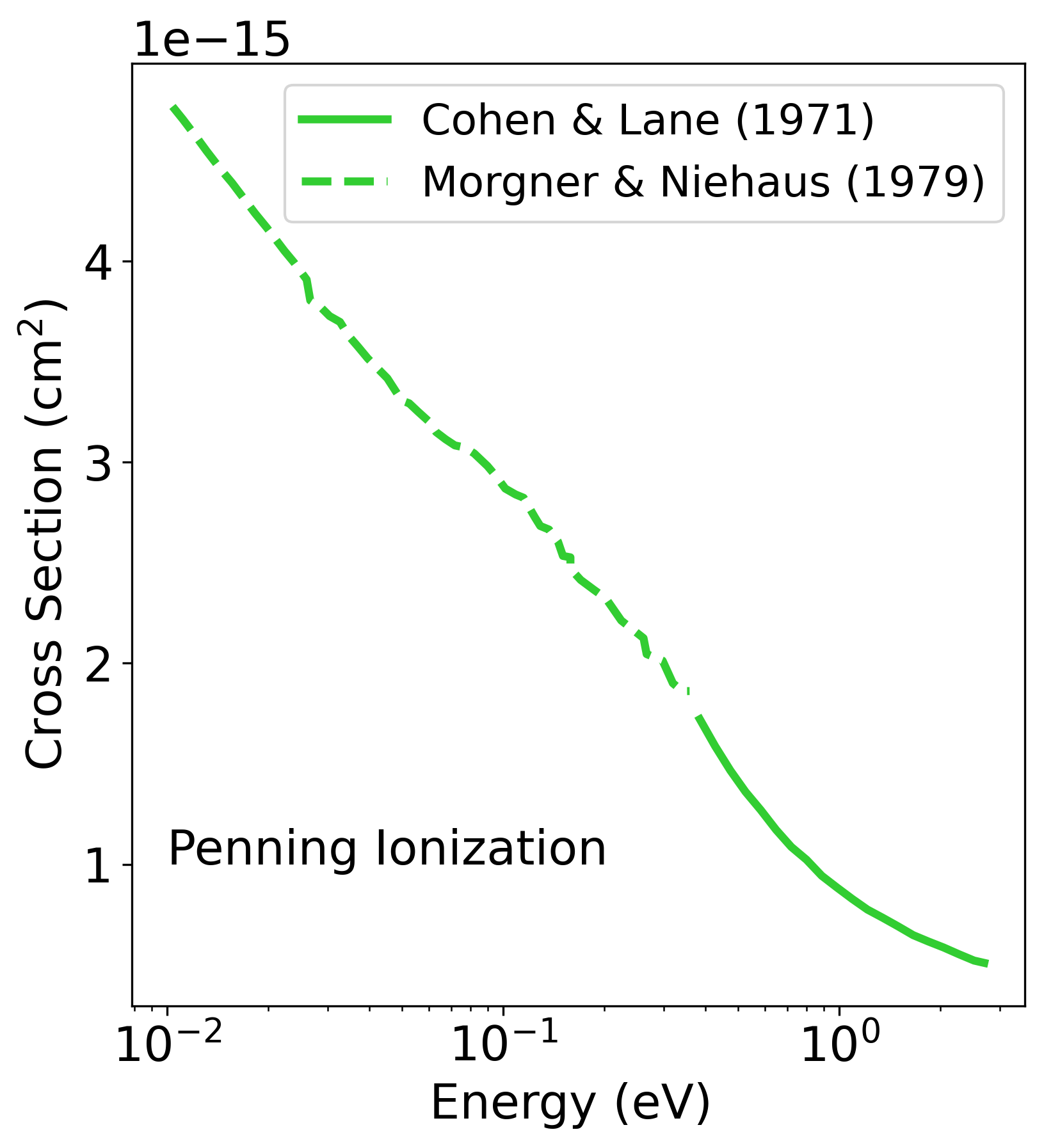}
\includegraphics[width=.4\textwidth]{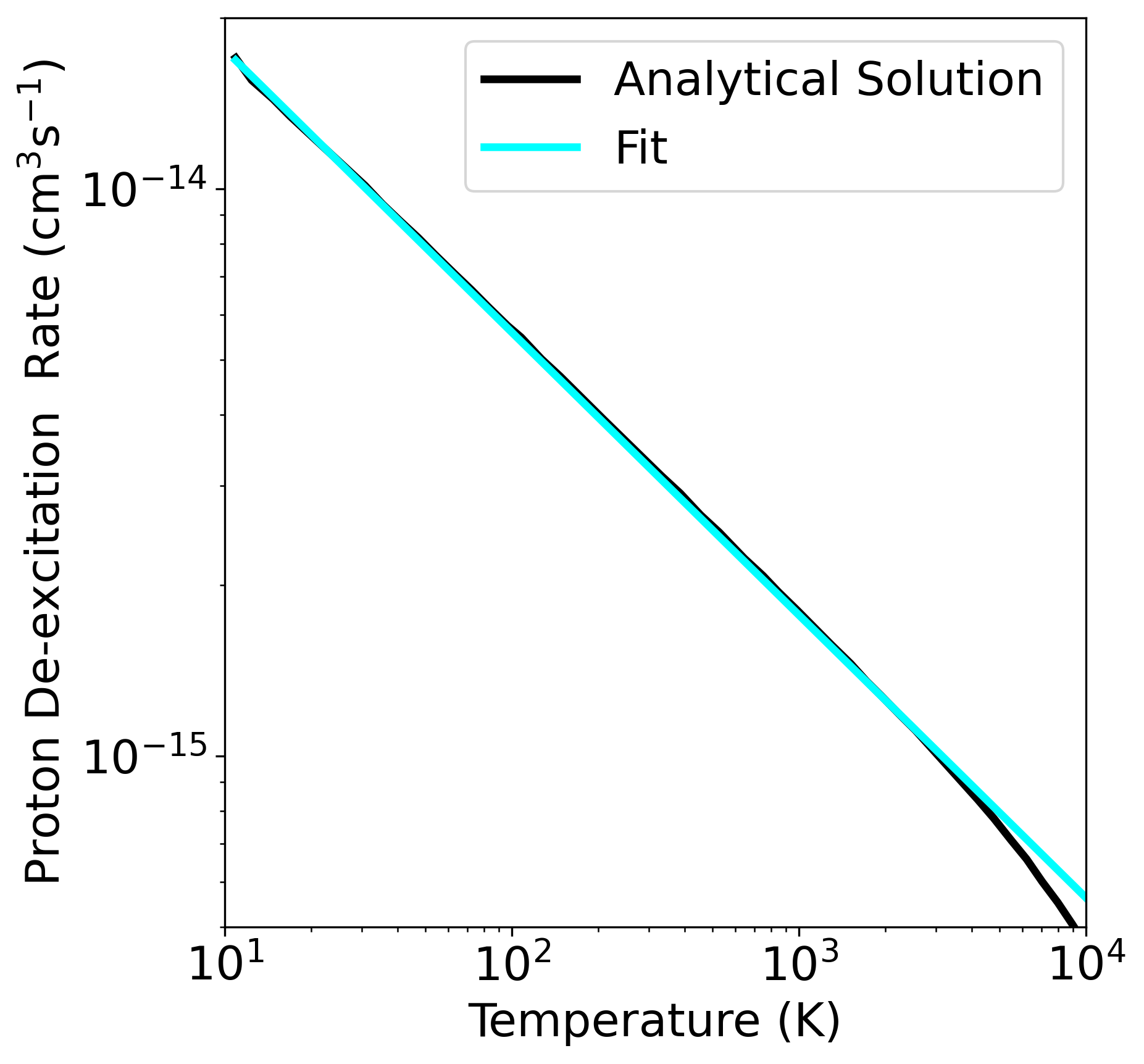} 
\includegraphics[width=.39\textwidth]{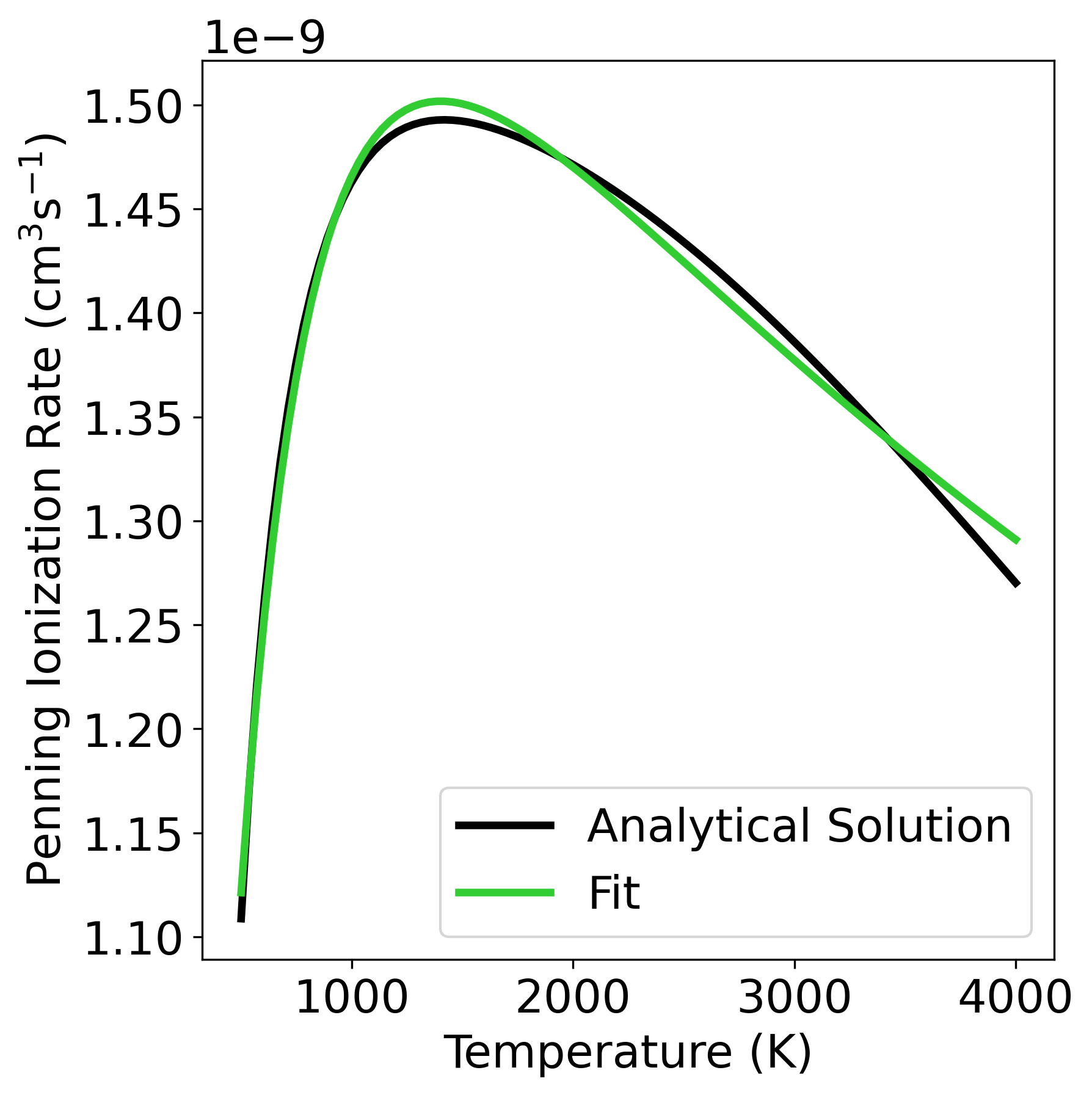}
\caption{Cross sections versus energy for  for proton de-excitation (\textit{top left}, \cite{AUGUSTOVICOVA201427}), and penning ionization (\textit{top right}, \cite{1971CPL....10..623C, 1979JPhB...12.1805M}). Reaction rates (cm$^3$ s$^{-1}$) versus temperature (K) for proton de-excitation (\textit{bottom left}), and penning ionization (\textit{bottom right}). The numerical solution for each rate is plotted in black, while the blue or green curves fit the rate.}
\label{fig:ratefits}
\end{figure}

For Penning ionization and proton de-excitation, we use cross sections to calculate a temperature-dependent rate to fit an equation already implemented into our KPP routine for simplicity using the Python package \texttt{scipy.optimize.curve\_fit} and implement it into our atmosphere code. Our cross sections are shown and cited in Figure \ref{fig:ratefits}. The per volume per time reaction rate, $N_{12}$, is given by \citep[e.g.,][]{Gombosi_1994},
\begin{equation}
    N_{12} = 4 \pi n_1  n_2 \left (\frac{m_{12}}{2 \pi k T}\right)^\frac{3}{2} \int_0^{\infty} v^3 \sigma (v) exp \left(- \frac{m_{12}v^2}{2kT}\right) dv
\end{equation}
where $n_1$ and $n_2$ are the number densities of the reactants, $v$ is velocity, $m_{12}$ is reduced mass, and $\sigma$ is the cross-section. Using this method, we derive a numerical solution for the temperature-dependent reaction rate and implement it into KPP by curve-fitting the data to an equation already implemented into our KPP routine for simplicity using the Python package \texttt{scipy.optimize.curve\_fit}. Figure \ref{fig:ratefits} shows the numerical solutions compared with the curve fits we have implemented into our upper atmosphere code.

\begin{figure}[h!]
\centering
\includegraphics[scale =0.35]{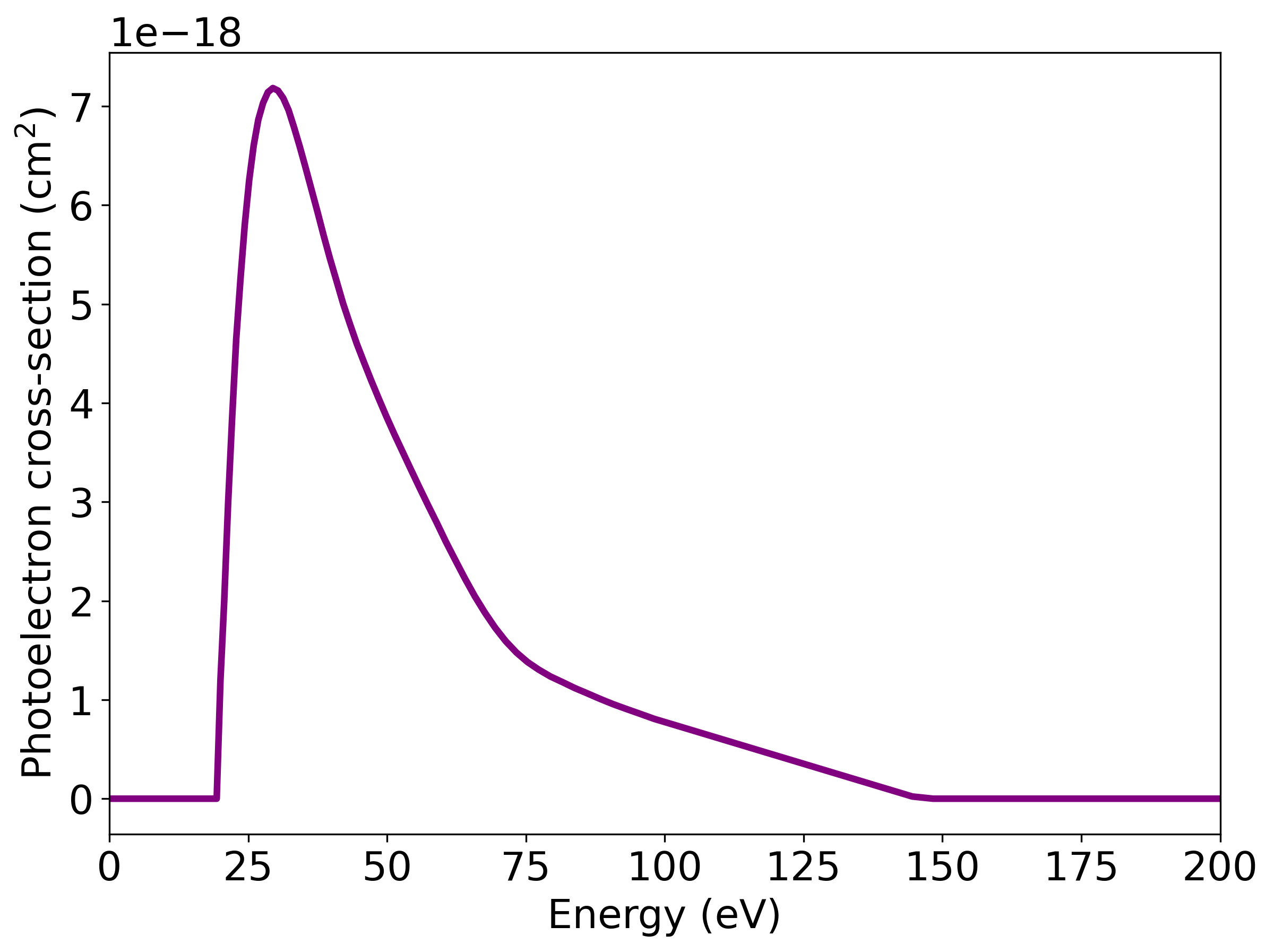}
\caption{Photoelectron cross-section as a function of energy for the impact reaction between photoelectrons and helium. }
\label{fig:photoecs}
\end{figure}

For the photoelectron calculation described in Section \ref{sec:chem}, for the cross-section between a photoelectron and He I (2$^3$S), we use discrete values given by \citep{1977JPhB...10.3741D} between the threshold energy at 19.82 eV and 80 eV, and an interpolated exponential function is used to describe the cross-section beyond 80 eV, calculated using \texttt{scipy.interpolate.interp1d}. Once the interpolation reaches 0 at 145 eV, we assume the cross-section is zero at all higher energies. Figure \ref{fig:photoecs} shows this cross-section versus energy.

\newpage

\bibliography{refs.bib}{}
\bibliographystyle{aasjournal}

\end{document}